\documentclass[pre,showpacs]{revtex4}
\usepackage{amsmath}
\usepackage{graphicx,graphics,color,epsfig}
\usepackage{epsfig}
\begin{document}

\title{Quasispecies Theory for Horizontal Gene Transfer and Recombination}
\author{Enrique Mu\~{n}oz$^{1}$, Jeong-Man Park$^{1,2}$, 
and Michael W. Deem$^{1}$}

\affiliation{
\hbox{}$^1$Department of Physics \& Astronomy,
Rice University, Houston,Texas 77005--1892, USA\\
\hbox{}$^2$Department of Physics, The Catholic University of
Korea, Bucheon 420-743, Korea}


\begin{abstract}
We introduce a generalization of the parallel, or Crow-Kimura, and Eigen
models of molecular evolution to represent the exchange of
genetic information between individuals in a population.
We study the effect of different schemes of genetic recombination on
the steady-state mean fitness and distribution of individuals in the
population, through an analytic field theoretic mapping.
We investigate both horizontal gene transfer from a population and
recombination between pairs of individuals. Somewhat surprisingly,
these nonlinear generalizations of quasi-species theory to
modern biology are analytically solvable. 
For two-parent recombination, we find two selected phases, one of which is
spectrally rigid.
We present exact analytical formulas for
the equilibrium mean fitness of the population, in terms
of a maximum principle, which are generally applicable to
any permutation invariant replication rate function.
For smooth fitness
landscapes, we show that when positive 
epistatic interactions are
present, recombination or horizontal
gene transfer introduces a mild load 
against selection. Conversely, if the fitness landscape exhibits
negative epistasis, horizontal gene transfer or recombination
introduce an advantage by enhancing selection towards the fittest
genotypes. These results prove that the mutational deterministic
hypothesis holds for quasi-species models.
For the discontinuous single sharp peak fitness landscape, 
we show that horizontal gene transfer has no effect on the fitness, while
recombination decreases the fitness, for both the parallel and the
Eigen models.
We present numerical and analytical results as well as phase
diagrams for the different cases.  
\end{abstract}

\maketitle

\section{Introduction}

It has been argued that genetic recombination provides a mechanism to
speed up evolution, at least in finite populations \cite{cohen05}.
Moreover, it has been suggested that recombination may provide
a way to escape from the phenomenon of ``Muller's ratchet'' \cite{Muller64},
or suboptimal fitness 
characteristic of finite populations with asexual reproduction. 
In bacteria, it has been proposed \cite{Lawrence97}
that horizontal gene
transfer allows for the gradual emergence of modularity, through the
formation of gene clusters and their eventual organization into operons. 
In in-vitro systems, protein engineering protocols by directed evolution
incorporate genetic recombination in the form of DNA shuffling
\cite{Patten97,Lutz00} to speed up the search for desired features such
as high binding constants among combinatorial libraries of mutants.
 
\begin{figure}[tbp]
\centering
\epsfig{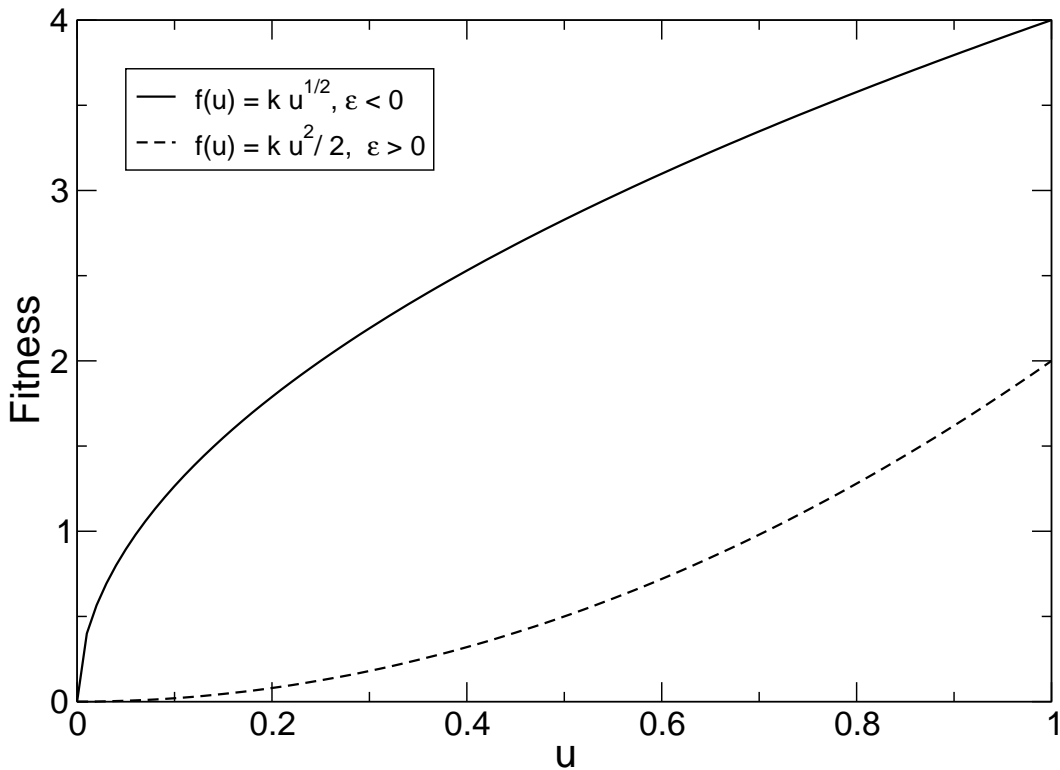}
\caption{Convention for the sign of epistasis, $\epsilon$. In the
figure are represented two smooth fitness landscapes, as a function of
$u = 2 l/N -1$, with $N$ the total length of the (binary) genetic sequences
and $0\leq l \leq N$ the number of beneficial mutations (number of '+' spins)
along the sequence. In this representation, positive (synergistic) epistasis
$\epsilon > 0$ corresponds to a positive curvature $f^{''}(u) > 0$, while
negative (antagonistic) epistasis $\epsilon < 0$ corresponds to a negative
curvarture $f^{''}(u) < 0$ \cite{Phillips,Azevedo06,Arjan07}. 
The examples shown are a quadratic fitness
landscape $f(u) = k u^2 /2$ (dashed line), with positive curvature and
$\epsilon>0$, and a square-root fitness landscape $f(u) = k\sqrt{u}$
(solid line), with negative curvature and $\epsilon < 0$. We set $k = 4.0$ in 
both examples. 
}
\label{fig0}
\end{figure}

Besides these inherently dynamical effects, it remains a matter of
debate 
if the exchange of genetic-encoding elements 
provides a long-term advantage to 
an infinite population in a nearly static environment. Indeed,
it is argued that \cite{Otto02} when advantageous 
genetic associations
have been generated as a result of selection in a given environment,
further random 
recombination is likely to disrupt these associations, thus decreasing
the overall fitness. This argument is less cogent
if we consider that recombination and horizontal gene transfer
preserve the modular structure of the genetic material \cite{Lawrence97}.
That is, entire operational and functional units are recombined, rather
than random pieces. It has also been proposed that for recombination
to introduce an advantage in infinite populations, 
negative linkage disequilibrium is
required \cite{Arjan07,Misevic06,Kondrashov82,Kondrashov93}. 
This situation means that particular allele
combinations are present in the population at a lower frequency
than predicted by chance. Negative linkage disequilibrium 
can result as a consequence of negative
epistasis: alleles with negative contributions to the fitness interact
synergistically, increasing their deleterious effect when
combined, 
and alleles with positive contributions to the fitness
interact antagonistically \cite{Azevedo06,Arjan07,Phillips}, see
Fig.\ \ref{fig0}. Under negative epistasis, the mutational
deterministic hypothesis 
\cite{Kimura66,Kondrashov82,Kondrashov88,Kondrashov93,Arjan07,Azevedo06,Kouyos07} 
postulates that recombination promotes
a more efficient removal of deleterious mutations, by bringing
them together into single genomes, 
and hence facilitating selection \cite{Kimura66,Rice01}
to discard those genotypes with low fitness.
It has been argued that the negative linkage
disequilibrium generated by negative epistatic interactions
is a factor to promote the evolution of recombination in nature 
\cite{Arjan07,Kouyos07,Kouyos06}, and conversely that recombination
may act as a mechanism to evolve epistasis 
\cite{Liberman05,Liberman07,Liberman08}. This later statement
is controversial, since it is intuitive that recombination
should contribute to weaken correlations between different genes
\cite{Malmberg77}. Despite these theoretical
arguments, experimental studies seem to indicate that negative
epistasis is not so common in nature \cite{Bonhoeffer04,Wloch01} 
as recombination and, moreover, both negative and positive epistasis
may coexist as different fitness components \cite{Arjan07}
within the same genome in natural organisms.   

To address some of these questions, we 
study the effect of transferring genetic information between different
organisms in an infinite population. We choose the conceptual
framework of ``quasi-species'' theory, represented by two
classical models of molecular evolution: the 
Eigen \cite{Eigen71,Eigen88,Eigen89,Biebricher05} model
and the parallel, or Crow-Kimura, model \cite{Kimura70,Baake01}. These 
classical models
include the basic processes of mutation, selection, and replication that
occur in biological evolution. Our goal is to solve these two standard
models of quasi-species theory, Crow-Kimura and Eigen, when horizontal
gene transfer or recombination are included. Since horizontal gene
transfer and recombination are essential features of evolutionary
biology, our solutions bring quasi-species theory closer to modern
biology. An operational definition of fitness
is provided in these models by the replication rate, which
is considered to be a function of the genotype. In their simplest
formulation quasi-species models 
consider a static environment, with
a deterministic mapping 
between individual genetic sequences and replication rate.
Both the Eigen \cite{Eigen71,Eigen88} and the parallel, or
Crow-Kimura model \cite{Kimura70}, are formulated in terms of a large
system of differential equations, describing the time evolution
of the relative frequencies of the different sequence types in
an infinite population, a mathematical language that is
common in the field of chemical kinetics \cite{Eigen71,Eigen88}. 
Sequences, representing information carrying molecules such as
RNA or DNA,
are assumed to be drawn from a binary alphabet (e.g. purines/pyrimidines).
The most remarkable property of these classical models is that when 
the mutation rate is below
a critical value 
they exhibit a phase transition in the infinite genome limit 
\cite{Eigen71,Eigen88,Eigen89,Tarazona92,Leuthausser87,Biebricher05,Franz97,Park06,Saakian06}, with
the emergence of a self-organized phase: the quasi-species 
\cite{Eigen71,Eigen88,Eigen89}. This organized phase, characterized
by a collection of nearly neutral mutants rather than by a single 
homogeneous sequence type, is mainly a consequence of the
auto-catalytic character of the evolution dynamics, which
tends to enrich exponentially the proportion of fittest
individuals in the population 
\cite{Eigen71,Eigen88,Eigen89,Biebricher05}. The quasi-species
concept, with its corresponding "error threshold" transition,
has been applied in the interpretation of 
experimental studies in RNA viruses 
\cite{Domingo78,Domingo05,Ortin80,Domingo85}.
In particular, the error-threshold transition has been proposed 
as a theoretical motivation for an antiviral strategy \cite{Eigen02},
termed "lethal mutagenesis", which drives an infecting population
of viruses towards extinction by enhancing their mutation 
rate \cite{Graci07,Loeb99,Loeb00}. It has
been argued, however, that the mechanism for lethal mutagenesis 
possesses a strong ecological component \cite{Bull07}, and that
perhaps the mean population fitness is simply driven negative, and so
the total number of viral particles in an infecting population
decreases in time towards extinction, in contrast with  
error-threshold theories that describe a randomization of the composition
of the quasi-species in genotype space.  

The existence of the error threshold transition has motivated
the attention of theoretical physicists, especially
since it was proved that the quasi-species theory can be exactly
mapped into an 2D Ising spin system \cite{Leuthausser87,Tarazona92},
with a phase transition that is first order for
a sharp peak fitness, and second or higher 
order for smooth fitness functions. 
More recently,
exact mappings into a quantum spin chain \cite{Baake97,Baake98,Saakian06b,
Saakian04a,Saakian04b}
or field theoretic representations \cite{Park06} have been developed. 
Analytical and numerical studies of these systems, in the
large genome limit, are possible when the fitness function
is considered to be permutation invariant 
\cite{Baake97,Baake98,Park06,Park07,Franz97},
or depending on the overlap with several peaks in sequence space 
\cite{Saakian06}. The mapping of the quasi-species models 
into a physical system 
allows for the application of the powerful mathematical techniques
of statistical mechanics, thus obtaining
exact analytical solutions which provide significant
insight over numerical studies \cite{Park06,Saakian06,Saakian06b}.
Most of the existing analytical solutions correspond
to the case when recombination is absent. 
Recombination and horizontal gene transfer have
been studied by computer simulations of
artificial gene networks \cite{Azevedo06} and
digital organisms \cite{Misevic06}, but
relatively few analytical approaches have been reported in the context of
quasi-species theory \cite{Boerlijst96,cohen05,Park07,Jacobi06}.
A numerical study of a mathematical model for viral super-infection
termed uniform crossover, and intermediate between horizontal gene
transfer and recombination, has been reported \cite{Boerlijst96}, 
with
numerical solutions based on relatively short viral sequences (N=15).
More recently, the
effect of incorporating horizontal gene transfer in quasi-species
theory has been studied in terms of the dynamics \cite{cohen05}, reporting
numerical studies and approximate analytical expressions.
Exact analytical expressions for the equilibrium properties
of the population in the presence of horizontal gene transfer
have been derived using the methods of quantum field theory \cite{Park07}. 

In this article, we 
study the effect of introducing different schemes of genetic recombination
in quasi-species theory.
Extending the results in \cite{Park07}, 
we present an exact field theoretical mapping
of the parallel and Eigen models. We remark that field theoretical
methods provide a unique and powerful set of tools 
for the analytical study of dynamical systems, such as
reaction-diffusion \cite{Lee95,Mattis98} 
or birth-death processes \cite{Peliti85}.
In this paper, we employ these theoretical tools to obtain
exact analytical expressions for the equilibrium mean fitness
and average composition of the population, for permutation
invariant but otherwise arbitrary replication rate functions. 
 
In Section 2 we consider the parallel model. We consider horizontal
gene transfer of non-overlapping blocks, as well as of blocks of
random size. We also consider a recombination process producing a 
daughter sequence symmetrically from two parents, as might occur
in viral super- or co-infection. In Section 3,
we study the effect of these different genetic recombination
schemes in the context of the Eigen model. 
In both models, 
recombination leads to two selected phases.  Interestingly, beyond a critical
recombination rate, the distribution of the population becomes independent
of the recombination rate.  Also interesting is that the steady-state
distribution is independent of the crossover probability.

To study the effect of epistasis, whose sign is determined
by the curvature of the fitness landscape (second derivative)
when represented as a function of the Hamming distance with
respect to the wild-type,
we considered two different examples
of smooth fitness functions: a quadratic function, representing
positive epistasis, and a square-root function representing
negative epistasis. We find that, for the quadratic fitness function,
horizontal gene transfer and recombination introduce
a mild load against selection. 
The opposite effect is observed for the square-root fitness, 
that is, horizontal gene transfer and recombination
introduce an advantage by enhancing selection towards fittest
genotypes. This results provide support for the mutational
deterministic hypothesis, which postulates that recombination
should be beneficial for negative epistasis fitness functions, and
deleterious for positive epistasis fitness functions. Moreover, 
we prove analytically in Appendix \ref{appendix12} 
that the mutational deterministic hypothesis
applies for the parallel model in the presence of horizontal gene
transfer. A similar proof is provided in Appendix \ref{appendix12_b}
for the Eigen model. We also show analytically that the
mutational deterministic hypothesis applies for the case of
two-parent recombination, as presented in Appendix \ref{appendix14}
for the parallel model, and in Appendix \ref{appendix15} for the
Eigen model.  

The effect of recombination becomes negligible
for discontinuous fitness landscapes, such as a single sharp
peak.
For all these cases,
we present exact analytical expressions that determine the phase structure
of the population at steady state. Results are explicit for any
microscopic fitness function: Eqs.\ (14), (31), and (62--63) for the parallel
model and Eqs.\ (82), (93), and (106--107) 
for the Eigen model. We evaluate these
expressions for three permutation invariant fitness functions: sharp peak, 
quadratic, and square root 
for the two common forms of quasi-species theory, parallel
and Eigen: Eqs.\ (22), (23), (33), (34), (68), (71), (85--87), (96--98),
(112), and (113). We also present
numerical tests supporting our analytical equations.

\section{The parallel model}

We consider a generalization \cite{Park07}
of the parallel, or Crow-Kimura \cite{Kimura70}, model
to take into account the transfer of genetic material between
pairs of individuals in an infinite population.
\begin{eqnarray}
\frac{d q_{i}}{dt}= r_{i} q_{i}+
\sum_{k=1}^{2^{N}}\mu_{ik}q_{k}+
\nu N \frac{\sum_{k,l}R^{i}_{kl}q_{k}q_{l}}{\sum_{k}q_{k}}
-\nu N q_{i}
\label{eq1}
\end{eqnarray}
Here, $q_{i}$ 
represents the (unnormalized) frequency of the sequence type 
$S_{i}=(s_{1}^{i},s_{2}^{i},\ldots,s_{N}^{i})$, with $s_{j}^{i}=\pm 1$,
for $1\leq i \leq 2^{N}$ and $1\leq j \leq N$. 
The normalized frequencies are obtained from
$p_{i}=q_{i}/\sum_{j=1}^{2^{N}}q_{j}$.
In Eq.\ (\ref{eq1}), $r_{i}$ is the replication rate of sequence $S_{i}$.
It is given that $r_{i}=N f\left(\frac{1}{N}\sum_{j=1}^{N}s_{j}^{i}\right)$.
The mutation rate from sequence $S_{j}$ into $S_{i}$ is
$\mu_{ij}=\mu \delta_{d_{ij},1}-N\mu\delta_{d_{ij},0}$. 
The Kronecker delta in this
expression ensures that mutations involve a single base substitution per
unit time (generation). Genetic recombination processes between pairs
of sequences in the population are represented by the nonlinear 
term. They are considered to occur with an overall rate $\nu$, while
the coefficient $R_{kl}^{i}$ represents the probability that a pair
of parental sequences $S_{k}$, $S_{l}$ produces an offspring $S_{i}$.
Depending on the particular recombination mechanism,
some of these coefficients will be identically zero. Also, these coefficients
must satisfy the condition
$\sum_{i=1}^{2^{N}}R_{kl}^{i}=1$,  $\forall$ $1\leq k,l \leq 2^{N}$.

For this generic process, we will present the analytical solutions
for the steady-state mean fitness by considering different schemes of 
genetic recombination.     

\subsection{Horizontal gene transfer of non-overlapping 
blocks}

In this recombination scheme, we consider the exchange of
blocks of genetic material between pairs of individuals.
We consider these blocks to be non-overlapping
in the parental sequences, and of a fixed size $\bar{M}$. Thus,
each sequence is made of $N/\bar{M}$ blocks.
The recombination coefficients in the differential Eq.\ (\ref{eq1}) 
are given for this horizontal gene transfer process by
\begin{eqnarray}
R_{kl}^{i}=\sum_{b=0}^{N/\bar{M}-1}\prod_{j_{b}=\bar{M}b+1}^{\bar{M}(b+1)}
\left(\frac{1+s_{j_{b}}^{l}s_{j_{b}}^{i}}{2}\right)
\prod_{j \ne \{j_{b}\}}^{N}
\left(\frac{1+s_{j}^{k}s_{j}^{i}}{2}\right).
\label{eq2}
\end{eqnarray}
Here, $0\leq b \leq N/\bar{M}-1$ represents the block index, while
$\bar{M}b+1\leq j_{b} \leq \bar{M}(b+1)$ 
represents the site index within block $b$.

\begin{figure}[tbp]
\centering
\epsfig{file=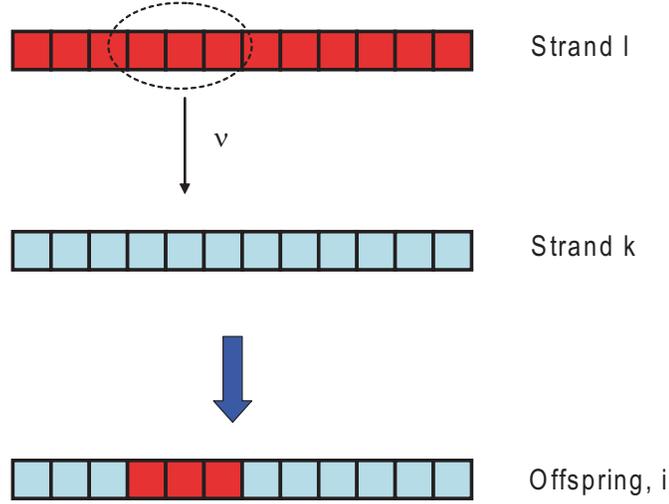,scale=0.6,clip=}
\caption{Pictorial representation of the horizontal gene transfer
process considered.}
\label{fig1}
\end{figure}

Generalizing the method presented in \cite{Park07}, we write the
non-linear term as 
\begin{eqnarray}
\frac{\sum_{l}q_{l}R_{kl}^{i}}{\sum_{m}q_{m}}=
\sum_{b=0}^{N/\bar{M}-1}\left\langle \prod_{j_{b}=\bar{M}b+1}^{\bar{M}(b+1)}
\left(\frac{1+s_{j_{b}}^{l}s_{j_{b}}^{i}}{2}\right)\right\rangle
\prod_{j\ne\{j_{b}\}}^{N}\left(\frac{1+s_{j}^{k}s_{j}^{i}}{2} \right).
\label{eq3}
\end{eqnarray}
Here, $\langle A_{l} \rangle = \sum_{l}q_{l}A_{l}/\sum_{m}q_{m}$
is a population average. At steady state, this average is independent
of the value of $b$, due to the symmetry of the fitness function.

The variance of the composition $u^{l}=\frac{1}{N}\sum_{j=1}^{N}s_{j}^{l}$
is given by $\frac{1}{N^{2}}\sum_{j,j'=1}^{N}
\langle\delta s_{j}^{l}\delta s_{j'}^{l}\rangle$. In the absence of
recombination or horizontal gene transfer this variance is 
$\mathcal{O}(N^{-1})$, which implies correlations along the sequence
are $\mathcal{O}(N^{-1})$ \cite{Park06}. We expect the same scaling of the
variance in the presence of recombination or horizontal gene transfer. 
Therefore, we introduce the  factorization
\begin{eqnarray}
\left\langle\prod_{j_{b}=\bar{M}b+1}^{\bar{M}(b+1)}
\frac{1+s_{j_{b}}^{l}s_{j_{b}}^{i}}{2}
\right\rangle
& \sim & \prod_{j_{b}=\bar{M}b+1}^{\bar{M}(b+1)}\left\langle 
\frac{1+s_{j_{b}}^{l}s_{j_{b}}^{i}}{2}
\right\rangle +\mathcal{O}(\bar{M}/N)\nonumber\\
& = & \prod_{j_{b}=\bar{M}b+1}^{\bar{M}(b+1)}\left(\delta_{s_{j_{b}}^{i},+1}
\frac{1+u(j_{b})}{2}
+\delta_{s_{j_{b}}^{i},-1}\frac{1-u(j_{b})}{2}\right)\nonumber\\
\label{eq4}
\end{eqnarray}
which becomes exact in the $N\rightarrow\infty$ limit.
Here, $u(j_{b})=\sum_{l}q_{l}s_{j_{b}}^{l}/\sum_{m}q_{m}$ is
the average base composition at site $j_{b}$.

We are interested in the long time behavior of the system,
when the average base composition 
becomes independent of time and position $u(j)\sim u$.
Thus, in the formalism of spin Boson operators \cite{Park07} 
$\vec{\hat{a}}(j)=(\hat{a}_{1}(j),\hat{a}_{2}(j))$, we define
the recombination operator describing this recombination term
by  
\begin{eqnarray}
\hat{R}=\frac{1}{N}\sum_{b=0}^{N/\bar{M}-1}
\left[\prod_{j_{b}=\bar{M}b+1}^{\bar{M}(b+1)}
[\rho_{+}\hat{a}_{1}^{\dagger}(j_{b})+
\rho_{-}\hat{a}_{2}^{\dagger}(j_{b})][\hat{a}_{1}(j_{b})+
\hat{a}_{2}(j_{b})]-\hat{I}\right]
\label{eq5}
\end{eqnarray}
Here, $\hat{I}$ is the identity operator.
The coefficients $\rho_{\pm}=(1\pm u)/2$ represent \cite{Park07} the
steady-state probability (per site) of having a ``+1'' or a ``-1''.
Defining the matrix
\begin{eqnarray}
D=\left(\begin{array}{cc}\rho_{+} & \rho_{+} \\ \rho_{-} 
& \rho_{-} \end{array}\right),
\label{eq6}
\end{eqnarray}
the recombination operator in Eq.\ (\ref{eq5}) can be 
expressed as
\begin{eqnarray}
\hat{R}=\frac{1}{N}\sum_{b=0}^{N/\bar{M}-1}
\left[\prod_{j_{b}=\bar{M}b+1}^{\bar{M}(b+1)} 
\vec{\hat{a}}^{\dagger}(j_{b})D\vec{\hat{a}}(j_{b})
-\hat{I}\right].
\label{eq7}
\end{eqnarray}

\subsubsection{The Hamiltonian}
Considering the recombination operator in Eq.\ (\ref{eq7}), we formulate the 
Hamiltonian describing the system 
\begin{eqnarray}
-\hat{H} =  N f\left[\frac{1}{N}\sum_{j=1}^{N}\vec{\hat{a}}^{\dagger}(j)
\sigma_{3}\vec{\hat{a}}(j)\right]+\mu\sum_{j=1}^{N}[\vec{\hat{a}}^{\dagger}(j)
\sigma_{1}\vec{\hat{a}}(j)-\hat{I}] +  \nu\sum_{b=0}^{N/\bar{M}-1}\left[\prod_{j_{b}=\bar{M}b+1}^{\bar{M}(b+1)}
\vec{\hat{a}}^{\dagger}(j_{b})D\vec{\hat{a}}(j_{b})
-\hat{I}\right].
\label{eq8}
\end{eqnarray}
Here, $\sigma_{3}=\left(\begin{array}{cc}1&0\\0&-1\end{array}\right)$ and
$\sigma_{1}=\left(\begin{array}{cc}0&1\\1&0\end{array}\right)$ are the Pauli
matrices.
We introduce a Trotter factorization
\begin{eqnarray}
e^{-\hat{H}t}=\lim_{M\rightarrow\infty}\int\left[\mathcal{D} \vec{z}^{*}
\mathcal{D} \vec{z}\right]
|\vec{z}_{M}\rangle\left(\prod_{k=1}^{M}\langle \vec{z}_{k}|
e^{-\epsilon \hat{H}}|\vec{z}_{k-1}\rangle \right)\langle \vec{z}_{0}|.
\label{eq9}
\end{eqnarray}
As shown in Appendix \ref{appendix1}, the partition function that
gives the mean population fitness is
\begin{eqnarray}
Z=\int\left[\mathcal{D}\bar{\xi}\mathcal{D}\xi
\mathcal{D}\bar{\phi}\mathcal{D}\phi\right]e^{-S\left[\bar{\xi},\xi,
\bar{\phi},\phi\right]} 
\sim e^{N f_m t} .
\label{eq20}
\end{eqnarray}
Here, the action in the continuous time limit is
\begin{eqnarray}
S\left[\bar{\xi},\xi,\bar{\phi},\phi\right]&=&-N\int_{0}^{t}dt\left[
-\bar{\xi}\xi-\bar{\phi}\phi-\mu
-\frac{\nu}{\bar{M}}+f(\xi)+\frac{\nu}{\bar{M}}\phi^{\bar{M}}\right]
-N\ln Q
\label{eq21}.
\end{eqnarray}

\subsubsection{The saddle point limit}
In the $N\rightarrow\infty$ limit, the saddle point is exact and we 
obtain an analytical expression
for the partition function Eq.\ (\ref{eq20}). We look for the steady-state
solution, when the fields become independent of time, $\xi_{c}$, 
$\bar{\xi}_{c}$, $\phi_{c}$, $\bar{\phi}_{c}$.
The trace defined by Eq.\ (\ref{eq19}) in the long time saddle-point
limit becomes
\begin{eqnarray}
\lim_{t\rightarrow\infty} \frac{\ln Q_{c}}{t}= 
\frac{\bar{\phi}_{c}}{2}+\left[\bar{\xi}_{c}
(\bar{\xi}_{c}+u\bar{\phi}_{c})+(\mu+\bar{\phi}_{c}/2)^{2}\right]^{1/2}
\label{eq22}
\end{eqnarray}
Hence, the saddle-point action is
\begin{eqnarray}
\lim_{N,t\rightarrow\infty}\frac{\ln Z}{N t} &=& 
\lim_{t\rightarrow\infty} \frac{-S_{c}}{N t}\nonumber\\
= f_{m} &=&
\max_{\xi_{c},\bar{\xi}_{c},\phi_{c},\bar{\phi}_{c}}
\left\{f(\xi_{c})-\bar{\xi}_{c}
\xi_{c}
-\bar{\phi}_{c}\phi_{c}-\mu-\frac{\nu}{\bar{M}}
+\frac{\nu}{\bar{M}}\phi_{c}^{\bar{M}}\right.\nonumber\\
&&+
\left.
\frac{\bar{\phi}_{c}}{2}+\left[\bar{\xi}_{c}
(\bar{\xi}_{c}+u\bar{\phi}_{c})+(\mu+\bar{\phi}_{c}/2)^{2}\right]^{1/2} 
\right\}.
\label{eq23}
\end{eqnarray}
As shown in Appendix \ref{appendix2}, the mean fitness of the population is
\begin{eqnarray}
f_{m}  = \max_{-1\leq\xi_{c}\leq 1}\bigg\{f(\xi_{c})-\mu-\frac{\nu}{\bar{M}}
+\frac{\nu}{\bar{M}}[\phi_{c}(\xi_{c})]^{\bar{M}} + \mu\sqrt{1-\xi_{c}^{2}}
\frac{\sqrt{1-u^{2}}(1+\frac{\nu}{2\mu}[\phi_{c}(\xi_{c})]^{\bar{M}-1})}
{\left[\left(1+\frac{\nu}{2\mu}(1-u^{2})[\phi_{c}(\xi_{c})]^{\bar{M}-1} 
\right)^{2}-u^{2}\right]^{1/2}} \bigg\}.
\nonumber\\
\label{eq31}
\end{eqnarray}
Here, $\phi_c$ is given by Eq.\ (\ref{eq30}), and
the surplus $u$ is obtained through the self-consistency condition
$f_{m}=f(u)$.
Equation (\ref{eq31}) represents an exact analytical expression
for the mean fitness of an infinite population experiencing
horizontal gene transfer. This expression is valid for an arbitrary, permutation
invariant replication rate $f(u)$.

It is worth to notice that Eq.\ (\ref{eq31}) is a natural generalization
of the single-site horizontal gene transfer
 process described in \cite{Park07}. 
Indeed, specializing the Eqs.\ (\ref{eq30}) and (\ref{eq31}) to the
particular case $\bar{M}=1$, after some algebra, we obtain
\begin{eqnarray}
f_{m}(\bar{M}=1) = \max_{-1\leq\xi_{c}\leq 1}\bigg\{f(\xi_{c})-\mu-
\frac{\nu}{2}+\frac{\nu u}{2}\xi_{c}
 + \sqrt{1-\xi_{c}^{2}}
\left[\left(\mu+\frac{\nu}{2}\right)^{2}-\left(\frac{u\nu}{2}\right)^{2}
\right]^{1/2} \bigg\},
\label{eq32}
\end{eqnarray}
which reproduces the analytical result in \cite{Park07}.

\subsubsection{Numerical tests and examples}
For numerical calculations, it is convenient to reformulate Eq.\ (\ref{eq1})
in terms of the fraction of the population at a distance $l$ from the
wild type, 
$P_{l}=\sum_{j\in\mathcal{C}_{l}}p_{j}$. Here, $\mathcal{C}_{l}$ is the class
of sequences with $l$ number of ``-1'' sites. The number of
sequences within this class is $\left(\begin{array}{c}N\\l\end{array}\right)$.

As an example, for the case $\bar{M}=3$, the differential equation representing
the time evolution of the probability distribution of classes within
an infinite population of binary sequences is
\begin{eqnarray}
\frac{dP_{l}}{dt} & = &  N\left[f(2l/N-1)-\sum_{l'=0}^{N}P_{l'}f(2l'/N-1)
-\mu \right]P_{l}
 +  \mu N \left[\frac{N-l+1}{N}P_{l-1}
+ \frac{l+1}{N}P_{l+1}\right]\nonumber\\
&& + 
\frac{\nu}{3}N\left\{\rho_{-}^{3}g_{3}(N-l+3)P_{l-3} +  
[\rho_{-}^{3}h(N-l+2)+3\rho_{-}^{2}\rho_{+}g_{3}(N-l+2)]P_{l-2}\right .
\nonumber\\
&& + 
\left . [\rho_{-}^{3}h(l-1)+3\rho_{-}\rho_{+}^{2}g_{3}(N-l+1)+
3\rho_{-}^{2}\rho_{+}h(N-l+1)]P_{l-1}\right . \nonumber\\
&& + 
\left .   [\rho_{+}^{3}h(N-l-1)+3\rho_{-}^{2}\rho_{+}g_{3}(l+1)+
3\rho_{-}\rho_{+}^{2}h(l+1)]P_{l+1}\right . \nonumber\\
&& + 
\left .   [\rho_{+}^{3}h(l+2)+3\rho_{-}\rho_{+}^{2}g_{3}(l+2)]P_{l+2}
+\rho_{+}^{3}g_{3}(l+3)P_{l+3}\right\} \nonumber \\
&& - \frac{\nu}{3}N\left\{(\rho_{-}^{3}+3\rho_{-}^{2}\rho_{+}+
3\rho_{-}\rho_{+}^{2})g_{3}(N-l)+(\rho_{-}^{3}+3\rho_{-}^{2}\rho_{+}+
\rho_{+}^{3})h(N-l)\right .\nonumber\\
&& + 
\left . (\rho_{-}^{3}+3\rho_{-}\rho_{+}^{2}+\rho_{+}^{3})h(l)+
(\rho_{+}^{3}+3\rho_{+}\rho_{-}^{2}+
3\rho_{-}\rho_{+}^{2})g_{3}(l)\right\}P_{l}\nonumber\\
\label{eq33}
\end{eqnarray}
In writing this equation we have made use of the only $\mathcal{O}(N^{-1})$
correlations between sites, which holds at long time as well as for short
time with suitable initial conditions.
Here, we defined
\begin{eqnarray}
\rho_{\pm}=\frac{1\pm u}{2}
\label{eq34}
\end{eqnarray}
where the average composition is  calculated as
\begin{eqnarray}
u = \sum_{l=0}^{N}\frac{N-2l}{N}P_{l}
\label{eq35}
\end{eqnarray}
and the functions
\begin{eqnarray}
g_{3}(l) &=& \frac{l(l-1)(l-2)}{N(N-1)(N-2)}
\nonumber \\
h(l)&=&3\frac{l(l-1)(N-l)}{N(N-1)(N-2)}
\label{eq36}
\end{eqnarray}

A comparison between the analytical expression Eq.\ (\ref{eq31}) and the
direct numerical solution of the differential Eq.\ (\ref{eq33}) for N = 1002
is presented in Table \ref{tab1}, where the quadratic fitness 
$f(u)=k u^{2}/2$ was considered. We notice that
the analytical method and the numerical solution 
provide the same results within $\mathcal{O}(N^{-1})$, as
expected from the saddle point limit.
\begin{table}
\begin{center}
\caption{Analytical versus numerical results for horizontal gene 
transfer in the parallel (Kimura) model for the quadratic
fitness $f(u)=k u^{2}/2$, with $\bar{M}=3$.
\label{tab1}}
\begin{tabular}{|c|c|c|c|}\hline
$k/\mu$ & $\nu/\mu$ & $u^{\rm{numeric}}$ & $u^{\rm{analytic}}$\\\hline
2.0 & 0.0 & 0.4993 & 0.5000\\
2.0 & 0.5 & 0.4830 & 0.4838\\
2.0 & 1.0 & 0.4668 & 0.4677\\
2.0 & 1.5 & 0.4510 & 0.4519\\\hline
2.5 & 0.0 & 0.5995 & 0.6000 \\
2.5 & 0.5 & 0.5915 & 0.5920\\
2.5 & 1.0 & 0.5838 & 0.5844\\
2.5 & 1.5 & 0.5766 & 0.5772\\\hline
5.0 & 0.0 & 0.7998 & 0.8000\\
5.0 & 0.5 & 0.7988 & 0.7990\\
5.0 & 1.0 & 0.7979 & 0.7981\\
5.0 & 1.5 & 0.7970 & 0.7972\\\hline  
\end{tabular}
\end{center}
\end{table}

The differential equation representing the horizontal gene transfer
of blocks of size $\bar{M}=4$ within an infinite population
of binary sequences is given by
\begin{eqnarray}
\frac{d}{dt}P_{l}& = & 
N\left[f(2l/N-1)-\sum_{l'=0}^{N}P_{l'}f(2l'/N-1)-\mu\right]P_{l}+\mu N
\left[\frac{N-l+1}{N}P_{l-1}+\frac{l+1}{N}P_{l+1}\right]\nonumber\\
&& + \frac{\nu}{4}N\left\{g_{4}(N-l+4)\rho_{-}^{4}P_{l-4}
+
[\rho_{-}^{4}h_{3}(N-l+3)+4\rho_{-}^{3}\rho_{+}g_{4}(N-l+3)]P_{l-3}
\right .\nonumber\\
&& +  \left . [\rho_{-}^{4}h_{2}(l-2)+4\rho_{-}^{3}\rho_{+}h_{3}(N-l+2)
\right.\nonumber\\
&&+
\left.6\rho_{-}^{2}\rho_{+}^{2}g_{4}(N-l+2)]P_{l-2}
 + 
 [\rho_{-}^{4}h_{3}(l-1)+4\rho_{-}^{3}\rho_{+}h_{2}(l-1) \right.\nonumber\\
&&+
\left.
6\rho_{-}^{2}\rho_{+}^{2}h_{3}(N-l+1)+4\rho_{-}\rho_{+}^{3}g_{4}(N-l+1)]P_{l-1}
\right .\nonumber\\
&& + 
\left.  [\rho_{+}^{4}h_{3}(N-l-1)+4\rho_{-}\rho_{+}^{3}h_{2}(l+1)+
6\rho_{-}^{2}\rho_{+}^{2}h_{3}(l+1)+4\rho_{-}^{3}\rho_{+}g_{4}(l+1)]
P_{l+1}\right
.\nonumber \\
&& + 
\left . [\rho_{+}^{4}h_{2}(l+2)+4\rho_{-}\rho_{+}^{3}h_{3}(l+2)+
6\rho_{-}^{2}\rho_{+}^{2}g_{4}(l+2) ]P_{l+2}\right . \nonumber\\
&& + 
\left . [\rho_{+}^{4}h_{3}(l+3)+4\rho_{-}\rho_{+}^{3}g_{4}(l+3)
]P_{l+3}+\rho_{+}^{4}g_{4}(l+4)P_{l+4}\right\}
\nonumber\\
&& -  \frac{\nu}{4}N\left\{[\rho_{-}^{4}+6\rho_{-}^{2}\rho_{+}^{2}+
4\rho_{-}^{3}\rho_{+}+\rho_{+}^{4}]h_{3}(N-l)+
[\rho_{-}^{4}+6\rho_{-}^{2}\rho_{+}^{2}+4\rho_{-}\rho_{+}^{3}
+\rho_{+}^{4}]h_{3}(l)\right .\nonumber\\
&& + 
\left.[4\rho_{-}^{3}\rho_{+}+6\rho_{-}^{2}\rho_{+}^{2}+4\rho_{-}\rho_{+}^{3}
+\rho_{-}^{4}]g_{4}(N-l)+
[4\rho_{-}^{3}\rho_{+}+6\rho_{-}^{2}\rho_{+}^{2}
+4\rho_{-}\rho_{+}^{3}+\rho_{+}^{4}]g_{4}(l)\right. \nonumber\\
&& + 
\left.[\rho_{-}^{4}+4\rho_{-}^{3}\rho_{+}+4\rho_{-}\rho_{+}^{3}+\rho_{+}^{4}
]h_{2}(l)\right\}P_{l}
\label{eq37}
\end{eqnarray}
Here, the parameters $\rho_{\pm}$ and $u$ are defined, as before,
by Eq.\ (\ref{eq34}) and Eq.\ (\ref{eq35}), respectively. 
We also define the functions
\begin{eqnarray}
g_{4}(l)&=&\frac{l(l-1)(l-2)(l-3)}{N(N-1)(N-2)(N-3)}\nonumber \\
h_{3}(l)&=&4\frac{l(l-1)(l-2)(N-l)}{N(N-1)(N-2)(N-3)}\nonumber\\
h_{2}(l)&=&6\frac{l(l-1)(N-l)(N-l-1)}{N(N-1)(N-2)(N-3)}
\label{eq38}
\end{eqnarray}

A comparison between the analytical expression Eq.\ (\ref{eq31}) and the
direct numerical solution of the differential Eq.\ (\ref{eq37}) 
for N = 1002 is presented in Table \ref{tab2}, for the quadratic
fitness $f(u)=k u^{2}/2$. As in the former case, the
numerical and analytical results agree to within $\mathcal{O}(N^{-1})$,
as expected.

\begin{figure}[tbp]
\centering
\epsfig{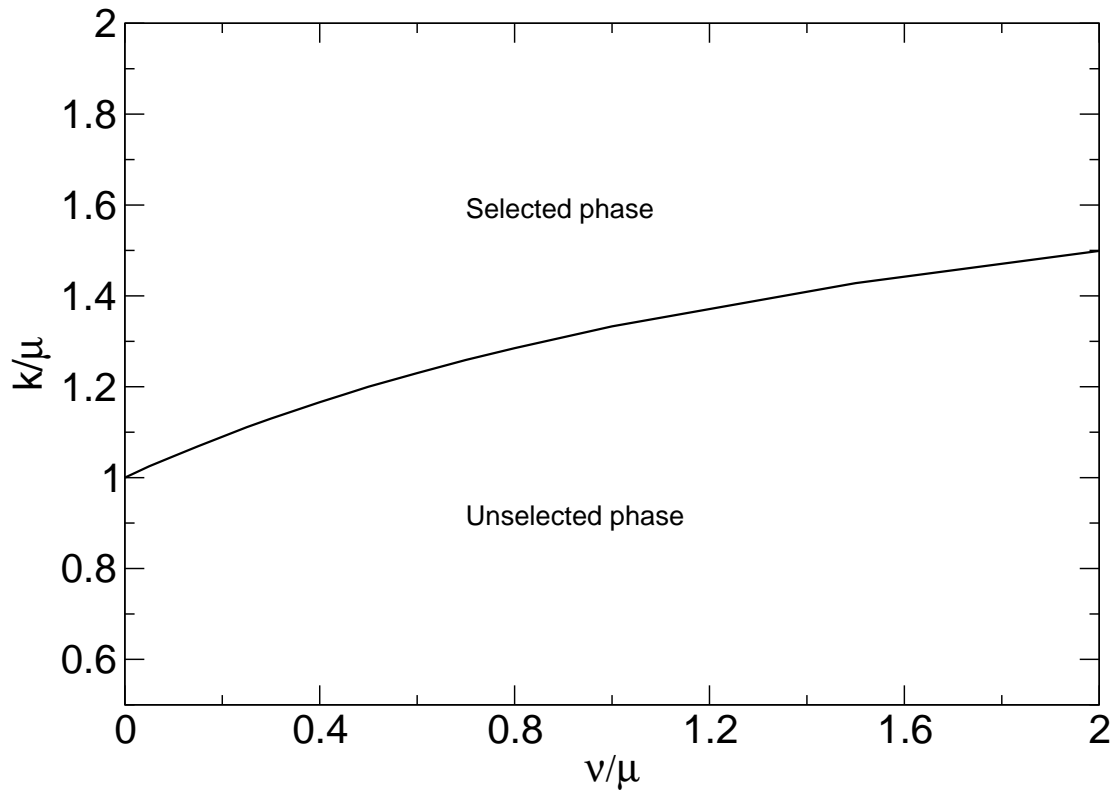}
\caption{Phase diagram of the parallel (Kimura) model for the quadratic
fitness $f(u)=k u^{2}/2$, with horizontal
gene transfer  of non-overlapping blocks of size $\bar{M}$.
The phase boundary of the error threshold phase
transition is given by the curve, and its
shape is independent of the block size $\bar{M}$. In the absence of
horizontal gene transfer, the phase transition occurs at $k/\mu = 1$.}
\label{fig2}
\end{figure}

\begin{table}
\begin{center}
\caption{Analytical versus numerical results for horizontal gene
transfer in the parallel model for the quadratic
fitness $f(u)=k u^{2}/2$, with $\bar{M}=4$.}
\label{tab2}
\begin{tabular}{|c|c|c|c|}\hline
$k/\mu$ & $\nu/\mu$ & $u^{\rm{numeric}}$ & $u^{\rm{analytic}}$\\\hline
2.0 & 0.0 & 0.4993 & 0.5000 \\
2.0 & 0.5 & 0.4832 & 0.4840\\
2.0 & 1.0 & 0.4672 & 0.4680\\
2.0 & 1.5 & 0.4510 & 0.4519\\\hline
2.5 & 0.0 & 0.5995 & 0.6000\\
2.5 & 0.5 & 0.5916 & 0.5921\\
2.5 & 1.0 & 0.5839 & 0.5845\\
2.5 & 1.5 & 0.5766 & 0.5773\\\hline
5.0 & 0.0 & 0.7998 & 0.8000 \\
5.0 & 0.5 & 0.7988 & 0.7990\\
5.0 & 1.0 & 0.7979 & 0.7981\\
5.0 & 1.5 & 0.7970 & 0.7973\\\hline
\end{tabular}
\end{center}
\end{table}

For the quadratic fitness case in the absence of recombination ($\nu=0$),
the exact analytical result predicts the existence of a ``selected''
organized phase, or quasi-species, when $k>\mu$. In this phase, the average
composition is given by $u=1-\mu/k$. For $k<\mu$, a phase
transition occurs and the quasi-species
disappears in favor of a disordered or ``unselected'' phase with $u=0$. 
In Figure \ref{fig2},
we display the phase structure in the presence of horizontal 
gene transfer. In agreement with the numerical results presented in
Table \ref{tab1} and Table \ref{tab2}, 
the recombination scheme considered in this model
introduces a mild mutational load. However, near
the critical region $k/\mu\sim 1$, one observes that horizontal
gene transfer distorts
the phase boundary which defines the error threshold, from the
horizontal line $k/\mu=1$, to a monotonically increasing curve
that saturates for large values of $\nu/\mu$. We obtain an analytical
expression for the phase boundary, by expanding Eqs.\ (\ref{eq30})
and (\ref{eq31})
near the critical region $\xi_{c}\sim 0$, $u\sim 0$. We find that the
boundary is defined by 
\begin{eqnarray}
k_{\rm{crit}} = \mu \frac{1+\nu/\mu}{1+\nu/2\mu}.
\label{eq39}
\end{eqnarray}
We notice from this expression that for small 
$\nu$, $k_{\rm{crit}}\sim \mu + \nu/2$,
whereas for large $\nu$ the phase boundary becomes asymptotically
independent of $\nu$, $k_{\rm{crit}}\sim 2\mu$. We also notice from this
formula that the phase boundary is independent of the block size $\bar{M}$.

As a second example, we consider a square-root fitness function
\begin{eqnarray}
f(u) = k\sqrt{|u|}
\label{eqsqrt1}
\end{eqnarray}
In Table \ref{tabsqrt1}, we present a comparison of our analytical
result, obtained from Eq.\ (\ref{eq31}), with the direct 
numerical solution of the differential Eq.\ (\ref{eq33}), for $\bar{M}=3$.
As in the quadratic fitness example, the analytical and numerical
results agree to order $\mathcal{O}(N^{-1})$, as expected.
\begin{table}
\begin{center}
\caption{Analytical versus numerical results for horizontal gene
transfer in the parallel (Kimura) model for the square-root
fitness $f(u)=k \sqrt{|u|}$, with $\bar{M}=3$, N = 801.}
\label{tabsqrt1}
\begin{tabular}{|c|c|c|c|}\hline
$k/\mu$ & $\nu/\mu$ & $u^{\rm{numeric}}$ & $u^{\rm{analytic}}$\\\hline
2.0 & 0.0 & 0.4858 & 0.4855 \\
2.0 & 0.5 & 0.4892 & 0.4889\\
2.0 & 1.0 & 0.4918 & 0.4915\\
2.0 & 1.5 & 0.4939 & 0.4936\\\hline
2.5 & 0.0 & 0.5399 & 0.5396\\
2.5 & 0.5 & 0.5428 & 0.5425\\
2.5 & 1.0 & 0.5450 & 0.5448\\
2.5 & 1.5 & 0.5469 & 0.5466\\\hline
4.0 & 0.0 & 0.6525 & 0.6523 \\
4.0 & 0.5 & 0.6542 & 0.6540\\
4.0 & 1.0 & 0.6556 & 0.6554\\
4.0 & 1.5 & 0.6568 & 0.6565\\\hline
\end{tabular}
\end{center}
\end{table}

From the results presented in Table \ref{tabsqrt1}, it is
remarkable that the average composition $u$, and correspondingly
the mean fitness of the population $f_{m}=k\sqrt{|u|}$, increase
when increasing the horizontal gene transfer rate $\nu$.

The mutational deterministic hypothesis states that recombination
is beneficial for negative epistasis fitness functions 
(see Fig.\ \ref{fig0}) $f^{''}(u)<0$,
and deleterious for positive epistasis fitness functions, $f^{''}(u)>0$
\cite{Kimura66,Arjan07,Azevedo06,Kondrashov82,Kondrashov88,Kondrashov93}.
Our results for the quadratic and square-root fitness functions, 
Eqs.\ (\ref{eq31})--(\ref{eq39}) and Tables \ref{tab1}, \ref{tab2}, and
\ref{tabsqrt1} provide support for this hypothesis. In fact, we can prove
the mutational deterministic hypothesis holds for the parallel
model in the presence of horizontal gene transfer, Appendix \ref{appendix12}. 
  
Horizontal gene transfer has less of an effect for the sharp peak fitness, 
$f(u)=A\delta_{u,1}$. For general $\bar{M}$, the maximum in 
Eq.\ (\ref{eq31})
is achieved for $\xi_{c}=1$, with $\phi_{c}(1)=(1+u)/2$ from 
Eq.\ (\ref{eq30}).
Thus, one obtains
\begin{eqnarray}
f_{m}=A - \mu 
- \frac{\nu}{\bar{M}}\left[1-\left(\frac{1+u}{2}\right)^{\bar{M}} \right].
\label{eq40}
\end{eqnarray}
The error threshold is given for $u=0$ by the condition 
$A>\mu+\frac{\nu}{\bar{M}}(1-2^{-\bar{M}})$. However, we notice from
Eq.\ (\ref{eq40}) that $f_m(u=1)=A-\mu > f_m(u=0)$. Therefore,
we have $u=1-\mathcal{O}(N^{-1})$ in the selected phase, with
the effect of horizontal gene transfer
being negligible for finite $\bar{M}$.
We obtain the fraction of the population located at the peak $P_{0}$, from the
self-consistency 
condition $P_{0}A = f_{m}$, which yields $P_{0}=1-\mu/A$. Thus, the
true error threshold is at $A_{\rm{crit}}=\mu$, with the condition
$A>\mu+\frac{\nu}{\bar{M}}(1-2^{-\bar{M}})$ defining the limit
of metastability for initial conditions with $u\sim 0$.
These results are similar to the ones obtained in the absence of 
horizontal gene transfer 
\cite{Park06,Park07,note}. Thus, we conclude that for the sharp
peak fitness, horizontal gene transfer does not spread out the
population in sequence space. This result differs from
the numerical studies presented in \cite{Boerlijst96}, where
a mathematical model for
'uniform crossover' recombination between viral strains
super-infecting a population of cells was described. We remark
that this model studied sequences of finite length (N = 15), where
the error threshold transition is not really sharp. Our results
correspond to the more realistic limit $N\rightarrow\infty$ (typical
viral genomes are $10^3-10^4$).  

In summary, from our exact analytical formula for the mean fitness
Eq.\ (\ref{eq31}),
which is valid for any permutation invariant replication rate, we
developed the explicit solution of three different examples: 
a quadratic fitness, a square-root fitness and a single sharp peak. For
the case of smooth fitness functions, from
our exact analytical formulas for the mean fitness $f_m$ and
average composition $u$, we conclude that in agreement with the
mutational deterministic hypothesis \cite{Kimura66,Arjan07,Kondrashov82,Kondrashov88,Kondrashov93}, a population whose
fitness represents positive epistasis (i.e. quadratic), will experience
an additional load against selection due to horizontal gene transfer.
On the contrary, when negative epistasis is present (e.g. square-root),
horizontal gene transfer is beneficial by enhancing selection. We provided
a mathematical proof for this effect, Appendix \ref{appendix12}.
When
the fitness is defined by a single sharp peak, the 
population steady-state distribution behaves more
rigidly in response to horizontal gene transfer. This fundamental
difference can be attributed to the structure of the quasi-species
distribution, which in the smooth fitness case is a Gaussian
centered at the mean fitness, while in the sharp peak it is a
fast decaying exponential, sharply peaked at the master 
sequence \cite{Park06}.
While the Gaussian distribution spreads its tails over a wide
region of sequence space, thus allowing for horizontal gene transfer 
effects to propagate
over a large diversity of mutants, the sharp exponential distribution
concentrates in a narrow neighborhood of the master sequence, acting
as a barrier to the propagation of such effects.

\subsection{Horizontal gene transfer for multiple-size blocks}

A natural extension to the model of horizontal gene transfer involving 
blocks of genes of a given size is to consider a process where
each site along the sequence may be transferred with probability 
$\gamma$, or
left intact with probability $1-\gamma$. The operator describing this
process is
\begin{eqnarray}
\hat{R}=\frac{1}{\langle\bar{M}\rangle}
\prod_{j=1}^{N}\left[(1-\gamma)\hat{I}_{j}+\gamma\hat{R}_{j}
\right]-\frac{1}{\langle\bar{M}\rangle}\hat{I}.
\label{eq41}
\end{eqnarray}
Here, $\hat{R}_{j}=\vec{\hat{a}}^{\dagger}(j)D\vec{\hat{a}}(j)$
is the single-site recombination operator defined in Eq.\ (\ref{eq5}), with
the matrix $D$ defined as in Eq.\ (\ref{eq6}). Notice that this
operator represents a binomial process, where an average number of sites
$\langle\bar{M}\rangle = \gamma N$ is transferred. If we consider,
as in the former finite block size case, 
that $N/\langle \bar{M}\rangle = \mathcal{O}(N)$, then we have
$\gamma = \langle\bar{M}\rangle/N$, and for very large $N$
Eq.\ (\ref{eq41}) reduces to
\begin{eqnarray}
\hat{R}=\frac{1}{\langle\bar{M}\rangle}
\prod_{j=1}^{N}\left[(1-\gamma)\hat{I}_{j}+\gamma\hat{R}_{j}
\right]-\frac{1}{\langle\bar{M}\rangle}\hat{I} 
\sim\frac{1}{\langle\bar{M}\rangle} e^{-\langle\bar{M}\rangle 
+ \frac{\langle\bar{M}\rangle}{N}
\sum_{j=1}^{N}\vec{\hat{a}}^{\dagger}(j)D\vec{\hat{a}}(j)}-
\frac{1}{\langle\bar{M}\rangle}\hat{I}.
\label{eq42}
\end{eqnarray}

Considering the recombination operator defined in Eq.\ (\ref{eq42}), the
spin Boson Hamiltonian for the Kimura model becomes
\begin{eqnarray}
-\hat{H}&=&N f\left[\frac{1}{N}\sum_{j=1}^{N}\vec{\hat{a}}^{\dagger}(j)
\sigma_{3}\vec{\hat{a}}(j)\right]+\mu\sum_{j=1}^{N}[\vec{\hat{a}}(j)^{\dagger}
\sigma_{1}\vec{\hat{a}}(j)-\hat{I}]
+\frac{\nu}{\langle\bar{M}\rangle} N e^{-\langle\bar{M}\rangle 
+ \frac{\langle\bar{M}\rangle}{N}
\sum_{j=1}^{N}\vec{\hat{a}}^{\dagger}(j)D\vec{\hat{a}}(j)}\nonumber\\
&&-\frac{\nu}{\langle\bar{M}\rangle} N \hat{I}.
\label{eq43}
\end{eqnarray}
We introduce a Trotter factorization
\begin{eqnarray}
e^{-\hat{H}t}=\lim_{M\rightarrow\infty}\int\left[\mathcal{D} \vec{z}^{*}
\mathcal{D} \vec{z}\right]
|\vec{z}_{M}\rangle\left(\prod_{k=1}^{M}\langle \vec{z}_{k}|
e^{-\epsilon \hat{H}}|\vec{z}_{k-1}\rangle \right)\langle \vec{z}_{0}|.
\label{eq44}
\end{eqnarray}
As shown in Appendix \ref{appendix3}, the partition function becomes
\begin{eqnarray}
Z=\int\left[\mathcal{D}\bar{\xi}\mathcal{D}\xi
\mathcal{D}\bar{\phi}\mathcal{D}\phi\right]e^{-S\left[\bar{\xi},\xi,
\bar{\phi},\phi\right]}
\sim e^{N f_m t} .
\label{eq55}
\end{eqnarray}
Here, the action in the continuous time limit is
\begin{eqnarray}
S\left[\bar{\xi},\xi,\bar{\phi},\phi\right]=-N\int_{0}^{t}dt'\left[
-\bar{\xi}\xi-\bar{\phi}\phi-\mu
-\frac{\nu}{\langle\bar{M}\rangle} + f(\xi)  
+ \frac{\nu}{\langle\bar{M}\rangle} 
e^{-\langle\bar{M}\rangle(1-\phi)}\right]
-N\ln Q
\label{eq56}
\end{eqnarray}

\subsubsection{The saddle point limit}

As in the previous model, the saddle point limit is exact as
$N\rightarrow\infty$ in Eq.\ (\ref{eq56}).

After a similar procedure as in section II.A.2, we find
the saddle-point equation for the mean fitness
\begin{eqnarray}
f_{m}&=&\max_{-1\leq\xi_{c}\leq 1 }\Biggl\{f(\xi_{c})-\mu
-\frac{\nu}{\langle\bar{M}\rangle}
+\frac{\nu}{\langle\bar{M}\rangle} 
e^{-\langle\bar{M}\rangle(1-\phi_{c}(\xi_{c}))}\nonumber\\
&&~~~~~~~~~+\mu\sqrt{1-\xi_{c}^{2}}\frac{\sqrt{1-u^{2}}(1
+\frac{\nu}{2\mu}
e^{-\langle\bar{M}\rangle(1-\phi_{c}(\xi_{c}))})}
{\left[\left(1+\frac{\nu}{2\mu}(1-u^{2})
e^{-\langle\bar{M}\rangle(1-\phi_{c}(\xi_{c}))}\right)^{2}-u^{2}
\right]^{1/2}}\Biggr\}
\label{eq57}
\end{eqnarray}
Here, $\phi_{c}(\xi_{c})$ is obtained from the equation
\begin{eqnarray}
\phi_{c}(\xi_{c})=\frac{1+u\xi_{c}}{2}+\frac{\sqrt{1-\xi_{c}^{2}}}{2}
\frac{\sqrt{1-u^{2}}}{\left[1-\left(\frac{u}{1+\frac{\nu}{2\mu}(1-u^{2})
 e^{-\langle\bar{M}\rangle(1-\phi_{c})}}\right)^{2}\right]^{1/2}}
\label{eq58}
\end{eqnarray}
Eq.\ (\ref{eq57}) represents an exact analytical expression
for the mean fitness $f_m$ of an infinite population experiencing
horizontal gene transfer of multiple size sequences. The formula is valid for
an arbitrary, permutation invariant replication rate function
$f(u)$.

\begin{table}
\begin{center}
\caption{Analytical results for horizontal gene transfer in the parallel
model for the quadratic
fitness $f(u)=\frac{k}{2}u^{2}$, with $\langle\bar{M}\rangle=3$.}
\label{tab3}
\begin{tabular}{|c|c|c|c|}\hline
$k$ & $\nu$ & $u^{\rm{analytic}}$\\\hline
2.0 & 0.0 & 0.50 \\
2.0 & 0.5 & 0.4840\\
2.0 & 1.0 & 0.4680\\
2.0 & 1.5 & 0.4522\\\hline
2.5 & 0.0 & 0.6000 \\
2.5 & 0.5 & 0.5921\\
2.5 & 1.0 & 0.5845\\
2.5 & 1.5 & 0.5773\\\hline
4.0 & 0.0 & 0.8000\\
4.0 & 0.5 & 0.7990\\
4.0 & 1.0 & 0.7981\\
4.0 & 1.5 & 0.7973\\\hline
\end{tabular}
\end{center}
\end{table}

We notice that recombination introduces an additional 
mutational load against selection. 
This load is mild at low values of the fitness
constant $k$, and becomes negligibly small at larger values.
Numerical evaluation of Eqs.\ (\ref{eq57}) and (\ref{eq58}) 
is presented in Table \ref{tab3} for the
quadratic fitness $f(u)=k u^{2}/2$, 
and average block size $\langle\bar{M}\rangle=3$.

An analytical expression for the phase boundary is obtained from
Eqs.\ (\ref{eq57}) and (\ref{eq58}), 
near the error threshold $u\sim 0$, $\xi_{c}\sim 0$. We find
\begin{eqnarray}
k_{\rm{crit}}=\mu\frac{1+\frac{\nu}{\mu}}
{1+\frac{\nu}{2\mu}}
\label{eq59}
\end{eqnarray}
We notice that for small $\nu$, the critical value is 
$k_{\rm{crit}}\sim \mu + \nu/2$, whereas for large
values of $\nu$ it becomes independent of recombination 
$k_{\rm{crit}}\sim 2\mu$.
This
behavior is  similar to the one previously observed
in Fig.\ \ref{fig2} for the case of horizontal gene transfer
 with blocks of fixed
size. The shape of the phase boundary is 
independent of the block size in the horizontal gene
transfer process, assuming
that the size of the blocks is finite.

As a second example, we consider the square root fitness
$f(u) = k\sqrt{|u|}$. Analytical results for the average
composition, obtained after Eq.\ (\ref{eq31}), are represented
in Table \ref{tabsqrt2} for blocks of average size 
$\langle\bar{M}\rangle = 3$. 
\begin{table}
\begin{center}
\caption{Analytical results for horizontal gene transfer in the
parallel model for the square-root
fitness $f(u)=k\sqrt{|u|}$, with $\langle\bar{M}\rangle=3$.}
\label{tabsqrt2}
\begin{tabular}{|c|c|c|c|}\hline
$k$ & $\nu$ & $u^{\rm{analytic}}$\\\hline
2.0 & 0.0 & 0.4855 \\
2.0 & 0.5 & 0.4889\\
2.0 & 1.0 & 0.4915\\
2.0 & 1.5 & 0.4936\\\hline
2.5 & 0.0 & 0.5396\\
2.5 & 0.5 & 0.5425\\
2.5 & 1.0 & 0.5448\\
2.5 & 1.5 & 0.5466\\\hline
5.0 & 0.0 & 0.6523\\
5.0 & 0.5 & 0.6540\\
5.0 & 1.0 & 0.6554\\
5.0 & 1.5 & 0.6566\\\hline
\end{tabular}
\end{center}
\end{table}
From the values displayed in Table \ref{tabsqrt2}, we notice
that horizontal gene transfer introduces a mild
increase in the average composition and, correspondingly,
in the mean fitness of the population $f_{m}=k\sqrt{|u|}$.
This trend, which is opposite to the quadratic fitness case,
can be attributed to the negative epistasis represented 
by the square root fitness, by similar arguments as in the
case of fixed block size.

Horizontal gene transfer does not affect the phase
boundary for the sharp peak fitness,
$f(u)=A\delta_{u,1}$. In this case, Eq.\ (\ref{eq57}) is maximized 
at $\xi_{c}=1$,
with $\phi_{c}=(1+u)/2$ from Eq.\ (\ref{eq58}). Thus, the mean
fitness becomes
\begin{eqnarray}
f_{m}=A-\mu-\frac{\nu}{\langle\bar{M}\rangle}
[1-e^{-\langle\bar{M}\rangle(1-u)/2}]
\label{eq60}
\end{eqnarray}
The error threshold is given, for $u=0$ in Eq.\ (\ref{eq60}), by
the condition $A>\mu+\frac{\nu}{\langle\bar{M}\rangle}
[1-e^{-\langle\bar{M}\rangle/2}]$. However, we notice that
$f_m(u=1) = A-\mu > f_m(u=0)$. Hence, in the selected phase 
$u=1-\mathcal{O}(N^{-1})$, and the recombination
effect becomes negligible for infinite N. From the self-consistency
condition $f_{m}=P_{0}A$, we obtain the fraction
of the population located at the peak $P_{0}=1-\mu/A$. Therefore,
the true error threshold is given by $A_{\rm{crit}}>\mu$, with
$A>\mu+\frac{\nu}{\langle\bar{M}\rangle}
[1-e^{-\langle\bar{M}\rangle/2}]$ the limit of metastability
for initial conditions with $u\sim 0$. 

Therefore, we conclude that horizontal gene transfer 
for multiple size blocks displays
a qualitatively similar behavior to the corresponding process
for fixed block size. A population evolving
under a smooth fitness function with positive epistasis 
(e.g. quadratic, see Fig.\ \ref{fig0}) experiences
an additional mutational load due to horizontal gene transfer, 
which modifies the
quasi-species structure, reducing the mean fitness, and hence shifting
the error threshold. On the contrary, when epistasis is negative (e.g.
square-root, see Fig.\ \ref{fig0}) 
a beneficial effect is induced by horizontal gene
transfer, in agreement with the mutational deterministic hypothesis,
as we demonstrate in Appendix \ref{appendix12}. 

A discontinuous sharp peak fitness function
does not change the quasi-species distribution or the mean fitness,
although it does introduce metastability.

\subsection{The parallel model with two-parent recombination}

\begin{figure}[tbp]
\centering
\epsfig{file=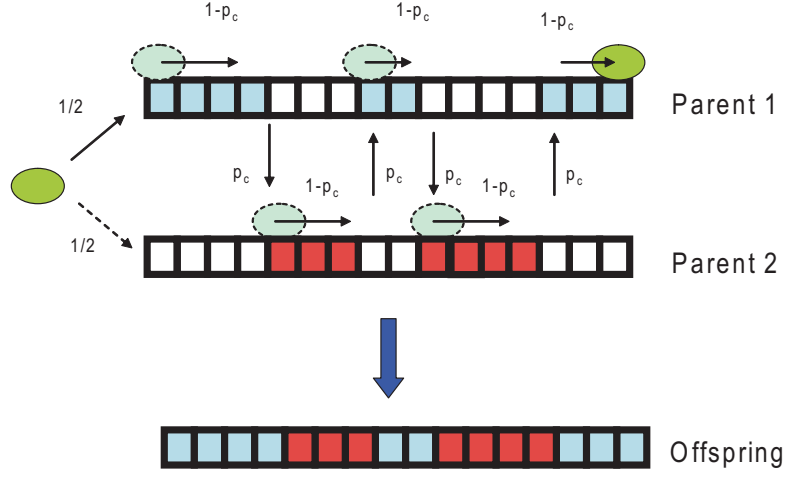,scale=0.6,clip=}
\caption{Pictorial representation of the two-parent genetic recombination
process considered in the theory.}
\label{fig3}
\end{figure}

Biological recombination, as occurs for example in viral
super- or co-infection or in sexual reproduction,
involves the crossing over of parental strands at  
random points along the sequence. The
copying process is carried out by the action of polymerase enzymes,
which move alternatively along one or the other parental strand.
An approximate representation of this process is to consider that
the polymerase enzyme starts, with probability 1/2 on either
parental strand, copying one base at a time. We consider the
crossovers to occur because there exists a probability $p_{c}$
per site that the polymerase ``jumps'' from its current position
towards the other parental strand. Alternatively, the enzyme progresses
along the current strand
with probability $1-p_{c}$. A pictorial representation is shown in
Fig.\ \ref{fig3}.

For this particular process representing the wandering path 
followed by the polymerase enzyme, the recombination coefficients
$R_{kl}^{i}$ in Eq.\ (\ref{eq1}) are given by the exact analytical
expression
\begin{eqnarray}
R_{kl}^{i}&=&\frac{1}{2}\sum_{\{\alpha_{j}=\pm 1\}}
\left(\frac{1+s_{1}^{k}s_{1}^{i}}{2}\right)^{\frac{1+\alpha_{1}}{2}}
\left(\frac{1+s_{1}^{l}s_{1}^{i}}{2}\right)^{\frac{1-\alpha_{1}}{2}}\nonumber\\
&&\times
[(1-p_{c})^{\frac{1+\alpha_{1}\alpha_{2}}{2}}
p_{c}^{\frac{1-\alpha_{1}\alpha_{2}}{2}}]
\left(\frac{1+s_{2}^{k}s_{2}^{i}}{2}\right)^{\frac{1+\alpha_{2}}{2}}
\left(\frac{1+s_{2}^{l}s_{2}^{i}}{2}\right)^{\frac{1-\alpha_{2}}{2}}\nonumber\\
&&\times[(1-p_{c})^{\frac{1+\alpha_{2}\alpha_{3}}{2}}
p_{c}^{\frac{1-\alpha_{2}\alpha_{3}}{2}}]\left(\frac{1+s_{3}^{k}s_{3}^{i}}{2}
\right)^{\frac{1+\alpha_{3}}{2}}
\left(\frac{1+s_{3}^{l}s_{3}^{i}}{2}\right)^{\frac{1-\alpha_{3}}{2}}\nonumber\\
&&\times\ldots \times [(1-p_{c})^{\frac{1+\alpha_{N-1}\alpha_{N}}{2}}
p_{c}^{\frac{1-\alpha_{N-1}\alpha_{N}}{2}}]
\left(\frac{1+s_{N}^{k}s_{N}^{i}}{2}\right)^{\frac{1+\alpha_{N}}{2}}
\left(\frac{1+s_{N}^{l}s_{N}^{i}}{2}\right)^{\frac{1-\alpha_{N}}{2}}
\nonumber\\
\label{eq61}
\end{eqnarray}
Here, the recombining parental sequences are 
$S_{k}=(s_{1}^{k},s_{2}^{k},\ldots,s_{N}^{k})$, 
$S_{l}=(s_{1}^{l},\ldots,s_{N}^{l})$ and the offspring sequence
is $S_{i}=(s_{1}^{i},s_{2}^{i},\ldots,s_{N}^{i})$, with $s_{j}=\pm 1$.
Using Eq.\ (\ref{eq61}), Eq.\ (\ref{eq1}) representing the time
evolution of an infinite population of binary sequences experiencing
replication, point mutations and two-parent recombination, exactly becomes
\begin{eqnarray}
\frac{dq_{i}}{dt}&=&r_{i}q_{i}+\sum_{k=1}^{2^{N}}\mu_{ik}q_{k}
+\nu N\sum_{k=1}^{2^{N}}\frac{1}{2}\sum_{\{\alpha_{j}=\pm 1\}}
\left\{\left[\prod_{j=2}^{N}p_{c}^{\frac{1-\alpha_{j-1}\alpha_{j}}{2}}
(1-p_{c})^{\frac{1+\alpha_{j-1}\alpha_{j}}{2}}\right]\right.\nonumber\\
&&\times
\left.
\prod_{j=1}^{N}\left(\frac{1+s_{j}^{k}s_{j}^{i}}{2}
\right)^{\frac{1+\alpha_{j}}{2}}
\sum_{l=1}^{2^{N}}p_{l}\left(\frac{1+s_{j}^{l}}{2}\delta_{s_{j}^{i},+1}
+\frac{1-s_{j}^{l}}{2}\delta_{s_{j}^{i},-1} \right)^{\frac{1-\alpha_{j}}{2}}
\right\}q_{k}- \nu N q_{i}
\label{eq62}
\end{eqnarray}
where, again, $p_{l}=q_{l}/\sum_{l=1}^{2^{N}}q_{l}$ is the normalized
probability for sequence $1\leq l \leq 2^{N}$.

From Eq.\ (\ref{eq62}), the recombination operator corresponding
to this recombination process in the spin Boson representation
is
\begin{eqnarray}
\hat{R}& = &\frac{1}{2}\sum_{l=1}^{2^{N}}p_{l}\sum_{\{\alpha_{i}=\pm
1\}}[\hat{I}_{1}^{\frac{1+\alpha_{1}}{2}}
\hat{R}_{l}(1)^{\frac{1-\alpha_{1}}{2}}] 
\times[(1-p_{c})^{\frac{1+\alpha_{1}\alpha_{2}}{2}}
p_{c}^{\frac{1-\alpha_{1}\alpha_{2}}{2}}]\nonumber\\
&&\times  [\hat{I}_{2}^{\frac{1+\alpha_{2}}{2}}
\hat{R}_{l}(2)^{\frac{1-\alpha_{2}}{2}}]
\times [(1-p_{c})^{\frac{1+\alpha_{2}\alpha_{3}}{2}}
p_{c}^{\frac{1-\alpha_{2}\alpha_{3}}{2}}]
\times[\hat{I}_{3}^{\frac{1+\alpha_{3}}{2}}
\hat{R}_{l}(3)^{\frac{1-\alpha_{3}}{2}}]
\nonumber \\
&&\times  \ldots\times[(1-p_{c})^{\frac{1+\alpha_{N-1}\alpha_{N}}{2}}
p_{c}^{\frac{1-\alpha_{N-1}\alpha_{N}}{2}}]
\times[\hat{I}_{N}^{\frac{1+\alpha_{N}}{2}}
\hat{R}_{l}(N)^{\frac{1-\alpha_{N}}{2}}]
-\hat{I}\nonumber\\
&& \equiv  g(\{\hat{R}_{l}(j)\}) - \hat{I}
\label{eq63}
\end{eqnarray}
Here, the local recombination operator is 
$\hat{R}_{l}(j)=\vec{\hat{a}}(j)^{\dagger}D_{j}^{l}\vec{\hat{a}}(j)$, with
\begin{eqnarray}
D_{j}^{l}=\left(\begin{array}{cc}\frac{1+s_{j}^{l}}{2}&\frac{1+s_{j}^{l}}{2}\\
\frac{1-s_{j}^{l}}{2}&\frac{1-s_{j}^{l}}{2}\end{array} \right).
\label{eq64}
\end{eqnarray}
The $\hat{I}_{j}$ are the identity operators acting 
on site $1\leq j \leq N$, whereas
$\hat{I}=\prod_{j=1}^{N}\hat{I}_{j}$ is the identity operator for the entire
sequence vector.

\subsubsection{The Hamiltonian}

The Hamiltonian describing the evolution of this system in the 
spin Boson
representation is given by
\begin{eqnarray}
-\hat{H} =  N f\left[\frac{1}{N}\sum_{j=1}^{N}\vec{\hat{a}}^{\dagger}(j)
\sigma_{3}\vec{\hat{a}}(j)\right]+\mu\sum_{j=1}^{N}\left[
\vec{\hat{a}}^{\dagger}(j)\sigma_{1}\vec{\hat{a}}(j)-\hat{I}\right] 
+  \nu N \left(g[\{\hat{R}_{l}(j)\}]-\hat{I}\right)
\label{eq65}
\end{eqnarray}
We introduce a Trotter factorization
\begin{eqnarray}
e^{-\hat{H}t}=\lim_{M\rightarrow\infty}\int\left[\mathcal{D}\vec{z}^{*}
\mathcal{D}\vec{z}\right]|\vec{z}_{M}\rangle\left(\prod_{k=1}^{M}
\langle\vec{z}_{k}|e^{-\epsilon\hat{H}}|\vec{z}_{k-1}\rangle\right)\langle
\vec{z}_{0}|
\label{eq66}
\end{eqnarray}
As shown in Appendix \ref{appendix6} the partition function is
\begin{eqnarray}
Z=\int\left[\mathcal{D}\bar{\xi}\mathcal{D}\xi
\mathcal{D}\bar{\phi}\mathcal{D}\phi\right]e^{-S\left[\bar{\xi},\xi,
\bar{\phi},\phi\right]}
\label{eq83}
\end{eqnarray}
Here, the action in the continuous time limit is given by
\begin{eqnarray}
S\left[\bar{\xi},\xi,\bar{\phi},\phi\right]=-N\int_{0}^{t}dt\left[
-\bar{\xi}\xi-\bar{\phi}\phi-\mu
-\nu  + f(\xi) + \nu g(\phi) \right]
-N\ln Q
\label{eq84}
\end{eqnarray}

As shown in Appendix \ref{appendix4}, 
the recombination term can be represented,
for $0\leq p_{c}\leq 1/2$, by the exact finite series
\begin{eqnarray}
g(\{\psi_{j}^{l}\})& = &\sum_{l=1}^{2^{N}}p_{l}\left\{
\prod_{j=1}^{N}\left(\frac{1+\psi_{j}^{l}}{2}\right)\right.\nonumber\\
&&+
\left.
\sum_{1\leq i<j}^{N}(1-2p_{c})^{j-i}\frac{1-\psi_{j}^{l}}{2}
\frac{1-\psi_{i}^{l}}{2}
\prod_{k\ne i,j}^{N}\left(\frac{1+\psi_{k}^{l}}{2}\right)\right.\nonumber\\
&&+
\left.
\sum_{1\leq i<j<k<n}^{N}(1-2p_{c})^{j-i+n-k}\frac{1-\psi_{j}^{l}}{2}
\frac{1-\psi_{i}^{l}}{2}\frac{1-\psi_{n}^{l}}{2}
\frac{1-\psi_{k}^{l}}{2}\right.\nonumber\\
&&\times
\left.\prod_{m\ne i,j,k,n}^{N}
\left(\frac{1+\psi_{m}^{l}}{2}\right)
+ \ldots +(1-2p_{c})^{\lfloor\frac{N}{2}\rfloor}
\prod_{j=1}^{N}\left(\frac{1-\psi_{j}^{l}}{2}\right)\right\}\nonumber\\
\label{eq70}
\end{eqnarray}
were we used the notation 
$\psi_{j}^{l}=\vec{z}^{*}_{k}(j)D_{j}^{l}\vec{z}_{k-1}(j)$, and $D_{j}^{l}$
is defined in Eq.\ (\ref{eq64}). 

We consider first the case when $p_c=1/2$ in the above expression. Then,
we have
\begin{eqnarray}
g(\{\psi_{j}^{l}\},p_{c}=1/2)
=\sum_{l=1}^{2^{N}}p_{l}\prod_{j=1}^{N}(1+\psi_{j}^{l})/2
\label{eq70a}
\end{eqnarray}
We notice that the recombination term in
the differential Eq.\ (\ref{eq62}) satisfies
$\sum_{l=1}^{2^{N}}p_{l}R_{kl}^{i}\leq 1,\:\forall\:k,i$,
because $R_{kl}^{i}\ge 0$ and $\sum_{i=1}^{2^{N}}R_{kl}^{i}=1$.
In our field theoretic representation of the model,
this condition is equivalent to
$g(\{\psi_{j}^{l}\})\leq 1$ for any physical state. We also have, for example,
$ \left\langle \prod_{j=1}^{N} 
\left(\frac{1+\psi_{j}^{l}}{2}\right)
\right\rangle_z \le 1$.  If we consider evaluating the $g$ interaction
term perturbatively,
 as in Appendix \ref{appendix1}, we obtain terms such as
\begin{eqnarray}
\langle g \rangle
&=& [ (1 + \langle \psi \rangle) / 2 ] ^ N + 
 [ (1 + \langle \psi \rangle) / 2 ] ^ {N-2} 
\nonumber \\ && \times
(1/8)\sum_l p_l \sum_{i \ne j}  \langle \delta \psi_i^l  \delta \psi_j^l \rangle_z
+  \ldots
\label{eq71}
\end{eqnarray}
where 
$\langle \psi \rangle = \sum_l p_l (1/N) \sum_j \langle \psi_j^l \rangle_z$.
Since both the correlations of the spins in the $D_i^l$
 matrix for typical, likely
$l$ and the correlations in the $z$ fields are each $\mathcal{O} (1/N)$, 
the interaction term $g$ in Eq.\ (\ref{eq70a}) contributes nothing, unless
$\langle \psi \rangle = 1 - \mathcal{O} (1/N)$, in which case
$\langle g \rangle = \mathcal{O} (N^0)$.

For the general case of $0<p_{c}<1/2$, we notice that $0<1-2p_{c}<1$.
Making the ansatz that correlations 
between $z$ fields and correlations between 
spins  of
typical, likely
sequences $l$  each remain
$\mathcal{O} (1 / N)$ at different sites,
terms other than the first in Eq.\ (\ref{eq70})
are at least $\mathcal{O} (1 / N)$ smaller
when 
$\langle \psi \rangle = 1 - \mathcal{O} (1/N)$.
Thus, when $\langle \psi \rangle \sim 1$, the first term 
dominates the series, and the others become arbitrarily small, thus
recovering the same expression as for $p_{c}=1/2$. 
On the other hand,
when $\langle \psi \rangle
\sim -1$, we notice that the dominant terms are the last
ones. However, those terms are proportional to powers of $1-2p_{c}$ of
order $N$, whereas the number of these terms is of just polynomial order
in $N$. Therefore, for N very large these terms become arbitrary small.
Thus, we conclude that in the limit $N\rightarrow\infty$,
regardless of the value of $p_c$, the function $g$ is represented by 
Eq.\ (\ref{eq70a}).

In the particular case of uniform
crossover $p_{c}=1/2$, and
when the fitness function is permutation invariant, i.e., it depends
only on the average composition of the sequence through the
average base composition $u$, it is possible to reformulate
the differential
equation Eq.\ (\ref{eq1}) for the evolutionary dynamics of
an infinite population of binary 
sequences in terms of the distribution of classes:
\begin{eqnarray}
P_{l} = \sum_{j \in \mathcal{C}_{l}}p_{j}
\label{eq100}
\end{eqnarray}
where $\mathcal{C}_{l}$ represents the class of sequences with
$l$, ``$-1$'' spins.
Although all the sequences in a given class do not have the same dynamics,
we can nonetheless calculate the class dynamics exactly:
\begin{eqnarray}
\frac{d P_{l}}{dt}&=&N\left[f(2l/N-1)-\sum_{l'=0}^{N}P_{l'}
f(2l'/N-1)\right]
P_{l}+\mu (N - l + 1) P_{l-1} + \mu (l + 1) P_{l + 1}
- N \mu P_{l}\nonumber\\
&&+\nu N
\sum_{l_{1},l_{2}}R(l|l_{1},l_{2})P_{l_{1}}P_{l_{2}}-N \nu P_{l}.
\label{eq101}
\end{eqnarray}
The coefficients $R(l|l_{1},l_{2})$ represent the probability that a pair
of parental
sequences in the classes $\mathcal{C}_{l_{1}}$, $\mathcal{C}_{l_{2}}$,
due to uniform crossover recombination, generate a child sequence
in the class $\mathcal{C}_{l}$. The number of sequences in these classes is
$\left(\begin{array}{c}N\\l_{1}\end{array}\right)$,
$\left(\begin{array}{c}N\\l_{2}\end{array} \right)$ and
$\left(\begin{array}{c}N\\l\end{array}\right)$, respectively. For a given pair
of parental sequences, let us consider the variables $n_{++}$, $n_{+-}$,
$n_{-+}$ and $n_{--}$, representing the number of pairs of $(+1,+1)$,
$(+1,-1)$, $(-1,+1)$ and $(-1,-1)$ spins respectively. These variables
satisfy the equation $N=n_{++}+n_{+-}+n_{-+}+n_{--}$. We further notice
that these variables also satisfy $n_{-+}=l_{1}-n_{--}$ and
$n_{+-}=l_{2}-n_{--}$. Considering that from each pair of $(+1,-1)$ or
$(-1,+1)$ spins in the parental sequences, the child sequence will inherit
a ``-1'' spin with probability $1/2$, while from a pair of the kind $(-1,-1)$
it will inherit a ``-1'' spin with probability $1$, we have
the explicit analytical expression
for these coefficients
\begin{eqnarray}
R(l|l_{1},l_{2})=\sum_{n=\max{\{0,l_{1}+l_{2}-N\}}}
^{\min{\{l_{1}+l_{2}-l,l_{1},l_{2}\}}}
\frac{\binom{N}{n,l_{1}-n,l_{2}-n}}{\binom{N}{l_{1}}\binom{N}{l_{2}}}
\binom{l_{1}+l_{2}-2n}{l-n}2^{-(l_{1}+l_{2}-2n)}\nonumber\\
\label{eq102}
\end{eqnarray}
The first factor is the probability for a configuration with $n\equiv n_{--}$,
given $l_{1}$, $l_{2}$ and $l$. The second factor is the number of
ways of picking $l-n_{--}$ ``-1'' spins among $n_{+-} + n_{-+}$. The third
factor is just $(1/2)^{n_{-+}}(1/2)^{n_{+-}}(1)^{n_{--}}$. 
These coefficients are different from zero only if
\begin{eqnarray}
\max{\{0,l_{1}+l_{2}-N\}\leq l \leq \min{\{N,l_{1}+l_{2}\}}}
\label{eq103}
\end{eqnarray}
They also satisfy the following properties:
\begin{eqnarray}
R(l|l_{1},l_{2})=R(l|l_{2},l_{1})
\label{eq104}
\end{eqnarray}
\begin{eqnarray}
\sum_{l=0}^{N}R(l|l_{1},l_{2})=1 & \forall~ l_{1},l_{2}
\label{eq105}
\end{eqnarray}
\begin{eqnarray}
R(N|N,N) = R(0|0,0) = 1
\label{eq106}
\end{eqnarray}
In the limit of large $N$, we find that the recombination coefficients
satisfy a Gaussian distribution in the variables 
$u_1 = 1-2l_1/N$, $u_2 = 1-2l_2/N$,
and $u = 1 - 2 l /N$ (see Appendix \ref{appendix5}):
\begin{eqnarray}
R_{u_{1},u_{2}}^{u} 
\sim \frac{e^{-N\left[(u_1+u_2)/2-u\right]^2/(1-u_*^2)}}
{\sqrt{\pi(1-u_*^2)/N}}
\label{eq80a}
\end{eqnarray}
where $f_m = f(u_*)$.

This form of the recombination operator, Eq.\ (\ref{eq80a}),
is equivalent to Eq.\ (\ref{eq70a})
with $s_{j}^{l}$ replaced by $u$ in the $D$ matrix. Alternatively, we
notice that when the singular behavior of the function $g$ can be described
as a delta function, we have
\begin{eqnarray}
g &=& \sum_{l=1}^{2^{N}}p_{l}\delta_{\frac{1}{N}\sum_{j=1}^{N}
\vec{z}_{k}^{*}(j)D_{j}^{l}\vec{z}_{k-1}(j),1}
\nonumber \\
&=&\sum_{l=1}^{2^{N}}p_{l}\int_{0}^{2\pi}\frac{d\lambda}{2\pi}
e^{i\lambda\left[\frac{1}{N}\sum_{j=1}^{N}\vec{z}_{k}^{*}(j)D_{j}^{l}
\vec{z}_{k-1}(j)-1\right]}\nonumber\\
&=&\sum_{l=1}^{2^{N}}p_{l}\int_{0}^{2\pi}
\frac{d\lambda}{2\pi}e^{-i\lambda}\left\{1
+\frac{i\lambda}{N}\sum_{j=1}^{N}\vec{z}_{k}^{*}(j)D_{j}^{l}\vec{z}_{k-1}(j)
\right.\nonumber\\
&&
+\left.
\frac{1}{2!}\left(\frac{i\lambda}{N}\right)^{2}
\sum_{j,m=1}^{N}\vec{z}_{k}^{*}(j)D_{j}^{l}\vec{z}_{k-1}(j)
\vec{z}^{*}_{k}(m)D_{m}^{l}\vec{z}_{k-1}(m)+\ldots\right\}
\label{eq73}
\end{eqnarray}
By noticing that correlations between compositions at different
sites along the sequence are of order
$\mathcal{O}(N^{-1})$, we have that for the second order correlation
\begin{eqnarray}
\langle D_{j}^{l}D_{m}^{l}\rangle - \langle D_{j}^{l} \rangle^{2}
\sim \mathcal{O}(N^{-1})
\label{eq74}
\end{eqnarray}
where
$\langle D_{j}^{l}\rangle = \sum_{l=1}^{2^{N}}p_{l}D_{j}^{l} \equiv D_{j}$
is the population average.
A similar analysis for the higher order correlations allows us to
factorize order by order the terms in the series Eq.\ (\ref{eq73}), to
obtain
\begin{eqnarray}
g \sim
\delta_{\frac{1}{N}\sum_{j=1}^{N}\vec{z}_{k}^{*}(j)D_{j}\vec{z}_{k-1}(j),1}
+\mathcal{O}(N^{-1})
\label{eq75}
\end{eqnarray}
We are interested in the long term, steady state distribution, when
the average base composition $u(j) = \langle s_{j}^{l}\rangle \sim u$
becomes independent of time. In this limit, the trace defined 
by Eq.\ (\ref{eq82}) becomes
\begin{eqnarray}
\lim_{t\rightarrow\infty}\frac{\ln Q_{c}}{t}= \frac{\bar{\phi}_{c}}{2}+\left[\bar{\xi}_{c}
(\bar{\xi}_{c}+u\bar{\phi}_{c})+(\mu+\bar{\phi}_{c}/2)^{2}\right]^{1/2}
\label{eq85}
\end{eqnarray}
Hence, from Eq.\ (\ref{eq84}), the saddle point action is
\begin{eqnarray}
\lim_{N,t\rightarrow\infty}\frac{\ln Z}{N t} &=&
\lim_{t\rightarrow\infty}\frac{-S_{c}}{N t} 
= f_m \nonumber\\
&=&
\max_{\xi_{c},\bar{\xi}_{c},\phi_{c},\bar{\phi}_{c}}
\bigg\{-\bar{\xi}_{c}\xi_{c}-
\bar{\phi}_{c}\phi_{c}
-\mu-\nu + \nu g(\phi_{c}) \nonumber\\
&&~~~~~~~~~~
+f(\xi_{c})
+\frac{\bar{\phi}_{c}}{2}+\left[\bar{\xi}_{c}
(\bar{\xi}_{c}+u\bar{\phi}_{c})+\left(\mu+\frac{\bar{\phi}_{c}}{2}
\right)^{2}\right]^{1/2}\bigg\}
\label{eq86}
\end{eqnarray}
As shown in Appendix \ref{appendix6a}, we find
\begin{eqnarray}
\frac{-S_{c}}{Nt}&=&\max_{\phi_{c},\xi_{c}}\left\{f(\xi_{c})-\mu-\nu
+\nu g(\phi_{c})
+\frac{\mu}{1-u^{2}}(2\phi_{c}-1-u\xi_{c})\right.\nonumber\\
&&
~~~~~~~~
- \left.\frac{\mu |u|}{1-u^{2}}\left[(2\phi_{c}-1-u\xi_{c})^{2}
-(1-u^{2})(1-\xi_{c}^{2})\right]^{1/2}\right\}
\label{eq91}
\end{eqnarray}
Because of the singular behavior of the function 
$g(\phi_{c})$,
to find the saddle point we need to consider three
separate cases: $\phi_c < 1$, $\phi_c = 1$, and
$\phi_c = 1-\mathcal{O}(1/N)$.
The existence of different expressions for the mean
fitness suggests the possibility of
different selected phases in certain conditions. 
We also notice
that the saddle point analysis may not apply exactly, unless
$g(\phi_{c}) = \delta_{\phi_{c},1}$. 

Case 1: $\phi_c < 1$.
For this case, we look for a saddle point in the field $\phi_{c}$, in the
interior of the domain, $\phi_{c}<1$ where $g(\phi_{c})=0$
\begin{eqnarray}
\frac{\delta}{\delta \phi_{c}}\left(\frac{-S_{c}}{Nt}\right)=
\frac{2\mu}{1-u^{2}}-\frac{\mu |u|}{1-u^{2}}
\frac{2(2\phi_{c}-1-u\xi_{c})}{\left[(2\phi_{c}-1-u\xi_{c})^{2}-
(1-u^{2})(1-\xi_{c}^{2})\right]^{1/2}}=0
\label{eq93}
\end{eqnarray}
From Eq.\ (\ref{eq93}), we solve for $\phi_{c}$ as a function of $\xi_{c}$
\begin{eqnarray}
\phi_{c}(\xi_{c})=\frac{1+u\xi_{c}}{2}
+\frac{1}{2}\sqrt{1-\xi_{c}^{2}}
\label{eq94}
\end{eqnarray}
Substituting Eq.\ (\ref{eq94}) in the saddle-point 
action Eq.\ (\ref{eq91}), we obtain
\begin{eqnarray}
f_{m}^{(1)}=\max_{-1\leq\xi_{c}\leq 1}\left\{f(\xi_{c})-\mu-\nu
+\mu\sqrt{1-\xi_{c}^{2}}\right\}
\label{eq95}
\end{eqnarray}

Case 2: $\phi_c =1$. The mean fitness is obtained from Eq.\ (\ref{eq91})
as
\begin{eqnarray}
f_{m}^{(2)} = \max_{-1\leq\xi_{c}\leq 1}\{f(\xi_{c})-\mu+
\frac{\mu}{1-u^{2}}(1-u\xi_{c}-|u\xi_{c}-u^{2}|)\}
\label{eq92}
\end{eqnarray}

Case 3: $\phi_{c}=1-\mathcal{O}(1/N)$.
In this case, additional analysis is necessary to calculate the 
mean fitness due to the singular behavior of the $g(\phi_c)$ function.
For a smooth fitness function, we can argue this case does not 
exist.  We first consider the Hamiltonian (\ref{eq65}) for the
case $g=0$.  The largest
eigenvalue, $f_m$, is shifted by $-\nu$ relative to the
$\nu = 0$ case.  This allows us to calculate the average
composition, $u_*$, from the implicit
relation $f_m(\nu) = f_m(\nu=0) - \nu = f(u_*)$. 
Alternatively, if we consider the differential 
equation for the unnormalized class probabilities, 
$dQ / dt = L Q$, we see that the 
differential operator $L$ looks like that in the absence of
recombination, save for a shift of $-\nu$ in the fitness function.
Thus, the variance of the population is given by \cite{Park06}
$\sigma^2_u/N = 2 \mu u_* / [N f'(u_*)]$.  Considering more carefully the
$g$ function, we find $\int d {u_1} d {u_2} R_{u_1 u_2}^u P(u_1) P(u_2)
= \exp[-N(u - u_*)^2 / (2 \sigma^2)] / \sqrt{2 \pi \sigma^2 N}$, with
$\sigma^2 = \sigma_u^2/2 + (1-u_*^2)/2$.  This term is 
exponentially negligible
compared to the $-\nu P(u)$ term when $\sigma^2 < \sigma_u^2$,
since $P(u) = \exp[-N(u - u_*)^2 / (2 \sigma_u^2)] / \sqrt{2 \pi \sigma_u^2 N}$.  In
other words, we must strictly be in case 1 when 
\begin{eqnarray}
1- u_*^2 < 2 \mu u_* / f'(u_*).
\label{eq92a}
\end{eqnarray}
We denote the value of $\nu$ at which 
\begin{eqnarray}
1-u_*^2 = 2 \mu u_* / f'(u_*) {\rm ~at~} \nu = \nu_*
\label{eq92b}
\end{eqnarray}
as $\nu_*$.
Now, at this value of $\nu_*$ we have 
 $\int d {u_1} d {u_2} R_{u_1 u_2}^u P(u_1) P(u_2)
= P(u)$. Thus, the term proportional to $\nu$ in 
Hamiltonian (\ref{eq65}), or differential equation
(\ref{eq101}), exactly vanishes.  Thus, we have $d f_m / d \nu
= 0$ and $d P(u) / d \nu = 0$
at this value of $\nu$.  
There is spectral rigidity.
This implies that for $\nu > \nu_*$, 
the distribution $P(u)$ is independent of $\nu$, and that the
value of $u_*$ is constant.  In other words, the 
value of $f_m$ in case 2 must be constant with $\nu$.  Assuming $f_m$
varies continuously with $\nu$ in case 1,  and that the
fitness values for case 1 and case 2 are equal at a single
value of $\nu$, therefore,
case 2 is simply case 1 with the value $\nu = \nu_*$
\begin{eqnarray}
f_m (\nu > \nu_*) = f_m(\nu = \nu_*)
\label{eq92c}
\end{eqnarray}

Eqs.\ (\ref{eq95}), (\ref{eq92}) provide an exact analytical
solution for the mean fitness of an infinite population,
for a general permutation invariant replication rate
represented by a continuous, smooth function $f(u)$.

For a non-smooth fitness function, additional analysis is necessary, since
$f'(u_*)$ is undefined, and $P(u)$ may no longer be Gaussian.

\subsubsection{Examples and numerical tests}
We investigate the phase diagrams, as predicted from our theoretical
equations Eqs.\ (\ref{eq95}), (\ref{eq92}) 
for three different fitness functions: A sharp peak, a quadratic
fitness landscape and a square-root fitness landscape.

For the sharp peak landscape $f(u)=A\delta_{u,1}$, we notice
that the maximum is achieved at $\xi_{c}=1$, with $u=1-\mathcal{O}(N^{-1})$.
From Eqs.\ (\ref{eq92}) and (\ref{eq95}), we obtain
\begin{eqnarray}
f_{m}^{(2)}=A-\mu > f_{m}^{(1)}=A-\mu-\nu
\label{eq96}
\end{eqnarray}
Therefore, for the sharp peak only a single selected phase is observed.
In this case, the function $g(\phi_c)$ is not exactly a Kronecker
delta $\delta_{\phi_{c},1}$, we are in case 3, 
and thus we find a small correction,
approximately linear in $\nu$,
to the saddle-point prediction.
In the selected phase, where the population is exponentially localized
near $u = 1$ for large $N$, Eq.\ (\ref{eq102}) becomes
$R(l \vert l_1, l_2) \sim  (l_1 + l_2)! 2^{-l_1 - l_2} / [l! (l_1 + l_2-l)!]$.
 By analyzing the differential
equation at zeroth-order in $\nu$ for large $N$, we find that the class
distribution is given by  $P_{l}^{(0)}=P_{0}^{(0)}(1-P_{0}^{(0)})^{l}$. 
Hence, we find
that at first order in $\nu$, the fraction of the population $P_{0}$ 
located at the peak is given by
\begin{eqnarray}
P_0 = 1 - \mu/A - \nu/A\left[1-4\frac{1-\frac{\mu}{A}}
{\left(2-\frac{\mu}{A}\right)^2}\right] + \mathcal{O}(\nu^2)
\label{eq105a}
\end{eqnarray}
We note that this value of $f_m = A P_0$ interpolates
between 
$f_m^{(1)}$ for $A/\mu = 1$ and
$f_m^{(2)}$ for $A /\mu = \infty$.
There is no dependence on $p_c$ because the -1 spins are
separated by $\mathcal{O}(N)$ sites.

\begin{figure}[ht]
\centering
\epsfig{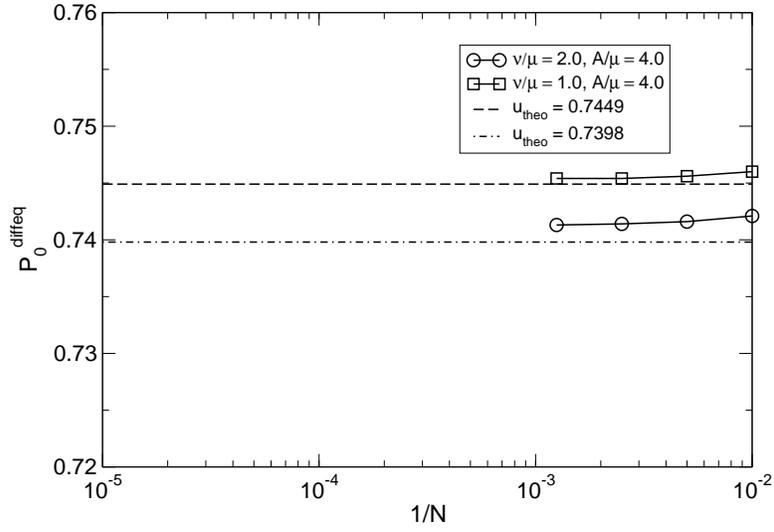}
\caption{Convergence of the numerical results towards the theoretical
value for two-parent recombination in the parallel (Kimura)
model for the sharp peak fitness. In this example, $A/\mu$ = 4.0.}
\end{figure}

As a second example, we consider the quadratic fitness landscape,
$f(u)=k u^{2}/2$. This smooth, continuous fitness function
allows for the use of the exact analytical formulas Eq.\ (\ref{eq95}),
(\ref{eq92}). 
By maximizing Eq.\ (\ref{eq95}) with respect to $\xi_{c}$,
when $\phi_{c}<1$ and hence $g(\phi_{c})=0$, we find
\begin{eqnarray}
f_{m}^{(1)}& = & \frac{k}{2}\left[\left(1-\frac{\mu}{k}\right)^{2}-
\frac{2\nu}{k}\right]
\label{eq98}
\end{eqnarray}
This mean fitness defines a selective phase S1.

According to our previous analysis, when $\phi_{c}=1$
and $g(\phi_{c})=1$, we maximize Eq.\ (\ref{eq92}) in $\xi_{c}$. 
Here, we consider that the order parameters
$\xi_{c}$ and $u$ have the same sign, $u\xi_{c}\ge0$. 
We then have $u\xi_{c} \ge u^{2}$  in Eq.\ (\ref{eq92}) \cite{note2}.
Hence, we find
\begin{eqnarray}
f_{m}^{(2)}& = & \frac{k}{2}\left(1-\frac{2\mu}{k}\right)
\end{eqnarray}
which defines a second selective phase S2.

By applying the self-consistency condition $f_{m}^{(1,2)}=k u^{2}/2$,
we find the following phases
\begin{eqnarray}
S1:\quad
u &=& \left[\left(1-\frac{\mu}{k}\right)^{2}-\frac{2\nu}{k}\right]^{1/2}, 
\quad \frac{2\nu}{\mu}<\frac{\mu}{k}<1
- \left[ \frac{2 \nu}{k} \right]^{1/2}
\nonumber\\
S2:\quad
u &=& \sqrt{1-\frac{2\mu}{k}},\quad \frac{2\nu}{\mu} > \frac{\mu}{k}
< \frac{1}{2}
\nonumber\\
NS: \quad
  u&=&0,\quad  {\rm otherwise}
\label{eq99}
\end{eqnarray}
We note that the phase transition between case 1 and case 2 is exactly
as predicted by Eq.\ (\ref{eq92b}).
We further note that the mean fitness is independent
of $\nu$ for $\nu > \nu_* = \mu^2 / (2 k)$, exactly
as predicted by Eq.\ (\ref{eq92c}).

\begin{table}
\begin{center}
\caption{Stochastic process versus analytical theory for
two-parent recombination in the parallel model for the 
quadratic fitness $f(u)=k u^{2}/2$, with $k/\mu$ = 4.0, 
$\nu/\mu=3.0$, and N = 100.}
\label{tab6a}
\begin{tabular}{|c|c|c|}\hline
$p_{c}$ & $u^{\rm{stochastic}}$ & $u^{\rm{analytic}}$\\\hline
0.1 & 0.7065 & 0.7071\\
0.3 & 0.7052 & 0.7071\\
0.5 & 0.7058 & 0.7071\\\hline
\end{tabular}
\end{center}
\end{table}
  
\begin{figure}[h]
\centering
\epsfig{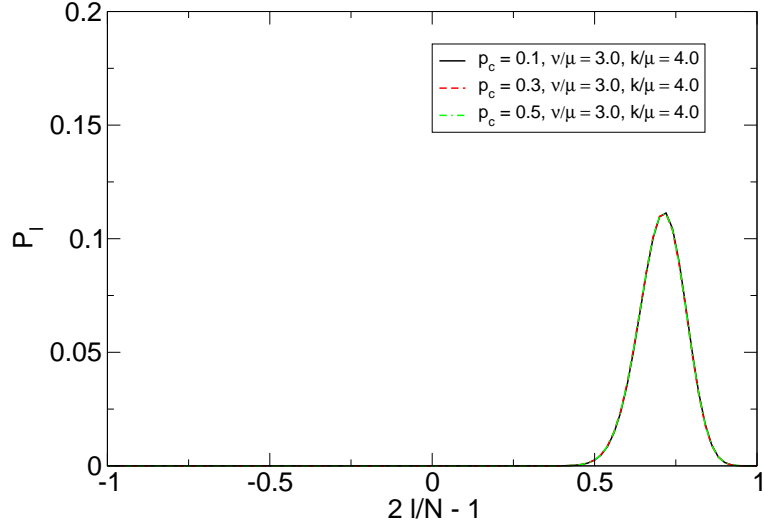}
\caption{Probability distributions for two-parent recombination
in the parallel model for the quadratic fitness
$f(u) = k u^2/2$, with $k/\mu=4.0$
and $\nu/\mu=3.0$, obtained from stochastic simulations with 
$M=10~000$ sequences of $N=100$ bases and different
values of $p_c$.}
\label{quadp}
\end{figure}

The system of differential equations (\ref{eq101}) provides an
exact representation of the evolution dynamics for an infinite
population, when uniform crossover probability $p_c=1/2$ is assumed.
On the other hand, our analytical equations Eq.\ (\ref{eq95}), 
Eq.\ (\ref{eq92})
for smooth fitness, or Eq.\ (\ref{eq105a}) for the discontinuous
sharp peak, 
predict that the equilibrium results should be independent
of the crossover probability $p_c$. To test this theory, 
we performed exact stochastic simulations  
based on a Lebowitz/Gillespie algorithm \cite{Lebowitz75,Gillespie76}.
We generate a population of $M=10\,000$ sequences initially in the wild-type. 
The size of the finite population represented in the simulation
was chosen large enough such that the results become independent
of size $M$. 
Then, the population is evolved in time
by point mutation, recombination and replication with 
rates proportional to $\mu$, $\nu$, and $f(u^{l})$
respectively, with $u^{l}=\frac{1}{N}\sum_{j=1}^{N}s_{j}^{l}$ the average
composition of sequence $S_{l}$, $1\leq l \leq M$. 
For that purpose, a list is generated by defining:
$\tau_{l}=\mu+\nu+f(u^{l})$, $\tau = \sum_{l=1}^{M}\tau_{l}$. With
probability $\tau_{l}/\tau$, a sequence $1\leq l \leq M$ 
is chosen from the
population to undergo either a single point mutation with probability
$\mu/\tau_{l}$, replication with probability $f(u^{l})/\tau_{l}$, or
recombination with another sequence with probability $\nu/\tau_{l}$
according to the process described in Fig.\ \ref{fig3}.

\begin{figure}[h]
\centering
\epsfig{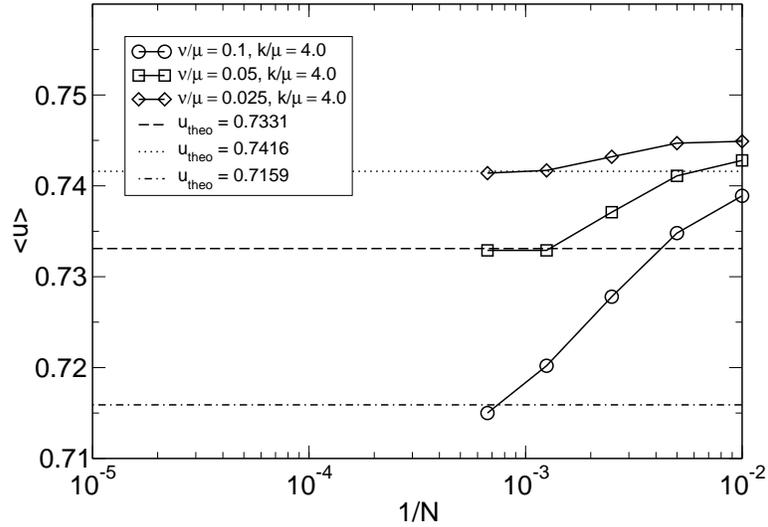}
\caption{Convergence of the numerical results towards the theoretical
value for two-parent recombination in the parallel model for 
the selective phase S1 in Eq.\ (\ref{eq99}). 
In this example, $k/\mu = 4.0$
and $\nu/\mu < 1/8$.}
\label{fig6_S1}
\end{figure}

To preserve
the size $M$ of the population, when replication or recombination is
performed, a sequence chosen at random from the population is
substituted with the offspring. The time increment after any
of these events is performed is calculated as $dt = -\log(w)/(N\tau)$,
with $w \in (0,1]$ a uniformly distributed random number. The 
results obtained from this stochastic simulation
are compared with the theoretical prediction in Table \ref{tab4}
for the sharp peak fitness landscape and uniform crossover $p_{c}=1/2$. 

\begin{figure}[t]
\centering
\epsfig{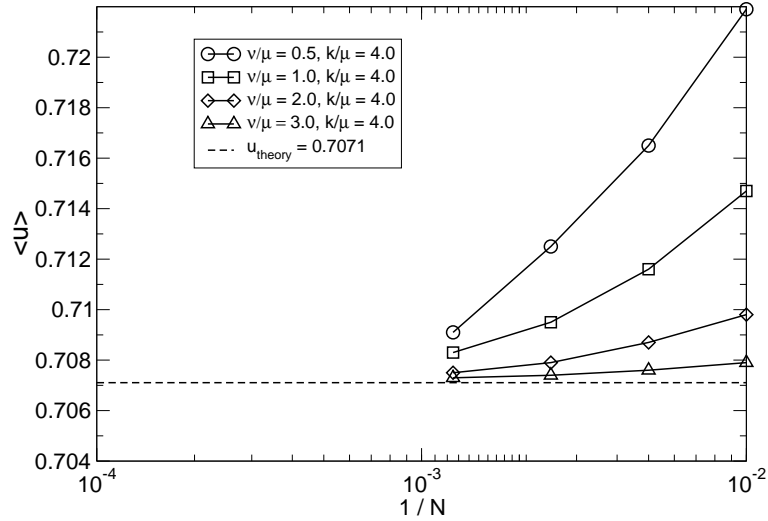}
\caption{Convergence of the numerical results towards the theoretical
value for two-parent recombination in the parallel model for 
the selective phase S2 in Eq.\ (\ref{eq99}). In this example,
$k/\mu = 4.0$ and $\nu/\mu > 1/8$.}
\label{fig4}
\end{figure}

In agreement with our theoretical prediction, as shown in Table 
\ref{tab6a} from stochastic simulations
in the quadratic fitness landscape, the effect of recombination
is independent of the polymerase crossover probability $p_c$.
The probability distributions obtained for the systems considered in 
Table \ref{tab6a}
are displayed in Fig.\ \ref{quadp}. Clearly, the distributions are independent
of $p_{c}$, in agreement with the theory.

We obtain a 
direct numerical solution of the deterministic system of differential
equations Eq. (\ref{eq101}), which provides an exact representation
of the evolution dynamics for an infinite population experiencing
uniform crossover recombination $p_c=1/2$. 
A comparison between these numerical solutions, and results
obtained from the stochastic simulation for a system large enough
to eliminate finite size effects, is displayed in 
Table \ref{tab4}
for the
sharp peak fitness. The theoretical prediction
from the analytical formula Eq.\ (\ref{eq105a}) is also shown for comparison.
It is evident from this table that the effect of recombination is
independent of the polymerase crossover probability $p_c$, in
agreement with our theoretical predictions.

\begin{table}
\begin{center}
\caption{Stochastic process versus differential equation for two-parent
recombination in the parallel model for the
sharp peak fitness, $A/\mu$ = 4.0, N = 400.}
\label{tab4}
\begin{tabular}{|c|c|c|c|c|c|c|c|}\hline
$\nu/\mu$ & $u^{\rm{stochastic}}$ & $u^{\rm{diffeq}}$ & 
$P_{0}^{\rm{stochastic}},\:p_{c}=0.1$ & 
$P_{0}^{\rm{stochastic}},\:p_{c}=0.3$ &
$P_{0}^{\rm{stochastic}},\:p_{c}=0.5$ &
$P_{0}^{\rm{diffeq}}$ & $P_{0}^{\rm{analytic}}$\\\hline
0.0 &0.998337 &0.998336 &0.75017 & 0.75017 & 0.75017  & 0.75016 & 0.75\\
1.0 & 0.998329 &  0.998326 &0.7455 &0.7454 & 0.74591 & 0.74544 & 0.7449\\
2.0 & 0.998312 & 0.998317 &0.7415 &0.7414 &0.74085 & 0.74140 & 0.7398\\\hline
\end{tabular}
\end{center}
\end{table}

From the data presented in Table \ref{tab4}, we
notice that the deterministic system of differential equations
provides an accurate representation of the underlying stochastic
dynamics for the case of uniform crossover, $p_{c}=1/2$. Thus,
the results obtained from the numerical solution of the deterministic
system of differential equations can be fairly compared with
the analytical theory. 

It is remarkable that the small, but finite, effect introduced by
recombination in the structure of the quasi-species distribution
for the sharp peak case, is not a consequence of the Muller's
ratchet phenomenon \cite{Muller64} 
characteristic of finite populations. Indeed,
the shift in the wild-type probability $P_0$ due to recombination, 
as predicted from our analytical equation Eq.\ (\ref{eq105a}),
was derived from the system of differential equations Eq.\ (\ref{eq101}), 
which describes the time evolution of an infinite population.
Moreover, this closed analytical result is in excellent agreement
with the numerical solution of the system of differential equations
Eq.\ (\ref{eq101}), 
as displayed in Fig.\ \ref{fig4} and Table \ref{tab5}. A good agreement
between our analytical and differential equation results, which
correspond to the infinite population case, and the
stochastic simulation is expected when the later is performed
in a large enough population. We determined that for the parameters
we consider, $M = 10\,000$
sequences provides simulation results that are independent of
the population size for the sharp peak fitness function, thus
allowing for a comparison with the infinite population
theory expressed by the differential equations Eq.\ (\ref{eq101})
and with our analytical solution Eq.\ (\ref{eq105a}).

\begin{table}
\begin{center}
\caption{Analytical theory versus numerical solution
 for two-parent recombination in the parallel model for the
quadratic fitness $f(u) = k u^2/2$ with $N=800$ and $k/\mu$=4.0.}
\label{tab5}
\begin{tabular}{|c|c|c|}\hline
$\nu/\mu$ & $u^{\rm{diff eq}}$ & $u^{\rm{analytic}}$\\\hline
0.0 & 0.7499 & 0.7500\\
0.025 & 0.7417 & 0.7416\\
0.05& 0.7329 & 0.7331\\
0.1 & 0.7202 & 0.7159\\
0.5 & 0.7091 & 0.7071\\
1.0 & 0.7083 & 0.7071\\
2.0 & 0.7075 & 0.7071\\
3.0 & 0.7073 & 0.7071\\\hline
\end{tabular}
\end{center}
\end{table}

Notice that for the quadratic fitness, 
the analytical theory reproduces the differential
equation results within $\mathcal{O}(N^{-1})$. The convergence
towards the theoretical value as a function
of the system size $1/N$, for parameters within the $S1$
phase defined in Eq.\ (\ref{eq99}), is displayed in Fig.\ \ref{fig6_S1}, 
and for the S2 phase in Fig.\ \ref{fig4}.

As a final example, we apply our analytical solution Eq.\ (\ref{eq95})
and Eq.\ (\ref{eq92}) to study the square-root fitness,
$f(u) = k\sqrt{|u|}$, as displayed in Table \ref{tabrecsqrt},
where analytical theory and direct numerical solution of the
differential equation agree to $\mathcal{O}(N^{-1})$.
\begin{table}
\begin{center}
\caption{Analytical theory versus numerical solution
 for two-parent recombination in the parallel model for the 
square-root fitness $f(u) = k \sqrt{|u|}$, 
with $N=400, 800, 1000$ 
and $k/\mu$=4.0.}
\label{tabrecsqrt}
\begin{tabular}{|c|c|c|c|c|}\hline
$\nu/\mu$ & $u^{\rm{diff eq}},N=400$ & $u^{\rm{diff eq}},N=800$ 
& $u^{\rm{diff eq}},N=1000$ & $u^{\rm{analytic}}$\\\hline
0.0 & 0.6527 & 0.6525 & 0.65249 & 0.6523\\
0.1 & 0.6650 & 0.6672 & 0.6678 & 0.6710\\
0.3 & 0.6686 & 0.6697  & 0.66993 & 0.6710\\
0.5 & 0.6696 & 0.6703  & 0.67043  & 0.6710\\
0.8 & 0.6703 & 0.6707  & 0.67073  & 0.6710\\
1.0 & 0.6705 & 0.6708  & 0.67083 & 0.6710\\\hline
\end{tabular}
\end{center}
\end{table}

As shown in Table \ref{tabrecsqrt}, two-parent recombination
in the square-root fitness landscape enhances selection towards
sequences which are on average more fit, as observed
by a slight increase of the average composition $u$, with respect
to the case when recombination is absent. This effect, which was
already observed for the square-root landscape in the presence
of horizontal gene transfer, can be attributed to the negative 
(see Fig.\ref{fig0}) 
epistatic interactions introduced by the square-root fitness,
in agreement with the mutational deterministic hypothesis,
Appendix \ref{appendix14}.

An additional interesting effect in two-parent recombination, which
was observed in the quadratic as well as in the square-root fitness
landscapes, is
the presence of spectral rigidity: the effect of recombination becomes
independent of the recombination rate for $\nu > 0$. 

In summary, from our generalization of the parallel or Crow-Kimura
model for an infinite population of evolving sequences Eq.\ (\ref{eq62}),
we conclude that two-parent recombination introduces a mild mutational load 
over discontinuous fitness functions, such as a single sharp peak, 
and thus it can shift the error-threshold transition.  
For smooth fitness functions, the effect of recombination 
depends on the sign of epistasis (see Fig.\ \ref{fig0}), in agreement with
the mutational deterministic hypothesis 
\cite{Kimura66,Kondrashov88,Kondrashov93,Kondrashov82}. 
We show this analytically in Appendix \ref{appendix14}.

In contrast
with horizontal gene transfer, recombination affects the structure of the
quasi-species (and the error threshold transition) for a sharp peak
fitness. We believe that this fundamental difference between 
horizontal gene transfer
and recombination is because of the fact that the latter can generate
a much larger diversity in the offspring per recombination event. Hence,
the diversity barrier that, as previously discussed in section II,
is imposed by the sharp exponential distribution in the sharp peak
case can be tunneled through due to the more radical mixing effects of
two-parent recombination. Our analytical 
theory, which provides explicit expressions for the
mean fitness $f_m$ and average composition $u$, 
is developed in the realistic regime ($N\rightarrow\infty$),
considering that typical viral genomes are $N\sim 10^3-10^4$.

\section{The Eigen model}
In this section, we present a generalization of the classical Eigen model
\cite{Eigen71,Eigen88,Eigen89},
including the exchange of genetic material between pairs
of individuals in an infinite population \cite{Park07},
\begin{eqnarray}
\frac{dq_{i}}{dt} = \sum_{j,k=1}^{2^{N}}[B_{ij}C_{jk}r_{k}-
\delta_{ij}\delta_{ik}D_{i}]q_{k}
\label{eq107}
\end{eqnarray}
Here, recombination as well as mutation are considered to be coupled
to the replication process. Recombination is represented 
by the coefficients $C_{jk}$, which in general will be functions
of the frequencies $q_{k}$, 
$C_{jk}\sim \delta_{jk} + \sum_{l}q_{l}\tilde{C}^{j}_{kl}/\sum_{k'}q_{k'}$. 

\subsection{Horizontal gene transfer of non-overlapping blocks}
In this recombination scheme, we consider the exchange of
blocks of genetic material between pairs of individuals in the
population. We consider the blocks to be non-overlapping, such that
we have $N/\bar{M}$ of them. We define a 
block index $0\leq b \leq N/\bar{M}-1$, and a site index
within each block to be $\bar{M}b+1\leq j_{b}\leq\bar{M}(b+1)$.
For this process, we have that the nonlinear 
recombination term in the differential
Eq.\ (\ref{eq107}) is
\begin{eqnarray}
C_{jk}&\sim& 
\left(1-\frac{\nu/\bar{M}}{N/\bar{M}}\right)\delta_{j,k} +  
\frac{\nu/\bar{M}}{N/\bar{M}}\nonumber\\
&&\times\prod_{b=0}^{N/\bar{M}-1}\left[
\prod_{j_{b}=\bar{M}b+1}^{\bar{M}(b+1)}
\delta_{s_{j_{b}}^{j},s_{j_{b}}^{k}}
\left(\delta_{s_{j_{b}},+1}\frac{1+u(j_{b})}{2}+
\delta_{s_{j_{b}},-1}\frac{1-u(j_{b})}{2}\right) \right]
\prod_{m\ne\{j_{b}\}}\delta_{s_{m}^{j},s_{m}^{k}}
\label{eq108}
\end{eqnarray}
The recombination operator representing this process, 
assuming the recombination
rate per block to be $\nu/\bar{M}$, becomes
\begin{eqnarray}
\hat{R} = \prod_{b=0}^{N/\bar{M}-1}\left[\left(1-\frac{\nu/\bar{M}}
{N/\bar{M}}\right)\prod_{j_{b}=\bar{M}b+1}^{\bar{M}(b+1)}\hat{I}_{j_{b}}
+\frac{\nu/\bar{M}}{N/\bar{M}}\prod_{j_{b}=\bar{M}b+1}^{\bar{M}(b+1)}
\hat{R}_{j_{b}}\right]
\label{eq109}
\end{eqnarray}
Here, we defined the single-site recombination operator as 
$\hat{R}_{j}=\vec{\hat{a}}^{\dagger}(j)D\vec{\hat{a}}(j)$, with the
matrix $D$ defined in Eq.\ (\ref{eq6}).
We consider the large N limit, while keeping $N/\bar{M}\simeq \mathcal{O}(N)$. 
Then,
the recombination operator defined in Eq.\ (\ref{eq109}) becomes, to order 
$\mathcal{O}(N^{-1})$
\begin{eqnarray}
\hat{R} = e^{-\frac{\nu}{\bar{M}}}e^{\frac{\nu}{N}\sum_{b=0}^{N/\bar{M}-1}
\prod_{j_{b}=\bar{M}b+1}^{\bar{M}(b+1)}
\vec{\hat{a}}^{\dagger}(j)D\vec{\hat{a}}(j)}
\label{eq110}
\end{eqnarray}

\subsubsection{The Hamiltonian}
The Hamiltonian operator for the Eigen model, including the horizontal
gene transfer process described by the operator Eq.\ (\ref{eq109})
is given by
\begin{eqnarray}
-\hat{H}&=&Ne^{-\mu+\frac{\mu}{N}\sum_{j=1}^{N}\vec{\hat{a}}^{\dagger}(j)
\sigma_{1}\vec{\hat{a}}(j)}e^{-\frac{\nu}{\bar{M}}
+\frac{\nu}{N}\sum_{b=0}^{N/\bar{M}-1}\prod_{j_{b}=\bar{M}b+1}^{\bar{M}(b+1)}
\left[\vec{\hat{a}}^{\dagger}(j_{b})D\vec{\hat{a}}
(j_{b})\right]}\nonumber\\
&&\times f\left[\frac{1}{N}\sum_{j=1}^{N}\vec{\hat{a}}^{\dagger}(j)\sigma_{3}
\vec{\hat{a}}(j)\right]-N d\left[\frac{1}{N}\sum_{j=1}^{N}
\vec{\hat{a}}^{\dagger}(j)\sigma_{3}\vec{\hat{a}}(j)\right]
\label{eq111}
\end{eqnarray}
The microscopic fitness function is $f(u)$ and degradation function
is $d(u)$. Here, the matrix $D$ is defined as in Eq.\ (\ref{eq6}).
We introduce a Trotter factorization of the evolution operator, in the 
basis of coherent states
\begin{eqnarray}
e^{-\hat{H}t}=\lim_{M\rightarrow\infty}\int\left[\prod_{k=1}^{M}\mathcal{D}
\vec{z}_{k}^{*}\mathcal{D}\vec{z}_{k}\right]|\vec{z}_{M}\rangle\left(
\prod_{k=1}^{M}\langle\vec{z}_{k}|e^{-\epsilon\hat{H}}|\vec{z}_{k-1}
\rangle\right)\langle\vec{z}_{0}|
\label{eq112}
\end{eqnarray}
As shown in Appendix \ref{appendix7}, the partition function is
\begin{eqnarray}
Z=\int\left[\mathcal{D}\bar{\xi}\mathcal{D}\xi\mathcal{D}\bar{\eta}
\mathcal{D}\eta\mathcal{D}\bar{\phi}\mathcal{D}\phi
\right]e^{-S\left[\bar{\xi},\xi,\bar{\eta},\eta,\bar{\phi},\phi\right]}
\label{eq124}
\end{eqnarray}
Here, the action is defined by
\begin{eqnarray}
S\left[\bar{\xi},\xi,\bar{\eta},\eta,\bar{\phi},\phi\right]=
-
N\int_{0}^{t}dt\left[-\bar{\xi}\xi-\bar{\eta}\eta-
\bar{\phi}\phi
+ e^{-\mu(1-\eta)-\nu/\bar{M}+\frac{\nu}{\bar{M}}\phi^{\bar{M}}}
f(\xi) - d(\xi)\right] - N\ln Q
\nonumber\\
\label{eq125}
\end{eqnarray}

\subsubsection{The saddle point limit}
We consider the saddle point limit of the 
action defined by Eq.\ (\ref{eq125}). 
In the saddle point limit, for long times, 
the trace defined by Eq.\ (\ref{eq123})
becomes
\begin{eqnarray}
\lim_{t\rightarrow\infty}\frac{\ln Q_{c}}{t}=\frac{\bar{\phi}_{c}}{2}+
\left[\bar{\xi}_{c}(\bar{\xi}_{c}+u\bar{\phi}_{c})+(\bar{\eta}_{c}
+\bar{\phi}_{c}/2)^{2}\right]^{1/2}
\label{eq126}
\end{eqnarray}
In this saddle-point limit, the action is given by
\begin{eqnarray}
\lim_{N,t\rightarrow\infty}
\frac{\ln Z}{Nt} &=& \lim_{t\rightarrow\infty}
\frac{-S_{c}}{Nt}\nonumber\\
&=&\max_{\xi_{c},\bar{\xi}_{c},\phi_{c},\bar{\phi}_{c},
\eta_{c},\bar{\eta}_{c}}\bigg\{f(\xi_{c})
e^{-\mu(1-\eta_{c})-\frac{\nu}{\bar{M}} +\frac{\nu}{\bar{M}}\phi_{c}^{\bar{M}}}
-d(\xi_{c})-\bar{\xi}_{c}\xi_{c}
-\bar{\eta}_{c}\eta_{c}-\bar{\phi}_{c}\phi_{c}\nonumber\\
&&
+
\frac{\bar{\phi}_{c}}{2}
+ \left[\bar{\xi}_{c}(\bar{\xi}_{c}+u\bar{\phi}_{c})+
(\bar{\eta}_{c}+\bar{\phi}_{c}/2)^{2}\right]^{1/2} 
\bigg\}\nonumber\\
\label{eq127}
\end{eqnarray} 
As shown in Appendix \ref{appendix8}
the mean fitness, defined from the saddle point action 
$f_{m}=\lim_{N,t\rightarrow\infty}\ln Z/Nt=-S_{c}/Nt$, is
\begin{eqnarray}
f_{m}=\max_{-1\leq \xi_{c}\leq 1}
\left\{e^{-\mu[1-\eta_{c}(\xi_{c})]-\frac{\nu}{\bar{M}}
\{1-[\phi_{c}(\xi_{c})]^{\bar{M}}\}}f(\xi_{c})-d(\xi_{c})\right\}
\label{eq140}
\end{eqnarray}
Here, the expressions $\phi_{c}(\xi_{c})$ and $\eta_{c}(\xi_{c})$ are
given by
\begin{eqnarray}
\phi_{c}(\xi_{c})=\frac{1+u\xi_{c}}{2}+\frac{\sqrt{1-\xi_{c}^{2}}}{2}
\frac{\mu+\frac{\nu}{2}(1-u^{2})\phi_{c}^{\bar{M}-1}}
{\left[\left(\mu+\frac{\nu}{2}\phi_{c}^{\bar{M}-1}\right)^{2}
-\frac{\nu^{2}u^{2}}{4}[\phi_{c}^{\bar{M}-1}]^{2}\right]^{1/2}}
\label{eq141}
\end{eqnarray}
\begin{eqnarray}
\eta_{c}(\xi_{c})=\sqrt{1-\xi_{c}^{2}}\frac{\mu+\frac{\nu}{2}
\phi_{c}^{\bar{M}-1}}
{\left[\left(\mu + \frac{\nu}{2}\phi_{c}^{\bar{M}-1}\right)^{2}-
\frac{\nu^{2}u^{2}}{4}[\phi_{c}^{\bar{M}-1}]^{2}\right]^{1/2}}
\label{eq142}
\end{eqnarray}
The average composition, $u$, is obtained from the self-consistency condition
$f_{m}=f(u)-d(u)$.

Eq.\ (\ref{eq140}) is an exact analytical expression for the
equilibrium mean fitness of an infinite population of evolving
sequences. This analytical expression is valid for
arbitrary permutation invariant replication rate $f(u)$ and
degradation rate $d(u)$.

\subsubsection{Examples}

We consider first the quadratic fitness case, 
$f(u)=k u^{2}/2+k_0$. By expanding the formulas 
Eqs.\ (\ref{eq140}), (\ref{eq141}) and (\ref{eq142})
near the error threshold
$\xi_{c}\sim 0$, $u \sim 0$, we obtain the phase boundary from the critical
condition
\begin{eqnarray}
k_{\rm{crit}}=\mu k_0\frac{1 + \nu/\mu}{1 + \nu/2\mu}
\label{eq143}
\end{eqnarray}
We notice that the phase boundary is qualitatively similar to the
horizontal gene transfer process analyzed in section II. A 2, Eq.\ (\ref{eq22}) 
for the parallel model. 
As in this former case, we notice that horizontal gene transfer 
introduces a mild mutational
load against selection for a smooth fitness (i.e. quadratic).

As a second example, we consider the square-root fitness
landscape $f(u)=k\sqrt{|u|} + 1$. 
In Table \ref{tabesqrt1}, we evaluate
our analytical Eqs.\ (\ref{eq140}--\ref{eq142}) for this particular case. 

\begin{table}
\begin{center}
\caption{Analytical results for horizontal gene transfer in the
Eigen model for the square-root
fitness $f(u)=k\sqrt{|u|}+1$, with $\bar{M}=3$.}
\label{tabesqrt1}
\begin{tabular}{|c|c|c|c|}\hline
$k$ & $\nu$ & $u^{\rm{analytic}}$\\\hline
3.0 & 0.0 & 0.3346 \\
3.0 & 0.5 & 0.3398\\
3.0 & 0.8 & 0.3422\\
3.0 & 1.5 & 0.3466\\\hline
5.0 & 0.0 & 0.3588\\
5.0 & 0.5 & 0.3642\\
5.0 & 0.8 & 0.3667\\
5.0 & 1.5 & 0.3713\\\hline
8.0 & 0.0 & 0.3741\\
8.0 & 0.5 & 0.3796\\
8.0 & 0.8 & 0.3822\\
8.0 & 1.5 & 0.3869\\\hline
\end{tabular}
\end{center}
\end{table}

From the results displayed in Table \ref{tabesqrt1}, we
notice that horizontal gene transfer increases
the average composition $u$ and therefore the mean
fitness of the population. This effect, which is
attributed to the negative epistasis introduced by
the square-root fitness (see Fig.\ \ref{fig0}), is in agreement with the
previous examples studied in the case of the
parallel model, and with the mutational deterministic hypothesis
\cite{Arjan07,Azevedo06,Phillips,Kondrashov93},
as we prove in Appendix \ref{appendix12_b}. 

As a third example, we consider the sharp peak fitness 
$f(u)=(A-A_{0})\delta_{u,1}+A_{0}$. In this case, the maximum
in Eq.\ (\ref{eq140}) 
corresponds to $\xi_{c}=1$.
From Eqs.\ (\ref{eq141}) and (\ref{eq142}), we have $\xi_{c}=(1+u)/2$, 
$\eta_{c}=0$, and hence after Eq.\ (\ref{eq140})
\begin{eqnarray}
f_{m}=A e^{-\mu -\frac{\nu}{\bar{M}}
\left[1-\left(\frac{1+u}{2}\right)^{\bar{M}}\right]}
\label{eq144}
\end{eqnarray}
The error threshold is given, for $u=0$ in Eq.\ (\ref{eq144}), by
the condition $Ae^{-\mu-\frac{\nu}{\bar{M}}[1-1/2^{\bar{M}}]}>A_0$.
However, we notice that $f_{m}(u=1)=A e^{-\mu}>f_m(u=0)$. Hence,
in the selected phase we have $u=1-\mathcal{O}(N^{-1})$.
The fraction of the population located at the peak $P_{0}$ is obtained
from the self-consistency condition $f_{m}=A P_{0} + A_{0}(1-P_{0})$
\begin{eqnarray}
P_{0} = \frac{A e^{-\mu}-A_{0}}{A - A_{0}}
\label{eq145}
\end{eqnarray}
After Eq.\ (\ref{eq145}), we find the true error threshold at 
$A_{\rm{crit}}=A_{0}e^{\mu}$, while the condition
$Ae^{-\mu-\frac{\nu}{\bar{M}}[1-2^{-\bar{M}}]}>A_0$ represents
the limit of metastability for initial conditions with $u\sim 0$.
 We notice that this result is
similar to the exact solution in the absence of horizontal
gene transfer \cite{Park06}.
Hence, as previously discussed in section I.A. for the
parallel model, we conclude that horizontal gene transfer 
does not affect the
structure of the quasi-species for a discontinuous, single
sharp peak fitness.

\subsection{Horizontal gene transfer for multiple-size blocks}

In analogy with the model treated in Section II.B, we consider the natural
extension of horizontal gene transfer of blocks with multiple size,
with average $\langle \bar{M} \rangle$ and 
$\langle \bar{M}\rangle/N = \mathcal{O}(N^{-1})$. 
Following a similar analysis as in the derivation of Eq.\ (\ref{eq41}), 
we define the recombination operator for multiple-size blocks as
\begin{eqnarray}
\hat{R} \sim e^{-\langle \bar{M} \rangle 
+ \frac{\langle\bar{M}\rangle}{N}\sum_{j=1}^{N}\vec{\hat{a}}^{\dagger}(j)D
\vec{\hat{a}}(j)}
\label{eq146}
\end{eqnarray}

\subsubsection{The Hamiltonian}

We consider horizontal gene transfer to be coupled to the replication process.
Moreover, we will consider that when replication occurs, a 
horizontal gene transfer event
also occurs with a probability $0\leq \nu/\langle\bar{M}\rangle\leq 1$. 
The Hamiltonian
operator for the Eigen model, including the horizontal gene transfer
process described by the operator Eq.\ (\ref{eq146}) is given by
\begin{eqnarray}
-\hat{H}&=& N e^{-\mu+\frac{\mu}{N}\sum_{j=1}^{N}\vec{\hat{a}}^{\dagger}(j)D
\vec{\hat{a}}(j)}\left(1-\frac{\nu}{\langle\bar{M}\rangle} 
+ \frac{\nu}{\langle\bar{M}\rangle} e^{-\langle\bar{M}\rangle+
\frac{\langle\bar{M}\rangle}{N}\sum_{j=1}^{N}\vec{\hat{a}}^{\dagger}(j)D
\vec{\hat{a}}(j)}\right)\nonumber\\
&&\times f\left[\frac{1}{N}\sum_{j=1}^{N}\vec{\hat{a}}^{\dagger}(j)\sigma_{3}
\vec{\hat{a}}(j)\right]-Nd\left[\frac{1}{N}\sum_{j=1}^{N}
\vec{\hat{a}}^{\dagger}(j)\sigma_{3}\vec{\hat{a}}(j) \right]\nonumber\\
\label{eq147}
\end{eqnarray}
We introduce a Trotter factorization
\begin{eqnarray}
e^{-\hat{H}t} = \lim_{M\rightarrow\infty}
\int [\mathcal{D}\vec{z}^{*}\mathcal{D}\vec{z}]|\vec{z}_{M}\rangle
\left(\prod_{j=1}^{M}
\langle\vec{z}_{k}|e^{-\epsilon\hat{H}}|\vec{z}_{k-1}\right)\langle\vec{z}_{0}|
\label{eq148}
\end{eqnarray}
As shown in Appendix \ref{appendix9}, the partition function is
\begin{eqnarray}
Z = \int\left[\mathcal{D}\bar{\xi}\mathcal{D}\xi\mathcal{D}\bar{\eta}
\mathcal{D}\eta\mathcal{D}\bar{\phi}\mathcal{D}\phi\right]
e^{-S\left[\bar{\xi},\xi,\bar{\eta},\eta,\bar{\phi},\phi\right]}
\label{eq159}
\end{eqnarray}
Here, the action in the continuous time limit is
\begin{eqnarray}
S\left[\bar{\xi},\xi,\bar{\eta},\eta,\bar{\phi},\phi\right]
&=&-N\int_{0}^{t}dt'\bigg\{-\bar{\xi}\xi-\bar{\eta}\eta-\bar{\phi}\phi
\nonumber\\
&&
+e^{-\mu(1-\eta)}[1-\frac{\nu}{\langle\bar{M}\rangle}
+\frac{\nu}{\langle\bar{M}\rangle} 
e^{-\langle\bar{M}\rangle(1-\phi)}]f(\xi)-
d(\xi)\bigg\}-N\ln Q
\label{eq160}
\end{eqnarray}

\subsubsection{The saddle point limit}

The saddle point limit is exact as $N\rightarrow\infty$ 
in Eq.\ (\ref{eq160}). After a similar 
procedure as in Section 3.A.2, we find the saddle point equation
for the mean fitness
\begin{eqnarray}
f_{m}=\max_{-1\leq\xi_{c}\leq 1}\left\{e^{-\mu(1-\eta_{c})}
[1-\frac{\nu}{\langle\bar{M}\rangle}+
\frac{\nu}{\langle\bar{M}\rangle} 
e^{-\langle\bar{M}\rangle(1-\phi_{c})}]f(\xi_{c})-
d(\xi_{c})\right\}
\label{eq161}
\end{eqnarray}
Here, the fields $\eta_{c}$ and $\phi_{c}$ are expressed as functions
of $\xi_{c}$
\begin{eqnarray}
\eta_{c}(\xi_{c})
=\sqrt{1-\xi_{c}^{2}}\frac{
\frac{\nu}{\langle\bar{M}\rangle}+
\left[1-\frac{\nu}{\langle\bar{M}\rangle}\right]
e^{\langle\bar{M}\rangle(1-\phi_{c})}+\frac{\nu}{2\mu}}{\left[
\left(\frac{\nu}{\langle\bar{M}\rangle}
+\left[1-\frac{\nu}{\langle\bar{M}\rangle} \right]
e^{\langle\bar{M}\rangle(1-\phi_{c})}+\frac{\nu}{2\mu}\right)^{2}
-\frac{u^{2}\nu^{2}}{4\mu^{2}}\right]^{1/2}}
\label{eq162}
\end{eqnarray}
\begin{eqnarray}
\phi_{c}(\xi_{c})=\frac{1+u\xi_{c}}{2}+\frac{\sqrt{1-\xi_{c}^{2}}}{2}
\frac{\frac{\nu}{\langle\bar{M}\rangle}
+\left[1-\frac{\nu}{\langle\bar{M}\rangle} \right]
e^{\langle\bar{M}\rangle(1-\phi_{c})}+\frac{\nu(1-u^{2})}{2\mu}}
{\left[
\left(\frac{\nu}{\langle\bar{M}\rangle}
+\left[1-\frac{\nu}{\langle\bar{M}\rangle} \right]
e^{\langle\bar{M}\rangle(1-\phi_{c})}+\frac{\nu}{2\mu}\right)^{2}
-\frac{u^{2}\nu^{2}}{4\mu^{2}}\right]^{1/2}}
\label{eq163}
\end{eqnarray}

Equations (\ref{eq161})--(\ref{eq163}) represent an exact analytical
solution for the equilibrium mean fitness of an infinite
population experiencing horizontal gene transfer 
of variable blocks size. This expression
is valid for arbitrary, permutation invariant replication rate
$f(u)$ and degradation rate $d(u)$. 

\subsubsection{Examples}

We consider first the sharp peak fitness 
$f(u)=(A-A_{0})\delta_{u,1}+A_{0}$. In this case, the maximum in 
Eq.\ (\ref{eq161}) is at $\xi_{c}=1$. 
From Eqs.\ (\ref{eq162}) and (\ref{eq163}), we
obtain $\eta_{c} = 0$ and $\phi_{c}=(1+u)/2$. Substituting these
values in Eq.\ (\ref{eq161}), we obtain for the mean fitness
\begin{eqnarray}
f_{m} = e^{-\mu}\left[1-\frac{\nu}{\langle\bar{M}\rangle}
+\frac{\nu}{\langle\bar{M}\rangle} 
e^{-\langle\bar{M}\rangle(1-u)/2}\right]A 
\label{eq164}
\end{eqnarray}
The error threshold for $u=0$ is obtained from Eq.\ (\ref{eq164})
by the condition $Ae^{-\mu}\left[1-\frac{\nu}{\langle\bar{M}\rangle}
+\frac{\nu}{\langle\bar{M}\rangle}
e^{-\langle\bar{M}\rangle/2}\right]>A_0$. However, we notice that
$f_{m}(u=1)=A e^{-\mu}> f_{m}(u=0)$. Therefore, in the
selected phase the average composition $u = 1 - \mathcal{O}(N^{-1})$, 
and the effect of
recombination becomes negligible for the sharp peak fitness. The fraction
of the population located at the peak $P_{0}$ is obtained from the
self-consistency condition $f_{m}=A P_{0} + A_{0}(1-P_{0})$
\begin{eqnarray}
P_{0}=\frac{A e^{-\mu}-A_{0}}{A-A_{0}}
\label{eq165}
\end{eqnarray}
From this expression, we find that the true error threshold for the sharp peak
fitness is $A_{\rm{crit}}=e^{\mu}A_{0}$, with the condition
$Ae^{-\mu}\left[1-\frac{\nu}{\langle\bar{M}\rangle}
+\frac{\nu}{\langle\bar{M}\rangle}
e^{-\langle\bar{M}\rangle/2}\right]>A_0$ representing the limit
for metastability for initial conditions with $u\sim 0$. 

As a second example, we consider the quadratic fitness 
$f(u)=k u^{2}/2+k_{0}$. An analytical expression for the
phase boundary is obtained from Eqs.\ (\ref{eq161}), (\ref{eq162}) and
(\ref{eq163}) near the error threshold $\xi_{c}\sim 0$, $u\sim 0$. We
find
\begin{eqnarray}
k_{\rm{crit}}=\mu k_{0}\frac{1+\frac{\nu}{\mu}}
{1+\frac{\nu}{2\mu}}
\label{eq166}
\end{eqnarray}
For small $\nu$, the critical value is 
$k_{\rm{crit}}\sim k_{0}(\mu + \nu/2)$.

As a final example, we consider the square-root fitness
$f(u)=k\sqrt{|u|} + 1$. Analytical results, as obtained
from Eqs.\ (\ref{eq161})--(\ref{eq163}) for this case,
are presented in
Table \ref{tabesqrt2}. 
\begin{table}
\begin{center}
\caption{Analytical results for horizontal gene transfer in the
Eigen model for the square-root
fitness $f(u)=k\sqrt{|u|}+1$, with $\langle\bar{M}\rangle=3$.}
\label{tabesqrt2}
\begin{tabular}{|c|c|c|c|}\hline
$k$ & $\nu$ & $u^{\rm{analytic}}$\\\hline
3.0 & 0.0 & 0.3346 \\
3.0 & 0.5 & 0.3409\\
3.0 & 0.8 & 0.3450\\
3.0 & 1.5 & 0.3546\\\hline
5.0 & 0.0 & 0.3588\\
5.0 & 0.5 & 0.3654\\
5.0 & 0.8 & 0.3695\\
5.0 & 1.5 & 0.3794\\\hline
8.0 & 0.0 & 0.3741\\
8.0 & 0.5 & 0.3809\\
8.0 & 0.8 & 0.3851\\
8.0 & 1.5 & 0.3950\\\hline
\end{tabular}
\end{center}
\end{table}

We notice that the results obtained for the horizontal gene transfer 
process with
variable block size agree with the corresponding ones when the
size of the recombination blocks is fixed. We recall that this
correspondence was also observed and discussed in the previous
section for the parallel model, so similar arguments apply
to the Eigen model as well. An analytical proof is provided
in Appendix \ref{appendix12_b}.

\subsection{The Eigen model with two-parent recombination}

For the Eigen model, we introduce the recombination process described
in Section II.C and illustrated in Fig.\ \ref{fig3}, 
which considers the exchange of genetic material
between pairs of sequences due to crossovers governed by the
polymerase switching from one parental chromosome to the other
with probability $p_{c}$ per site.
For the Eigen model, mutation and recombination are considered to be coupled
to the recombination process, as stated in the generic differential
equation Eq.\ (\ref{eq107}). We will consider that 
during replication, a sequence can recombine with probability
$\nu \leq 1$, or just replicate without recombining with
probability $1-\nu$.
This process is represented by the coefficients in Eq.\ (\ref{eq107})
\begin{eqnarray}
C_{jk}& = &(1-\nu)\delta_{j,k} + 
\frac{\nu}{2}\sum_{\{\alpha_{n}=\pm 1\}}\left\{
\left[\prod_{n=2}^{N}p_{c}^{\frac{1-\alpha_{n-1}\alpha_{n}}{2}}
(1-p_{c})^{\frac{1+\alpha_{n-1}\alpha_{n}}{2}}\right]\right.\nonumber\\
&&\times
\left.
\sum_{l=1}^{2^{N}}p_{l}
\prod_{n=1}^{N}\left(\frac{1+s_{n}^{k}s_{n}^{j}}{2}
\right)^{\frac{1+\alpha_{n}}{2}}
\left(\frac{1+s_{n}^{l}}{2}\delta_{s_{n}^{j},+1}
+\frac{1-s_{n}^{l}}{2}\delta_{s_{n}^{j},-1}\right)^{\frac{1-\alpha_{n}}{2}}
\right\}\nonumber\\
\label{eq167}
\end{eqnarray}
Here, again, $p_{l}=q_{l}/\sum_{l=1}^{2^{N}}q_{l}$ 
is the normalized probability
for the sequence $1\leq l \leq 2^{N}$.

In the spin Boson representation, 
we express the Eigen model Hamiltonian by the operator
\begin{eqnarray}
-\hat{H}&=& Ne^{-\mu+\frac{\mu}{N}\sum_{j=1}^{N}\vec{\hat{a}}^{\dagger}(j)
\sigma_{1}\vec{\hat{a}}(j)}\left[(1-\nu)\hat{I}+\nu 
g[\{\vec{\hat{a}}^{\dagger}(j)D_{j}^{l}
\vec{\hat{a}}(j)\}]\right]\nonumber\\
&&\times  f\left[\frac{1}{N}\sum_{j=1}^{N}
\vec{\hat{a}}^{\dagger}(j)\sigma_{3}\vec{\hat{a}}(j)\right]
-N d\left[\frac{1}{N}\sum_{j=1}^{N}\vec{\hat{a}}^{\dagger}(j)
\sigma_{3}\vec{\hat{a}}(j)\right]
\label{eq168}
\end{eqnarray}
Here, $g[\{\hat{R}_{j}^{l}\}]$ was defined in Eq.\ (\ref{eq63}), 
and the matrices $D_{j}^{l}$ were defined in Eq.\ (\ref{eq64}).
We introduce a Trotter factorization
\begin{eqnarray}
e^{-\hat{H}t}=\lim_{M\rightarrow\infty}\int[\mathcal{D}\vec{z}^{*}
\mathcal{D}\vec{z}]|\vec{z}_{M}\rangle\left(\prod_{k=1}^{M}
\langle\vec{z}_{k}|e^{-\epsilon\hat{H}}|\vec{z}_{k-1}\rangle 
\right)\langle\vec{z}_{0}|
\label{eq169}
\end{eqnarray}
As shown in Appendix \ref{appendix10}, the partition function is
\begin{eqnarray}
Z = \int\mathcal{D}\bar{\xi}\mathcal{D}\xi\mathcal{D}\bar{\eta}
\mathcal{D}\eta\mathcal{D}\bar{\phi}
\mathcal{D}\phi e^{-S[\bar{\xi},\xi,\bar{\eta},\eta,\bar{\phi},\phi]}
\label{eq178}
\end{eqnarray}
Here, the action is defined by
\begin{eqnarray}
S[\bar{\xi},\xi,\bar{\eta},\eta,\bar{\phi},\phi]
&=& -N\int_{0}^{t}dt'\left[-\bar{\xi}\xi-\bar{\eta}\eta 
-\bar{\phi}\phi\right.\nonumber\\
&&+ 
\left.
e^{-\mu(1-\eta)}(1-\nu+\nu g(\phi))f(\xi) - d(\xi)\right]
-N\ln Q\nonumber\\
\label{eq179}
\end{eqnarray}

\subsubsection{The saddle point limit}

For long times, a steady state condition is achieved. Then, the fields
become time-independent, and we have
\begin{eqnarray}
\lim_{t\rightarrow\infty}\frac{\ln Q_{c}}{t}= \frac{\bar{\phi}_{c}}{2}+
\left[\bar{\xi}_{c}(\bar{\xi}_{c}+u\bar{\phi}_{c})+
\left(\bar{\eta}_{c}+\frac{\bar{\phi}_{c}}{2}\right)^{2}\right]^{1/2}
\label{eq181}
\end{eqnarray}
We look for the saddle point solution from the action
\begin{eqnarray}
\lim_{N,t\rightarrow\infty}\frac{\ln Z}{Nt}=\lim_{t\rightarrow\infty} 
\frac{-S_{c}}{Nt}& =&\max_{\bar{\xi}_{c},\xi_{c},\bar{\eta}_{c},\eta_{c},
\bar{\phi}_{c},\phi_{c}}\bigg\{ 
-\bar{\xi}_{c}\xi_{c}-\bar{\eta}_{c}
\eta_{c}-\bar{\phi}_{c}\phi_{c}
+ e^{-\mu(1-\eta_{c})}(1-\nu+\nu g(\phi_{c}))
f(\xi_{c})\nonumber\\
&&-
d(\xi_{c})
+\frac{\bar{\phi}_{c}}{2}+
\left[\bar{\xi}_{c}(\bar{\xi}_{c}+u\bar{\phi}_{c})
+\left(\bar{\eta}_{c}+\frac{\bar{\phi}_{c}}{2}\right)^{2}\right]^{1/2}\bigg\}
\label{eq182}
\end{eqnarray}

Because of the singular behavior of the function $g(\phi_c)$, 
to find the saddle point we need to consider
three separate cases: $\phi_c < 1$, $\phi_c = 1$,
and $\phi_c = 1-\mathcal{O}(1/N)$.
We notice that the saddle
point analysis may not apply exactly, unless $g(\phi_c)=\delta_{\phi_c,1}$.

Case 1: $\phi_c < 1$.  The mean fitness is given by
\begin{eqnarray}
f_{m}^{(1)}=\max_{-1\leq\xi_{c}\leq 1}\{(1-\nu)
e^{-\mu\left[1-\sqrt{1-\xi_{c}^{2}}\right]}f(\xi_{c})-d(\xi_{c})\}
\label{eq184}
\end{eqnarray}
We note $\phi_c$ is  still given by Eq.\ (\ref{eq94}).

Case 2: $\phi_{c}=1$.  
The mean fitness is given by
\begin{eqnarray}
f_{m}^{(2)}=\max_{-1\leq\xi_{c}\leq 1}\{e^{-\mu
\left[1-\frac{1-u\xi_{c}-|u\xi_{c}-u^2|}
{1-u^{2}}\right]}f(\xi_{c})-d(\xi_{c})\}
\label{eq183}
\end{eqnarray}

Case 3: $\phi_c = 1-\mathcal{O}(1/N)$. In this case, additional analysis
is necessary to calculate the mean fitness due to the singular
behavior of the $g(\phi_c)$ function.
For a smooth fitness function, we can argue this case does not 
exist.  We first consider the Hamiltonian (\ref{eq168}) for the
case $g=0$.  In this case, the fitness function is
simply multiplied by $(1-\nu)$. If the degradation
function is zero, the largest
eigenvalue, $f_m$ is simply multiplied by $(1-\nu)$ relative to the
$\nu = 0$ case.  
Without degradation,  this result allows us to calculate the average
composition, $u_*$, from the implicit
relation $f_m(\nu) = (1- \nu) f_m(\nu=0) = f(u_*)$. 
With a non-zero degradation function, the equation for
$f_m(\nu)$ will be a bit more involved.
Alternatively, if we consider the differential 
equation for the unnormalized class probabilities, 
$dQ / dt = L Q$, we see that the 
differential operator $L$ looks like that in the absence of
recombination, save for a multiplication of $(1-\nu)$ in the fitness function.
Thus, the variance of the population is given by \cite{Park06}
$\sigma^2_u/N = 2 \mu u_* (1-\nu) f(u_*)/ [N ((1-\nu) f'(u_*) - d'(u_*))]$.
 Considering more carefully the
$g$ function, we find as before
this term is exponentially negligible
compared to the $-\nu P(u)$ term when $\sigma^2 < \sigma_u^2$.
In other words, we must strictly be in case 1 when 
\begin{eqnarray}
1- u_*^2 < 2 \mu u_* (1-\nu) f(u_*)/ [(1-\nu) f'(u_*) - d'(u_*)]
\label{eq192a}
\end{eqnarray}
We denote the value of $\nu$ at which 
\begin{eqnarray}
1-u_*^2 = 2 \mu u_* (1-\nu) f(u_*)/ [(1-\nu) f'(u_*) - d'(u_*)] {\rm ~at~} \nu = \nu_*
\label{eq192b}
\end{eqnarray}
as $\nu_*$.
Now, at this value of $\nu_*$ we have 
 $\int d {u_1} d {u_2} R_{u_1 u_2}^u P(u_1) P(u_2)
= P(u)$. Thus, the term proportional to $\nu$ in 
Hamiltonian (\ref{eq168}) exactly vanishes.  Thus, we have $d f_m / d \nu
= 0$ and
$dP(u) / d \nu = 0$
at this value of $\nu$.  
There is spectral rigidity.
This result implies that for $\nu > \nu_*$, 
the distribution $P(u)$ is independent of $\nu$, and that the
value of $u_*$ is constant.  In other words, the 
value of $f_m$ in case 2 must be constant with $\nu$.  Assuming $f_m$
varies continuously with $\nu$ in case 1,  and that the
fitness values for case 1 and case 2 are equal at a single
value of $\nu$, which mathematically may be negative,
case 2 is simply case 1 with the value $\nu = \nu_*$
\begin{eqnarray}
f_m (\nu > \nu_*) = f_m(\nu = \nu_*)
\label{eq192c}
\end{eqnarray}
Equations (\ref{eq184}), (\ref{eq183}) constitute an exact
analytical expression for the equilibrium mean fitness
of an infinite population of sequences evolving under
the dynamics of the Eigen model, and experiencing two-parent
recombination. These equations are exact for a smooth, permutation
invariant replication rate $f(u)$ and degradation rate $d(u)$.

For a non-smooth fitness function, additional analysis is necessary, since
$f'(u_*) - d'(u_*)$ is undefined, and $P(u)$ may no longer be Gaussian.

\subsubsection{Examples}

We investigate the phase diagrams, as predicted from our theoretical
equations, for two different fitness functions: A sharp peak and a quadratic
fitness landscape.

As an example, we consider the sharp peak fitness, 
$f(u)=(A-A_{0})\delta_{u,1}+A_{0}$. The maximum is obtained at $\xi_{c}=1$,
$u=1-\mathcal{O}(N^{-1})$. From Eqs.\ (\ref{eq183}) and (\ref{eq184}) we have
\begin{eqnarray}
f_{m}^{(2)}=A e^{-\mu} > f_{m}^{(1)} = (1-\nu)A e^{-\mu}
\label{eq185}
\end{eqnarray}
Hence, for the sharp peak fitness a single selective phase is observed.
In this case, the function $g(\phi_c)$ is not exactly a Kronecker
delta $\delta_{\phi_c,1}$, we are in case 3, and then we expect to observe
a small correction, approximately linear in $\nu$ from the prediction
of the saddle point analysis. By considering the differential equations
for the sharp peak case at zeroth-order in $\nu$, we find that the class
distributions satisfy
$e^{-\mu/2} \sum_k (r_k/N) P^{(0)}_k/2^k = f_{m}^{(0)}
 \sum_l P^{(0)}_l / 2^l$
 with $P^{(0)}_0 = (Ae^{-\mu}-A_0) / (A - A_0)$ and 
$f^{(0)}_{m} = A P^{(0)}_0 +
 A_0 (1-P^{(0)}_0) = A e^{-\mu}$.
Thus we find $S = \sum_l P^{(0)}_l / 2^l = (A - A_0) P^{(0)}_0 e^{-\mu/2} 
/ (f^{(0)}_{m} - A_0 e^{-\mu/2}) = (A e^{-\mu} - A_0) e^{-\mu/2}
/ (A e^{-\mu} - A_0 e^{-\mu/2}) $.
Thus, we find the recombination term 
$\sum_k (r_k/N) P^{(0)}_k/2^k \sum_l P^{(0)}_l / 2^l
= A e^{-\mu/2} S^2$.
Hence, we find that at first order in $\nu$, the fraction of the
population located at the peak is given by
\begin{eqnarray}
P_0 = \frac{Ae^{-\mu}-A_0}{A-A_0}-\nu e^{-\mu} 
\left[
\frac{A}{A-A_0}
-
A e^{-\mu/2} \frac{ A e^{-\mu} - A_0}
{( A e^{-\mu/2} - A_0)^2}
\right]
+ \mathcal{O}(\nu^2)
\label{eq186}
\end{eqnarray}
We note that this value of $f_m = A P_0 + A_0 (1-P_0)$ interpolates
between 
$f_m^{(1)}$ for $A e^{-\mu}/A_0 = 1$ and
a value intermediate to $f_m^{(1)}$  and
$f_m^{(2)}$ for $A e^{-\mu}/A_0 = \infty$.

As a second example, we consider the quadratic fitness 
$f(u)=k u^2/2+k_0$. By maximizing expressions Eq.\ (\ref{eq183}) 
\cite{note3} and
Eq.\ (\ref{eq184}), we obtain two selective phases S1 and S2, and a 
non-selective phase NS, defined by the equations
\begin{eqnarray}
S1: \; u &=& \left[
2 (1-\nu)
e^{-\mu\left[1-\sqrt{1-\xi_{c}^{2}}\right]} (\xi_c^2 / 2 + k_0/k)
- 2 k_0/k
\right]^{1/2}, \quad \nu  < \min(\nu_*, \nu_c)
\nonumber\\
S2: \; u &=& \left[\frac{1-2\mu k_0/k}{1 +\mu  }\right]^{1/2},
\quad \nu_c > \nu_* < \nu
\nonumber \\
NS: \; u&=&0, \quad {\rm otherwise}
\label{eq187} 
\end{eqnarray}
where in the S1 phase
\begin{eqnarray}
\xi_c^2 = 2 [  \sqrt{1 + \mu^2 (1 + 2k_0/k )}
-  1 - \mu^2 k_0/k ]/ \mu^2
\label{eq187b} 
\end{eqnarray}
and we have defined
\begin{eqnarray} 
\nu_c &=& 1 - \frac{k_0}{k}
e^{\mu\left[1-\sqrt{1-\xi_{c}^{2}}\right]}/ (\xi_c^2 / 2 + k_0/k)
\nonumber \\
\nu_* &=& 1 - \frac{ k + 2 k_0}{2 k (1 + \mu  )}
e^{\mu\left[1-\sqrt{1-\xi_{c}^{2}}\right]}/ (\xi_c^2 / 2 + k_0/k)
\label{eq187a}
\end{eqnarray}
where $\xi_c^2$ is given by Eq.\ (\ref{eq187b}).
The phase structure is defined by the conditions:
For $2\mu k_0 / k \ge 1$, 
the system is in S1 if $\nu < \nu_c$, or in
NS if $\nu\ge\nu_c$; for $2\mu k_0/k < 1$,
the system is in S1 if $\nu\leq\nu_*$, 
or in S2 if $\nu>\nu_*$.
From Eq.\ (\ref{eq187a}), we notice that at $2\mu k_0/k= 1$,
$\nu_c = \nu_*$.

We note that the phase transition between case 1 and case 2 is exactly
as predicted by Eq.\ (\ref{eq192b}).
We further note that the mean fitness is independent
of $\nu$ for $\nu > \nu_*$, exactly
as predicted by Eq.\ (\ref{eq192c}).

As a final example, we consider the square-root fitness 
$f(u)=k \sqrt{|u|}+1$. By maximizing expressions Eq.\ (\ref{eq183}) 
\cite{note3} and
Eq.\ (\ref{eq184}) for the square-root fitness landscape, we
obtain the results presented in Table \ref{taberecsqrt}
\begin{table}
\begin{center}
\caption{Analytical results for two-parent recombination in the
Eigen model for the square-root
fitness $f(u)=k \sqrt{|u|}+1$.}
\label{taberecsqrt}
\begin{tabular}{|c|c|c|}\hline
$k/\mu$ & $\nu/\mu$ & $u^{\rm{analytic}}$\\\hline
4.0 & 0.0 & 0.3493 \\
4.0 & 0.1 & 0.3892 \\
4.0 & 0.2 & 0.3892 \\
4.0 & 0.5 & 0.3892 \\\hline
3.0 & 0.0 & 0.3346 \\
3.0 & 0.1 & 0.3892 \\
3.0 & 0.2 & 0.3892 \\
3.0 & 0.5 & 0.3892 \\\hline
\end{tabular}
\end{center}
\end{table}
From the results displayed in Table \ref{taberecsqrt}, we
observe a similar qualitative behavior as in the two-parent recombination
for the parallel case, Table \ref{tabrecsqrt}. In the square-root fitness,
recombination introduces a favorable effect over selection, which
can be attributed to negative epistasis (see Fig.\ \ref{fig0}) according to the
mutational deterministic hypothesis 
\cite{Arjan07,Azevedo06,Phillips,Kondrashov93}, as shown in 
Appendix \ref{appendix15}.
Spectral rigidity is also observed in this case when $\nu>0$. 

\section{Conclusion}

We have generalized two classical models of evolutionary biology, the
Crow-Kimura and Eigen models.  We have introduced inter-individual transfer
of genetic information to these models, bringing them closer to the modern
understanding of evolutionary biology.  For both models, we showed how to
incorporate horizontal gene transfer.  We showed that these generalized
models may be written in an equivalent field-theoretic formulation. 
This mapping allows us to apply the powerful mathematical techniques
of quantum field theory to obtain exact analytical solutions.
For
fitness landscapes that depend only on distance from a wild-type genome and
for long genome lengths, we are able to solve for the mean population
fitness for arbitrary functional forms of the fitness.  Horizontal gene
transfer of $\bar M$ genetic units was shown to be analogous to horizontal
gene transfer of one genetic unit, with a suitably scaled horizontal gene
transfer rate.

We also showed how to incorporate recombination to these classical models, as
might occur in viral super- or co-infection.  This case seems at first
glance far more non-linear, since on average half of the genetic material is
taken from each parent to make the child, rather than $O(1)$ genes as in
horizontal gene transfer. Somewhat surprisingly, we were able to exactly
solve the two-parent 
recombination case for both the Eigen and Crow-Kimura model as
well.  In the limit of a long genome and for fitness landscapes that depend
on the distance from a wild-type genome, we find that
the mean population fitness is independent of the average cross-over length
in the recombination process.
We also find two selected phases.  The phase for large recombination rates
is spectrally rigid, with the mean fitness and population distribution
independent of the rate of recombination.

We proved the mutational deterministic hypothesis holds for horizontal
gene transfer or recombination in both the parallel (Kimura) 
and Eigen models. That is, horizontal gene transfer and recombination reduce 
the mean fitness in the presence of positive epistasis 
and increase the fitness in the presence of negative 
epistasis (see Fig.\ \ref{fig0} and 
Appendices \ref{appendix12}, \ref{appendix12_b}, \ref{appendix14}, and
\ref{appendix15}).  

For a discontinuous, sharp peak fitness landscape, we found
that horizontal gene transfer does not affect the structure
of the quasi-species distribution or the error threshold
transition. For the sharp peak fitness function, 
the only appreciable effect of horizontal gene transfer is related to
the potential emergence of metastability depending
on the initial conditions, and we analytically determined the region
of parameters space in which this situation may occur. On the other
hand, even for the sharp peak fitness function, 
two-parent recombination induces enough mixing to enhance
diversity in systems evolving under a sharp peak replication rate,
thus changing the quasi-species distribution and shifting the
error threshold transition. We found explicit analytical
expressions for this shift.

For smooth fitness landscapes, these genetic transfers affect the
steady-state population distribution and mean fitness. Recombination and
horizontal gene transfer may, of course, dramatically change the dynamics of
the evolution process as well.  The most dramatic impact of these exchanges
of genetic material is expected for fitness landscapes that have a
correlated, biological structure that is conjugate to these exchanges
\cite{Sun07}.  Analytic investigation of such correlated fitness
landscapes is perhaps one of the next steps in the development of modern
theories of evolution.

\section*{Acknowledgments}
This work was supported by DARPA under the FunBio program and by the
Korea Research Foundation.

\appendix

\section{}
\label{appendix1}

We consider Eq.\ (\ref{eq9}) for horizontal gene transfer of blocks
of fixed length $\bar M$ in the parallel model.
For $\epsilon=t/M$ and $M\rightarrow\infty$, we have
\begin{eqnarray}
\langle \vec{z}_{k}|e^{-\epsilon \hat{H}}|\vec{z}_{k-1}\rangle \simeq
\langle \vec{z}_{k}|\vec{z}_{k-1}\rangle -\epsilon\langle \vec{z}_{k}|
\hat{H}|\vec{z}_{k-1}\rangle\simeq \langle \vec{z}_{k}|\vec{z}_{k-1}\rangle
e^{-\epsilon\frac{\langle \vec{z}_{k}|
\hat{H}|\vec{z}_{k-1}\rangle}{\langle \vec{z}_{k}|\vec{z}_{k-1}\rangle}}.
\label{eq10}
\end{eqnarray}
For the Hamiltonian matrix elements in the coherent states basis, we obtain
to order $\mathcal{O}(N^{0})$
\begin{eqnarray}
-\frac{\langle \vec{z}_{k}|\hat{H}|\vec{z}_{k-1}\rangle}
{\langle \vec{z}_{k}|\vec{z}_{k-1}\rangle} 
& = & N f\left[\frac{1}{N}
\sum_{j=1}^{N}\vec{z}^{*}_{k}(j)\sigma_{3}\vec{z}_{k-1}(j)\right]+
\mu\sum_{j=1}^{N}[\vec{z}_{k}^{*}(j)\sigma_{1}\vec{z}_{k-1}(j)
-1]\nonumber\\
&& +  \nu\sum_{b=0}^{N/\bar{M}-1}\left[\prod_{j_{b}=\bar{M}b+1}^{\bar{M}(b+1)}
\vec{z}^{*}_{k}(j_{b})D\vec{z}_{k-1}(j_{b})-1\right]
\label{eq11}
\end{eqnarray}
We introduce the auxiliary field
\begin{eqnarray}
\xi_{k}=\frac{1}{N}\sum_{j=1}^{N}\vec{z}_{k}^{*}(j)\sigma_{3}\vec{z}_{k-1}(j)
\label{eq12}
\end{eqnarray}
and the conjugate field $\bar{\xi}_{k}$ to 
enforce the constraint via a Laplace representations of the delta
function. Substituting into Eq.\ (\ref{eq11}) into Eq.\ (\ref{eq9}), we obtain
\begin{eqnarray}
e^{-\hat{H}t} & = & \lim_{M\rightarrow\infty}
\int\left[\mathcal{D}\vec{z}^{*}\mathcal{D}\vec{z}\right]
\int\left[\prod_{k=0}^{M}\frac{i\epsilon 
N d\bar{\xi}_{k}d\xi_{k}}{2\pi}\right]
|\vec{z}_{M}\rangle\langle\vec{z}_{0}|\nonumber \\
&&\times e^{\sum_{k=1}^{M}\sum_{j=1}^{N}\left\{-1/2[\vec{z}_{k}^{*}(j)\cdot
\vec{z}_{k}(j)+\vec{z}_{k-1}^{*}(j)\cdot\vec{z}_{k-1}(j)-2\vec{z}_{k}^{*}(j)
\cdot\vec{z}_{k-1}(j)]+\epsilon [\vec{z}_{k}^{*}(j)(\bar{\xi}_{k}\sigma_{3}+
\mu\sigma_{1})\vec{z}_{k-1}(j)]\right\}}\nonumber \\
&& \times  e^{-\epsilon N \sum_{k=1}^{M}\left[\bar{\xi}_{k}\xi_{k}
+\mu+
\frac{\nu}{\bar{M}}-f(\xi_{k})
-\frac{\nu}{N}\sum_{b=0}^{N/\bar{M}-1}
\prod_{j_{b}=\bar{M}b+1}^{\bar{M}(b+1)}
\vec{z}_{k}^{*}(j_{b})D\vec{z}_{k-1}(j_{b})\right]}.
\nonumber\\
\label{eq13}
\end{eqnarray}

The contribution of the 
interaction term $\frac{\nu}{N}\sum_{b=0}^{N/\bar{M}-1}
\prod_{j_{b}=\bar{M}b+1}^{\bar{M}(b+1)}
\vec{z}_{k}^{*}(j_{b})D\vec{z}_{k-1}(j_{b})$ to the
partition function
can be treated to arbitrary order in perturbation theory
using the formula 
$Z = Z_{0}\langle e^{-\delta S} \rangle_{0}$, and
its contribution shown 
to be site-independent. Moreover, this reference perturbation theory has
$\mathcal{O}(N^{-1})$ fluctuations. Thus, it can be
shown that with an
error $\mathcal{O}(\bar{M}/N)$ at all orders in perturbation theory, 
we obtain the same partition function when substituting this interaction
term by 
$\frac{\nu}{\bar{M}}
\left(1/N\sum_{j=1}^{N}\vec{z}_{k}^{*}(j)D\vec{z}_{k-1}(j)\right)^{\bar{M}}$. 
Therefore, we define the auxiliary field
\begin{eqnarray}
\phi_{k} = \frac{1}{N}\sum_{j=1}^{N}\vec{z}_{k}^{*}(j)D\vec{z}_{k-1}(j).
\label{eq14}
\end{eqnarray}
We obtain the partition function from the trace of the evolution
operator, Eq.\ (\ref{eq13}), projected onto physical states \cite{Park06}
\begin{eqnarray}
Z  = {\rm{Tr}}
\left[e^{-\hat{H}t}\hat{P}\right]
 =  \int_{0}^{2\pi}\left[\prod_{j=1}^{N}
\frac{d\lambda_{j}}{2\pi}e^{-i\lambda_{j}}\right]
\lim_{M\rightarrow\infty}
\int\left[\prod_{k=0}^{M}\mathcal{D}\vec{z}^{*}_{k}\mathcal{D}
\vec{z}_{k}\right]
\left . e^{-S[\vec{z}^{*},\vec{z}]}\right|_{\vec{z}_{0}
=e^{i\lambda}\vec{z}_{M}}.
\label{eq15}
\end{eqnarray}
By inserting Eq.\ (\ref{eq14}), we obtain
\begin{eqnarray}
Z &=&\lim_{M\rightarrow\infty}
\int\left[\mathcal{D}\bar{\xi}\mathcal{D}\xi
\mathcal{D}\bar{\phi}\mathcal{D}\phi\right] e^{-N\epsilon\sum_{k=1}^{M}
[\bar{\xi}_{k}\xi_{k}+
\bar{\phi}_{k}\phi_{k}-f(\xi_{k})+\mu+\frac{\nu}{\bar{M}}
-\frac{\nu}{\bar{M}}
\phi_{k}^{\bar{M}}]}\nonumber\\
&&\times\int_{0}^{2\pi}\left[\prod_{j=1}^{N}\frac{d\lambda_{j}}{2\pi}
e^{-i\lambda_{j}}\right] 
\int\left[\prod_{k=0}^{M}\mathcal{D}\vec{z}_{k}^{*}
\mathcal{D}\vec{z}_{k}\right]\left . e^{-\sum_{j=1}^{N}\sum_{k,l=1}^{M}
\vec{z}_{k}^{*}(j)S_{kl}(j)\vec{z}_{l}(j)}\right|_{\vec{z}_{M}=e^{i\lambda}
\vec{z}_{0}}\nonumber\\
\label{eq16}
\end{eqnarray}
The matrix $S(j)$ in Eq.\ (\ref{eq16}) is defined by
\begin{eqnarray}
S(j)=\left(\begin{array}{ccccc}I & 0 & 0 & \ldots & -e^{i\lambda_{j}}
A_{1}\\
-A_{2} & I & 0 & \ldots & 0\\ 0 & -A_{3} & I & \ldots & 0\\
\vdots &\ldots &\ldots & \ldots & 0\\0 & \ldots & 0 & -A_{M} & I\end{array}
\right) \label{eq17}
\end{eqnarray}
Here $A_{k}=I+\epsilon(\bar{\xi}_{k}\sigma_{3}+\mu\sigma_{1}+
\bar{\phi}_{k}D)$.

After calculating the Gaussian integral over the coherent state fields, we
obtain
\begin{eqnarray}
& & \lim_{M\rightarrow\infty}
\int_{0}^{2\pi}\prod_{j=1}^{N}\frac{d\lambda_{j}}{2\pi}e^{-i\lambda_{j}} 
\int\left[\prod_{k=0}^{M}\mathcal{D}\vec{z}^{*}_{k}\mathcal{D}\vec{z}_{k}
\right]e^{-\sum_{j=1}^{N}\sum_{k,l=1}^{M}\vec{z}^{*}_{k}(j)S_{kl}(j)
\vec{z}_{l}(j)}\nonumber\\
& = & \lim_{M\rightarrow\infty}
\int_{0}^{2\pi}\prod_{j=1}^{N}\frac{d\lambda_{j}}{2\pi}
e^{-i\lambda_{j}}\left[\det S(j)\right]^{-1}\nonumber\\
& = &\lim_{M\rightarrow\infty}
\int_{0}^{2\pi}\prod_{j=1}^{N}\frac{d\lambda_{j}}{2\pi}e^{-i\lambda_{j}}
e^{-{\rm{Tr}}\ln[I-e^{i\lambda_{j}}\hat{T}
\exp(\epsilon\sum_{k=1}^{M}\bar{\xi}_{k}
\sigma_{3}+\mu\sigma_{1}+\bar{\phi}_{k}D)]}\nonumber\\
& = &\lim_{M\rightarrow\infty}\prod_{j=1}^{N}{\rm{Tr}}\:\hat{T}
e^{\epsilon\sum_{k=1}^{M}(\bar{\xi}_{k}\sigma_{3}+
\mu\sigma_{1}+\bar{\phi}_{k}D)}= Q^{N},
\label{eq18}
\end{eqnarray}
where $\hat{T}$ is the time ordering operator and
\begin{eqnarray}
Q={\rm{Tr}}\:\hat{T}e^{\int_{0}^{t}
dt'(\bar{\xi}\sigma_{3}+\mu\sigma_{1}+\bar{\phi}D)}.
\label{eq19}
\end{eqnarray}
With this result the partition function
in Eq.\ (\ref{eq16}) becomes Eq.\ (\ref{eq20}).

\section{}
\label{appendix2}
From Eq.\ (\ref{eq23}), we obtain the saddle-point equations with respect
to the fields $\bar{\xi}_{c}$, $\bar{\phi}_{c}$ for
horizontal gene transfer of blocks of fixed length $\bar M$
in the parallel model:
\begin{eqnarray}
\frac{\delta}{\delta\bar{\xi}_{c}}\left(\frac{-S_{c}}{Nt} \right)
=-\xi_{c}+\frac{2\bar{\xi}_{c}+u\bar{\phi}_{c}}
{2\left[\bar{\xi}_{c}(\bar{\xi}_{c}+u\bar{\phi}_{c})+
\left(\mu+\frac{\bar{\phi}_{c}}{2}\right)^{2}\right]^{1/2}}=0
\label{eq24}
\end{eqnarray}

\begin{eqnarray}
\frac{\delta}{\delta\bar{\phi}_{c}}\left(\frac{-S_{c}}{Nt}\right)
 = -\phi_{c}
+\frac{1}{2}
 +  \frac{u\bar{\xi}_{c}+\mu+\frac{\bar{\phi}_{c}}{2}}
{2\left[\bar{\xi}_{c}(\bar{\xi}_{c}+u\bar{\phi}_{c})+
\left(\mu+\frac{\bar{\phi}_{c}}{2}\right)^{2}\right]^{1/2}}=0.
\label{eq25}
\end{eqnarray}
Then, the system of Eqs.\ (\ref{eq24}) and (\ref{eq25}) reduces to
\begin{eqnarray}
\xi_{c}=\frac{\bar{\xi}_{c}+\frac{u}{2}\bar{\phi}_{c}}
{\left[\bar{\xi}_{c}(\bar{\xi}_{c}+u\bar{\phi}_{c})+
\left(\mu+\frac{\bar{\phi}_{c}}{2}\right)^{2}\right]^{1/2} }
\label{eq26}
\end{eqnarray}
\begin{eqnarray}
\phi_{c}-\frac{1}{2}=\frac{u\bar{\xi}_{c}+\mu+\frac{\bar{\phi}_{c}}{2}}
{2\left[\bar{\xi}_{c}(\bar{\xi}_{c}+u\bar{\phi}_{c})+\left(
\mu+\frac{\bar{\phi}_{c}}{2} \right)^{2}\right]^{1/2}}.
\label{eq27}
\end{eqnarray}
We eliminate $\bar{\xi}_{c}$, $\bar{\phi}_{c}$, to obtain
\begin{eqnarray}
\frac{-S_{c}}{Nt}&=&\max_{\xi_{c},\phi_{c}}\left\{f(\xi_{c})-\mu-
\frac{\nu}{\bar{M}}+\frac{\nu}{\bar{M}}\phi_{c}^{\bar{M}}
+\frac{\mu}{1-u^{2}}(2\phi_{c}-1-u\xi_{c})\right.\nonumber\\
&&-
\left.
\frac{\mu |u|}{1-u^{2}}\left[(2\phi_{c}-1-u\xi_{c})^{2}-
(1-u^{2})(1-\xi_{c}^{2})\right]^{1/2}
\right\}.
\label{eq28}
\end{eqnarray}
Finally, we look for an extremum in $\phi_{c}$,
\begin{eqnarray}
\frac{\delta}{\delta\phi_{c}}
\left(\frac{-S_{c}}{Nt} \right)  =  \frac{\nu}{\bar{M}}
\bar{M}\phi_{c}^{\bar{M}-1}+\frac{2\mu}{1-u^{2}} -  \frac{\mu |u|}{1-u^{2}}\frac{2(2\phi_{c}-1-u\xi_{c})}
{\left[(2\phi_{c}-1-u\xi_{c})^{2}-(1-u^{2})(1-\xi_{c}^{2})\right]^{1/2}}=0.
\nonumber\\
\label{eq29}
\end{eqnarray}
We solve for $\phi_{c}$ as a function of $\xi_{c}$ from this equation
\begin{eqnarray}
\phi_{c}(\xi_{c})=\frac{1+u\xi_{c}}{2}+\frac{\sqrt{1-\xi_{c}^{2}}}{2}
\frac{\sqrt{1-u^{2}}}{\left[1-\left(\frac{u}
{1+\frac{\nu}{2\mu}(1-u^{2})\phi_{c}^{\bar{M}-1}}\right)^{2}\right]^{1/2}}.
\label{eq30}
\end{eqnarray}
Substituting into Eq.\ (\ref{eq28}), we obtain for the mean fitness or average
replication rate Eq.\ (\ref{eq31}).

\section{}
\label{appendix3}
We consider Eq.\ (\ref{eq44}) for horizontal gene transfer of 
blocks of variable length in the parallel model.
For $\epsilon=t/M$ and $M\rightarrow\infty$, we have
\begin{eqnarray}
\langle \vec{z}_{k}|e^{-\epsilon \hat{H}}|\vec{z}_{k-1}\rangle \simeq
\langle \vec{z}_{k}|\vec{z}_{k-1}\rangle -\epsilon\langle \vec{z}_{k}|
\hat{H}|\vec{z}_{k-1}\rangle\simeq \langle \vec{z}_{k}|\vec{z}_{k-1}\rangle
e^{-\epsilon\frac{\langle \vec{z}_{k}|
\hat{H}|\vec{z}_{k-1}\rangle}{\langle \vec{z}_{k}|\vec{z}_{k-1}\rangle}}.
\label{eq45}
\end{eqnarray}
For the Hamiltonian matrix elements in the coherent states basis, we obtain
\begin{eqnarray}
-\frac{\langle \vec{z}_{k}|\hat{H}|\vec{z}_{k-1}\rangle}
{\langle \vec{z}_{k}|\vec{z}_{k-1}\rangle}&=&N f\left[\frac{1}{N}
\sum_{j=1}^{N}\vec{z}^{*}_{k}(j)\sigma_{3}\vec{z}_{k-1}(j)\right]+
\mu\sum_{j=1}^{N}[\vec{z}_{k}^{*}(j)\sigma_{1}\vec{z}_{k-1}(j)
-1]\nonumber\\
&&
+\frac{\nu}{\langle\bar{M}\rangle} N e^{-\langle\bar{M}\rangle 
+ \frac{\langle\bar{M}\rangle}{N}\sum_{j=1}^{N}
\vec{z}^{*}_{k}(j)D\vec{z}_{k-1}(j)} - \frac{\nu}{\langle\bar{M}\rangle} N.
\label{eq46}
\end{eqnarray}
We introduce the fields
\begin{eqnarray}
\xi_{k}=\frac{1}{N}\sum_{j=1}^{N}\vec{z}_{k}^{*}(j)\sigma_{3}\vec{z}_{k-1}(j)
\label{eq47}
\end{eqnarray}
\begin{eqnarray}
\phi_{k} = \frac{1}{N}\sum_{j=1}^{N}\vec{z}_{k}^{*}(j)D\vec{z}_{k-1}(j)
\label{eq48}
\end{eqnarray}
and the conjugate fields $\bar{\phi}_{k}$ and $\bar{\xi}_{k}$ to 
enforce the constraints via Laplace representations of the Dirac
delta functions. Substituting into Eq.\ (\ref{eq44}), we obtain
\begin{eqnarray}
e^{-\hat{H}t}&=&\lim_{M\rightarrow\infty}
\int\left[\mathcal{D}\vec{z}^{*}\mathcal{D}\vec{z}\right]
\int\left[\prod_{k=1}^{M}\frac{i\epsilon 
N d\bar{\xi}_{k}d\xi_{k}}{2\pi}
\frac{i\epsilon N d\bar{\phi}_{k}d\phi_{k}}{2\pi} \right]
|\vec{z}_{M}\rangle\langle\vec{z}_{0}|\nonumber \\
&&\times e^{\sum_{k=1}^{M}\sum_{j=1}^{N}\left\{-1/2[\vec{z}_{k}^{*}(j)\cdot
\vec{z}_{k}(j)+\vec{z}_{k-1}^{*}(j)\cdot\vec{z}_{k-1}(j)-2\vec{z}_{k}^{*}(j)
\cdot\vec{z}_{k-1}(j)]+\epsilon [\vec{z}_{k}^{*}(j)(\bar{\xi}_{k}\sigma_{3}+
\mu\sigma_{1}+\bar{\phi}_{k}D)\vec{z}_{k-1}(j)]\right\}}\nonumber \\
&&\times  e^{-\epsilon N \sum_{k=1}^{M}\left[\bar{\xi}_{k}\xi_{k}
+\bar{\phi}_{k}\phi_{k}+\mu+
\frac{\nu}{\langle\bar{M}\rangle}-f(\xi_{k})
-\frac{\nu}{\langle\bar{M}\rangle} 
e^{-\langle\bar{M}\rangle (1-\phi_{k})}\right]}
\label{eq49}
\end{eqnarray} 
We obtain the partition function from the trace of the evolution
operator Eq.\ (\ref{eq49})
\begin{eqnarray}
Z &=&{\rm{Tr}}\left[e^{-\hat{H}t}\hat{P}\right]\nonumber\\
&=&\int_{0}^{2\pi}\left[\prod_{j=1}^{N}
\frac{d\lambda_{j}}{2\pi}e^{-i\lambda_{j}}\right]\lim_{M\rightarrow\infty}
\int\left[\prod_{k=1}^{M}D\vec{z}^{*}_{k}D\vec{z}_{k}\right]
\left . e^{-S[\vec{z}^{*},\vec{z}]}\right|_{\vec{z}_{0}
=e^{i\lambda}\vec{z}_{M}}
\label{eq50}
\end{eqnarray}
By inserting Eq.\ (\ref{eq49}), we obtain
\begin{eqnarray}
Z &=&\lim_{M\rightarrow\infty}\int\left[\mathcal{D}\bar{\xi}\mathcal{D}\xi
\mathcal{D}\bar{\phi}\mathcal{D}\phi\right] e^{-N\epsilon\sum_{k=1}^{M}
[\bar{\xi}_{k}\xi_{k}+
\bar{\phi}_{k}\phi_{k}-f(\xi_{k})+\mu+\frac{\nu}{\langle\bar{M}\rangle}
-\frac{\nu}{\langle\bar{M}\rangle} 
e^{-\langle\bar{M}\rangle (1-\phi_{k})}]}\nonumber\\
&&\times\int_{0}^{2\pi}\left[\prod_{j=1}^{N}\frac{d\lambda_{j}}{2\pi}
e^{-i\lambda_{j}}\right] 
\int\left[\prod_{k=1}^{M}\mathcal{D}\vec{z}_{k}^{*}
\mathcal{D}\vec{z}_{k}\right]\left . e^{-\sum_{j=1}^{N}\sum_{k,l=1}^{M}
\vec{z}_{k}^{*}(j)S_{kl}(j)\vec{z}_{l}(j)}\right|_{\vec{z}_{M}=e^{i\lambda}
\vec{z}_{0}}
\label{eq51}
\end{eqnarray}
The matrix $S(j)$ in Eq.\ (\ref{eq51}) is defined by
\begin{eqnarray}
S(j)=\left(\begin{array}{ccccc}I & 0 & 0 & \ldots & -e^{i\lambda_{j}}
A_{1}\\
-A_{2} & I & 0 & \ldots & 0\\ 0 & -A_{3} & I & \ldots & 0\\
\vdots &\ldots &\ldots & \ldots & 0\\0 & \ldots & 0 & -A_{M} & I\end{array}
\right) 
\label{eq52}
\end{eqnarray}
where $A_{k}=I+\epsilon(\bar{\xi}_{k}\sigma_{3}+\mu\sigma_{1}+
\bar{\phi}_{k}D)$.

After calculating the Gaussian integral over the coherent state fields, we
obtain
\begin{eqnarray}
&&\lim_{M\rightarrow\infty}
\int_{0}^{2\pi}\prod_{j=1}^{N}\frac{d\lambda_{j}}{2\pi}
e^{-i\lambda_{j}}
\int\left[\prod_{k=1}^{M}\mathcal{D}\vec{z}^{*}_{k}\mathcal{D}\vec{z}_{k}
\right]e^{-\sum_{j=1}^{N}\sum_{k,l=1}^{M}\vec{z}^{*}_{k}(j)S_{kl}(j)
\vec{z}_{l}(j)}\nonumber\\
&=&\lim_{M\rightarrow\infty}
\int_{0}^{2\pi}\prod_{j=1}^{N}\frac{d\lambda_{j}}{2\pi}
e^{-i\lambda_{j}}\left[\det S(j)\right]^{-1}\nonumber\\
&=&\lim_{M\rightarrow\infty} 
\int_{0}^{2\pi}\prod_{j=1}^{N}\frac{d\lambda_{j}}{2\pi}e^{-i\lambda_{j}}
e^{-{\rm{Tr}}
\ln[I-e^{i\lambda_{j}}\hat{T}\exp(\epsilon\sum_{k=1}^{M}\bar{\xi}_{k}
\sigma_{3}+\mu\sigma_{1}+\bar{\phi}_{k}D)]}\nonumber\\
&=&\lim_{M\rightarrow\infty}
\prod_{j=1}^{N}{\rm{Tr}}\:
\hat{T}e^{\epsilon\sum_{k=1}^{M}(\bar{\xi}_{k}\sigma_{3}+
\mu\sigma_{1}+\bar{\phi}_{k}D)}=Q^{N}
\label{eq53}
\end{eqnarray}
where
\begin{eqnarray}
Q ={\rm{Tr}}\:\hat{T}e^{\int_{0}^{t}
dt'(\bar{\xi}\sigma_{3}+\mu\sigma_{1}+\bar{\phi}D)}
\label{eq54}
\end{eqnarray}
With this result, in the limit $M\rightarrow\infty$, the partition function
in Eq.\ (\ref{eq51}) becomes Eq.\ (\ref{eq55}).

\section{}
\label{appendix6}
We consider recombination in the parallel model.
For the Hamiltonian matrix elements in the coherent states basis, we obtain
to order $\mathcal{O}(N^{0})$
\begin{eqnarray}
-\frac{\langle\vec{z}_{k}|\hat{H}|\vec{z}_{k-1}\rangle}
{\langle\vec{z}_{k}|\vec{z}_{k-1}\rangle} &=&  
Nf\left[\frac{1}{N}\sum_{j=1}^{N}
\vec{z}_{k}^{*}(j)\sigma_{3}\vec{z}_{k-1}(j)\right]+
\mu\sum_{j=1}^{N}\left[\vec{z}_{k}^{*}(j)\sigma_{1}\vec{z}_{k-1}(j)
-1\right]\nonumber\\
&&+\nu N (g[\{\vec{z}_{k}^{*}(j)D_{j}^{l}\vec{z}_{k-1}(j)\}]-1) 
\label{eq67}
\end{eqnarray}
where the matrices $D_{j}^{l}$ are defined by Eq.\ (\ref{eq64}). 
We introduce the auxiliary fields
\begin{eqnarray}
\xi_{k}=\frac{1}{N}\sum_{j=1}^{N}\vec{z}_{k}^{*}(j)\sigma_{3}\vec{z}_{k-1}(j)
\label{eq68}
\end{eqnarray}
and the conjugate fields $\bar{\xi}_{k}$ to 
enforce the constraints via a Laplace representations of the
delta functions. Substituting into Eq.\ (\ref{eq66}), we obtain
\begin{eqnarray}
e^{-\hat{H}t}&=&\lim_{M\rightarrow\infty}
\int\left[\mathcal{D}\vec{z}^{*}\mathcal{D}\vec{z}\right]
\int\left[\prod_{k=1}^{M}\frac{i\epsilon 
N d\bar{\xi}_{k}d\xi_{k}}{2\pi}\right]
|\vec{z}_{M}\rangle\langle\vec{z}_{0}|\nonumber \\
&&\times  e^{\sum_{k=1}^{M}\sum_{j=1}^{N}\left\{-(1/2)[\vec{z}_{k}^{*}(j)\cdot
\vec{z}_{k}(j)+\vec{z}_{k-1}^{*}(j)\cdot\vec{z}_{k-1}(j)-2\vec{z}_{k}^{*}(j)
\cdot\vec{z}_{k-1}(j)]+\epsilon N[\vec{z}_{k}^{*}(j)(\bar{\xi}_{k}\sigma_{3}+
\mu\sigma_{1})\vec{z}_{k-1}(j)]\right\}}\nonumber \\
&& \times   e^{-\epsilon  \sum_{k=1}^{M}\left[\bar{\xi}_{k}\xi_{k}+\mu+
\nu - f(\xi_{k})
-\nu g(\{\vec{z}_{k}^{*}(j)D_{j}^{l}\vec{z}_{k-1}(j)\})\right]}
\label{eq69}
\end{eqnarray}
We obtain the partition function from the trace of the evolution operator,
Eq.\ (\ref{eq69}), for recombination in the parallel model
\begin{eqnarray}
Z = {\rm{Tr}}
\left[e^{-\hat{H}t}\hat{P}\right]=\int_{0}^{2\pi}\left[\prod_{j=1}^{N}
\frac{d\lambda_{j}}{2\pi}e^{-i\lambda_{j}}\right]\lim_{M\rightarrow\infty}
\int\left[\prod_{k=1}^{M}D\vec{z}^{*}_{k}D\vec{z}_{k}\right]
\left . e^{-S[\vec{z}^{*},\vec{z}]}\right|_{\vec{z}_{0}
=e^{i\lambda}\vec{z}_{M}}
\label{eq77}
\end{eqnarray}
It is convenient to define the auxiliary field
\begin{eqnarray}
\phi_{k}=\frac{1}{N}\sum_{j=1}^{N}\vec{z}^{*}_{k}(j)D\vec{z}_{k}(j)
\label{eq78}
\end{eqnarray}
and the corresponding $\bar{\phi}_{k}$ to enforce the constraint by
a Laplace representation of the Dirac delta function.
From Eq.\ (\ref{eq77}), we have
\begin{eqnarray}
Z &=&\lim_{M\rightarrow\infty}
\int\left[\mathcal{D}\bar{\xi}\mathcal{D}\xi
\mathcal{D}\bar{\phi}\mathcal{D}\phi\right]e^{-N\epsilon\sum_{k=1}^{M}
[\bar{\xi}_{k}\xi_{k}+
\bar{\phi}_{k}\phi_{k}-f(\xi_{k})+\mu+\nu-
\nu g(\phi_{k})]}\nonumber\\
&&\times \int_{0}^{2\pi}\left[\prod_{j=1}^{N}\frac{d\lambda_{j}}{2\pi}
e^{-i\lambda_{j}}\right] 
\int\left[\prod_{k=1}^{M}\mathcal{D}\vec{z}_{k}^{*}
\mathcal{D}\vec{z}_{k}\right]\left . e^{-\sum_{j=1}^{N}\sum_{k,l=1}^{M}
\vec{z}_{k}^{*}(j)S_{kl}(j)\vec{z}_{l}(j)}\right|_{\vec{z}_{M}=e^{i\lambda}
\vec{z}_{0}}\nonumber\\
\label{eq79}
\end{eqnarray}
Here, for large $N$ the function $g(\phi)$ has the singular behavior
$g(\phi) =0$ unless $\phi = 1 - \mathcal{O}(1/N)$. We also
notice $g(1)=1$.
The matrix $S(j)$ in Eq.\ (\ref{eq79}) is defined by
\begin{eqnarray}
S(j)=\left(\begin{array}{ccccc}I & 0 & 0 & \ldots & -e^{i\lambda_{j}}
A_{1}\\
-A_{2} & I & 0 & \ldots & 0\\ 0 & -A_{3} & I & \ldots & 0\\
\vdots &\ldots &\ldots & \ldots & 0\\0 & \ldots & 0 & -A_{M} & I\end{array}
\right) \label{eq80}
\end{eqnarray}
Here, $A_{k}=I+\epsilon(\bar{\xi}_{k}\sigma_{3}+\mu\sigma_{1}+
\bar{\phi}_{k}D)$.
After calculating the Gaussian integral over the coherent states fields, we
obtain
\begin{eqnarray}
&&\lim_{M\rightarrow\infty}
\int_{0}^{2\pi} \prod_{j=1}^{N} \frac{d\lambda_{j}}{2\pi}
e^{-i\lambda_{j}}
\int\left[\prod_{k=1}^{M}\mathcal{D}\vec{z}^{*}_{k}\mathcal{D}\vec{z}_{k}
\right]e^{-\sum_{j=1}^{N}\sum_{k,l=1}^{M}\vec{z}^{*}_{k}(j)S_{kl}(j)
\vec{z}_{l}(j)}\nonumber\\
&=&\lim_{M\rightarrow\infty}
\int_{0}^{2\pi}\prod_{j=1}^{N}\frac{d\lambda_{j}}{2\pi}
e^{-i\lambda_{j}}\left[\det S(j)\right]^{-1}\nonumber\\
&=&\lim_{M\rightarrow\infty}
\int_{0}^{2\pi}\prod_{j=1}^{N}\frac{d\lambda_{j}}{2\pi}e^{-i\lambda_{j}}
e^{-{\rm{Tr}}
\ln[I-e^{i\lambda_{j}}\hat{T}\exp(\epsilon\sum_{k=1}^{M}\bar{\xi}_{k}
\sigma_{3}+\mu\sigma_{1}+\bar{\phi}_{k}D)]}\nonumber\\
&=&\lim_{M\rightarrow\infty}
\prod_{j=1}^{N}{\rm{Tr}}\:
\hat{T}e^{\epsilon\sum_{k=1}^{M}(\bar{\xi}_{k}\sigma_{3}+
\mu\sigma_{1}+\bar{\phi}_{k}D)}=Q^{N}\nonumber\\
\label{eq81}
\end{eqnarray}
where in the continuous limit
\begin{eqnarray}
Q={\rm{Tr}}\:\hat{T}e^{\int_{0}^{t}dt'
(\bar{\xi}\sigma_{3}+\mu\sigma_{1}+\bar{\phi}D)}
\label{eq82}
\end{eqnarray}
With this result, the partition function
in Eq.\ (\ref{eq79}) becomes Eq.\ (\ref{eq91}).

\section{}
\label{appendix4}
\subsection*{The recombination operator}

For the recombination process, we consider that in the first step, 
the polymerase enzyme starts the copying path 
in either of both parental chains 
with equal probability $1/2$. Then, at each site, it can 
jump to the other chain with probability $0<p_{c}\leq 1/2$ or continue
along the same chain with probability $1-p_{c}$.

As presented in Section II.C, this process is represented
in the general differential Eq.\ (\ref{eq1}) by the coefficients
in Eq.\ (\ref{eq61})
\begin{eqnarray}
R_{kl}^{i}&=&\frac{1}{2}\sum_{\{\alpha_{j}=\pm 1\}}
\left(\frac{1+s_{1}^{k}s_{1}^{i}}{2}\right)^{\frac{1+\alpha_{1}}{2}}
\left(\frac{1+s_{1}^{l}s_{1}^{i}}{2}\right)^{\frac{1-\alpha_{1}}{2}}\nonumber\\
&&\times
[(1-p_{c})^{\frac{1+\alpha_{1}\alpha_{2}}{2}}
p_{c}^{\frac{1-\alpha_{1}\alpha_{2}}{2}}]
\left(\frac{1+s_{2}^{k}s_{2}^{i}}{2}\right)^{\frac{1+\alpha_{2}}{2}}
\left(\frac{1+s_{2}^{l}s_{2}^{i}}{2}\right)^{\frac{1-\alpha_{2}}{2}}\nonumber\\
&&\times[(1-p_{c})^{\frac{1+\alpha_{2}\alpha_{3}}{2}}
p_{c}^{\frac{1-\alpha_{2}\alpha_{3}}{2}}]\left(\frac{1+s_{3}^{k}s_{3}^{i}}{2}
\right)^{\frac{1+\alpha_{3}}{2}}
\left(\frac{1+s_{3}^{l}s_{3}^{i}}{2}\right)^{\frac{1-\alpha_{3}}{2}}\nonumber\\
&&\times\ldots\times [(1-p_{c})^{\frac{1+\alpha_{N-1}\alpha_{N}}{2}}
p_{c}^{\frac{1-\alpha_{N-1}\alpha_{N}}{2}}]
\left(\frac{1+s_{N}^{k}s_{N}^{i}}{2}\right)^{\frac{1+\alpha_{N}}{2}}
\left(\frac{1+s_{N}^{l}s_{N}^{i}}{2}\right)^{\frac{1-\alpha_{N}}{2}}
\nonumber\\
\label{eq188}
\end{eqnarray}
The operator for this process in the Schwinger-boson 
representation is presented in Eq.\ (\ref{eq63}) 
\begin{eqnarray}
\hat{R}&=&\frac{1}{2}\sum_{l=1}^{2^{N}}p_{l}\sum_{\{\alpha_{i}=\pm
1\}}[\hat{I}_{1}^{\frac{1+\alpha_{1}}{2}}
\hat{R}_{l}(1)^{\frac{1-\alpha_{1}}{2}}] 
\times[(1-p_{c})^{\frac{1+\alpha_{1}\alpha_{2}}{2}}
p_{c}^{\frac{1-\alpha_{1}\alpha_{2}}{2}}]\nonumber\\
&&\times[\hat{I}_{2}^{\frac{1+\alpha_{2}}{2}}
\hat{R}_{l}(2)^{\frac{1-\alpha_{2}}{2}}]
\times[(1-p_{c})^{\frac{1+\alpha_{2}\alpha_{3}}{2}}
p_{c}^{\frac{1-\alpha_{2}\alpha_{3}}{2}}]
\times[\hat{I}_{3}^{\frac{1+\alpha_{3}}{2}}
\hat{R}_{l}(3)^{\frac{1-\alpha_{3}}{2}}]
\nonumber \\
&&\times\ldots \times[(1-p_{c})^{\frac{1+\alpha_{N-1}\alpha_{N}}{2}}
p_{c}^{\frac{1-\alpha_{N-1}\alpha_{N}}{2}}]
\times[\hat{I}_{N}^{\frac{1+\alpha_{N}}{2}}
\hat{R}_{l}(N)^{\frac{1-\alpha_{N}}{2}}]-\hat{I}
\nonumber\\
&&\equiv  g(\{\hat{R}_{l}(j)\})-\hat{I}
\label{eq189}
\end{eqnarray}
Here, we define the single-site recombination operator as 
$\hat{R}_{l}(j)=\vec{\hat{a}}^{\dagger}(j)D_{j}^{l}\vec{\hat{a}}(j)$,
with
\begin{eqnarray}
D_{j}^{l}=\left(\begin{array}{cc}\frac{1+s_{j}^{l}}{2}&
\frac{1+s_{j}^{l}}{2}\\
\frac{1-s_{j}^{l}}{2}&\frac{1-s_{j}^{l}}{2}\end{array}\right)
\label{eq190}
\end{eqnarray} 
and $p_{l}=q_{l}/\sum_{l=1}^{2^{N}}q_{l}$ is the normalized
probability for sequence $1\leq l \leq 2^{N}$.

It is possible to group the different terms in 
the form of Ising-like traces, by using the definition 
$J=-(1/2)\ln[p_{c}/(1-p_{c})]$,
\begin{eqnarray}
g(\{\hat{R}_{l}(j)\})=
\frac{1}{2}\left[2\cosh(J)\right]^{-(N-1)}
\sum_{l=1}^{2^{N}}p_{l}  \sum_{\{\alpha_{j}=\pm 1\}}
e^{J\sum_{j=2}^{N}\alpha_{j}\alpha_{j-1}}\prod_{j=1}^{N}
\left[\frac{1+\alpha_{j}}{2}\hat{I}_{j}
+\frac{1-\alpha_{j}}{2}\hat{R}_{l}(j)\right]
\nonumber\\
\label{eq191}
\end{eqnarray}
After the representation in terms of coherent states fields, we have
$\hat{R}_{l}(j)\rightarrow \vec{z}_{k}^{*}(j)D_{j}^{l}\vec{z}_{k-1}(j) \equiv
\psi_{j}^{l}$, and correspondingly $g \rightarrow g(\{\psi_{j}^{l}\})$
\begin{eqnarray}
g(\{\psi_{j}^{l}\})= \frac{1}{2}\left[2\cosh(J)\right]^{-(N-1)}
\sum_{l=1}^{2^{N}}p_{l}\sum_{\{\alpha_{j}=\pm 1\}}
e^{J\sum_{j=2}^{N}\alpha_{j}\alpha_{j-1}}\prod_{j=1}^{N}
\left[\frac{1+\alpha_{j}}{2}
+\frac{1-\alpha_{j}}{2}\psi_{j}^{l}\right]
\label{eq192}
\end{eqnarray}
It is convenient to reorganize this expression as
\begin{eqnarray}
g(\{\psi_{j}^{l}\})=\frac{1}{2}\left[2\cosh(J)\right]^{-(N-1)}
\sum_{l=1}^{2^{N}}p_{l}\prod_{j=1}^{N}\left(\frac{1+\psi_{j}^{l}}{2}\right)
\sum_{\{\alpha_{j}=\pm 1\}}
e^{J\sum_{j=2}^{N}\alpha_{j}\alpha_{j-1}}\prod_{j=1}^{N}
\left[1+\alpha_{j}\frac{1-\psi_{j}^{l}}{1+\psi_{j}^{l}}\right]
\nonumber\\
\label{eq193}
\end{eqnarray}
We define the transfer matrix
\begin{eqnarray}
T = \left(\begin{array}{cc}e^{J} & e^{-J}\\e^{-J} & e^{J}\end{array}\right)
\label{eq194}
\end{eqnarray}
with eigenvalues $\lambda_{+}=2\cosh(J)$ and $\lambda_{-}=2\sinh(J)$.

The Ising trace in Eq.\ (\ref{eq193}) is given by
\begin{eqnarray}
\sum_{\{\alpha_{j}=\pm 1\}}e^{J\sum_{j=2}^{N}\alpha_{j}\alpha_{j-1}}
&=& 
\sum_{\{\alpha_{1}=\pm 1\}}\left(\langle\alpha_{1}|T^{N-1}|\alpha_{1}\rangle
+\langle\alpha_{1}|T^{N-1}|-\alpha_{1}\rangle\right)\nonumber\\
&=& {\rm{Tr}}[T^{N-1}]+{\rm{Tr}}[T^{N-1}\sigma_{1}]\nonumber\\
&=& 
\lambda_{+}^{N-1} + \lambda_{-}^{N-1} + \lambda{+}^{N-1} - \lambda_{-}^{N-1}
\nonumber\\
&=& 2\lambda_{+}^{N-1}=2\left[2\cosh(J)\right]^{N-1}
\label{eq195}
\end{eqnarray}
By considering this formula, and expanding the product in 
Eq.\ (\ref{eq193}), we obtain
\begin{eqnarray}
g(\{\psi_{j}^{l}\})&=&\sum_{l=1}^{2^{N}}p_{l}
\prod_{j=1}^{N}\left(\frac{1+\psi_{j}^{l}}{2}\right)
\left\{1 + \sum_{j=1}^{N}\langle\alpha_{j}\rangle\frac{1-\psi_{j}^{l}}
{1+\psi_{j}^{l}}+
\sum_{1\leq k < m}^{N}\langle\alpha_{k}\alpha_{m}\rangle
\frac{1-\psi_{k}^{l}}{1+\psi_{k}^{l}}
\frac{1-\psi_{m}^{l}}{1+\psi_{m}^{l}}\right.\nonumber\\
&&
+
\left.\sum_{1\leq k <m<n}\langle \alpha_{k}\alpha_{m}\alpha_{n}\rangle
\frac{1-\psi_{k}^{l}}{1+\psi_{k}^{l}}\frac{1-\psi_{m}^{l}}{1+\psi_{m}^{l}}
\frac{1-\psi_{n}^{l}}{1+\psi_{n}^{l}}
+\ldots+\langle\alpha_{1}\alpha_{2}\ldots\alpha_{N}\rangle\prod_{j=1}^{N}
\frac{1-\psi_{j}^{l}}{1+\psi_{j}^{l}}\right\}
\nonumber\\
\label{eq196}
\end{eqnarray}
In this notation, we defined the averages
\begin{eqnarray}
\langle \alpha_{k}\alpha_{l}\ldots\rangle\equiv 
\frac{1}{2\lambda_{+}^{N-1}}\sum_{\{\alpha_{j}=\pm 1\}}
e^{J\sum_{j=2}^{N}\alpha_{j}\alpha_{j-1}}\alpha_{k}\alpha_{l}\ldots
\label{eq197}
\end{eqnarray}
We present the first and second order averages, to illustrate
the general technique to obtain the higher orders.

The first order average is
\begin{eqnarray}
\langle \alpha_{k}\rangle & = & \frac{1}{2\lambda_{+}^{N-1}}
\sum_{\alpha_{j}=\pm 1}e^{J\sum_{j=2}^{N}\alpha_{j}\alpha_{j-1}}\alpha_{k}
\nonumber\\
&=&\frac{1}{2\lambda_{+}^{N-1}}{\rm{Tr}}\left\{\left(\begin{array}{cc}1&1\\1&1
\end{array}\right)T^{k-1}\sigma_{3}T^{N-k}\right\}\nonumber\\
&=&\frac{1}{2\lambda_{+}^{N-1}}{\rm{Tr}}\left\{P^{-1}\left(\begin{array}{cc}
1&1\\1&1\end{array}\right)P P^{-1}T^{k-1}P P^{-1}\sigma_{3}P 
P^{-1} T^{N-k} P\right\}\nonumber\\
\label{eq198}
\end{eqnarray}
To evaluate the trace, we introduced the matrix $P$ 
which diagonalizes the transfer matrix $T$

\begin{eqnarray}
P = \frac{1}{\sqrt{2}}\left(\begin{array}{cc}1&1\\-1&1\end{array}\right)
\label{eq199}
\end{eqnarray}
We use the identities
\begin{eqnarray}
P^{-1}T P = \left(
\begin{array}{cc}\lambda_{-}&0\\0&\lambda_{+}\end{array}\right), & 
P^{-1}\left(\begin{array}{cc}1&1\\1&1\end{array}\right)P =
\left(\begin{array}{cc}0&0\\0&2\end{array}\right), &
P^{-1}\sigma_{3} P = \sigma_{1}\nonumber\\
\label{eq200}
\end{eqnarray}
Substituting into Eq.\ (\ref{eq198}), we obtain
\begin{eqnarray}
\langle\alpha_{k}\rangle = \frac{1}{\lambda_{+}^{N-1}}
\rm{Tr}\left\{\left(\begin{array}{cc}0&0\\0&1\end{array}\right)
\left(\begin{array}{cc}\lambda_{-}^{k-1}&0\\0&\lambda_{+}^{k-1}
\end{array}\right)\sigma_{1}
\left(\begin{array}{cc}\lambda_{-}^{N-k}&0\\0&\lambda_{+}^{N-k}
\end{array}\right) \right\}
=0
\label{eq201}
\end{eqnarray}
a result we expect due to the symmetry of the Hamiltonian in
Eq.\ (\ref{eq198}).
Following a similar procedure, we can express the second order
correlation in the form
\begin{eqnarray}
\langle\alpha_{k}\alpha_{m}\rangle &=&\frac{1}{2\lambda_{+}^{N-1}}
{\rm Tr}\left\{\left(\begin{array}{cc}1&1\\1&1\end{array}\right)
T^{k-1}\sigma_{3}T^{m-k}\sigma_{3}T^{N-m}\right\}\nonumber\\
&=&\frac{1}{\lambda_{+}^{N-1}}
{\rm Tr}\left\{\left(\begin{array}{cc}0&0\\0&1\end{array}\right)
\left(\begin{array}{cc}\lambda_{-}^{k-1}&0\\0&\lambda_{+}^{k-1}
\end{array}\right)\sigma_{1}
\left(\begin{array}{cc}\lambda_{-}^{m-k}
&0\\0&\lambda_{+}^{m-k}\end{array}\right)\sigma_{1}
\left(\begin{array}{cc}\lambda_{-}^{N-m}&0\\0&\lambda_{+}^{N-m}
\end{array}\right)\right\}\nonumber\\
&=&\frac{\lambda_{+}^{k-1+N-m}\lambda_{-}^{m-k}}{\lambda_{+}^{N-1}}
=\left(\frac{\lambda_{-}}{\lambda_{+}}\right)^{m-k}\nonumber\\
&=&\left(\tanh(J)\right)^{m-k}=(1-2 p_{c})^{m-k}
\label{eq202}
\end{eqnarray}

From the same analysis, we prove that the correlations
for an odd number of $\alpha's$ vanish, whereas those for an even
number become
\begin{eqnarray}
\langle\alpha_{k}\alpha_{l}\alpha_{m}\alpha_{n}\ldots \rangle
=\left(\frac{\lambda_{-}}{\lambda_{+}}\right)^{l-k+n-m+\ldots}
=\left(\tanh(J)\right)^{l-k+n-m+\ldots}=(1-2p_{c})^{l-k+n-m+\ldots}
\label{eq203}
\end{eqnarray}
Substituting into Eq.\ (\ref{eq196}), 
we obtain the finite series representation
\begin{eqnarray}
g(\{\psi_{j}^{l}\})&=&\sum_{l=1}^{2^{N}}p_{l}
\prod_{j=1}^{N}\left(\frac{1+\psi_{j}^{l}}{2}\right)
\left\{1 + \sum_{1\leq k < m}^{N}(1-2p_{c})^{m - k}
\frac{1-\psi_{k}^{l}}{1+\psi_{k}^{l}}
\frac{1-\psi_{m}^{l}}{1+\psi_{m}^{l}}\right.
\nonumber\\
&&+
\left.\sum_{1\leq k<m<n<q}^{N}(1-2 p_{c})^{m-k + q - n}
\frac{1-\psi_{k}^{l}}{1+\psi_{k}^{l}}\frac{1-\psi_{m}^{l}}
{1+\psi_{m}^{l}}
\frac{1-\psi_{n}^{l}}{1+\psi_{n}^{l}}
\frac{1-\psi_{q}^{l}}{1+\psi_{q}^{l}}\right.\nonumber\\
&&
+
\left.\ldots+(1-2 p_{c})^{\lfloor \frac{N-1}{2} \rfloor }
\prod_{j=1}^{N}\left(\frac{1-\psi_{j}^{l}}
{1+\psi_{j}^{l}}\right)\right\}
\label{eq204}
\end{eqnarray}
Finally, we can obtain the alternative representation
\begin{eqnarray}
g(\{\psi_{j}^{l}\})&=&\sum_{l=1}^{2^{N}}p_{l}\left\{
\prod_{j=1}^{N}\left(\frac{1+\psi_{j}^{l}}{2}\right)\right.\nonumber\\
&&+
\left. 
\sum_{1\leq k < m}(1-2 p_{c})^{m-k}\frac{1-\psi_{k}^{l}}{2}
\frac{1-\psi_{m}^{l}}{2}
\prod_{j\ne k,l}\frac{1+\psi_{j}^{l}}{2}\right.\nonumber\\
&&+
\left.
\sum_{1\leq k<m<n<q}^{N}(1-2 p_{c})^{m-k+q-n}\frac{1-\psi_{k}^{l}}{2}
\frac{1-\psi_{m}^{l}}{2}\frac{1-\psi_{n}^{l}}{2}\frac{1-\psi_{q}^{l}}{2}
\right.\nonumber\\
&&\times
\left.
\prod_{j\ne k,m,n,q}^{N}\frac{1+\psi_{j}^{l}}{2}
+\ldots+(1-2p_{c})^{\lfloor\frac{N}{2}\rfloor}
\prod_{j=1}^{N}\frac{1-\psi_{j}^{l}}{2}
\right\}\nonumber\\
\label{eq205}
\end{eqnarray}

\section{}
\label{appendix5}

For the case of uniform crossover recombination, $p_c=1/2$, a
simplified analysis can be carried out to obtain the large N,
or Gaussian limit, of the recombination coefficients $R_{u_1,u_2}^{u}$
because permutation symmetry is exactly obeyed.
For the child sequence created from parental sequences 
with number of ``+1'' sites as $n_1$ and $n_2$, 
the number of child sequences, $n$, with ``+1'' sites is given
by the expression
\begin{eqnarray}
n = \sum_{i=1}^{N}\left(\frac{1+\alpha_i}{2}\frac{1+s_i^1}{2} +
\frac{1-\alpha_i}{2}\frac{1+s_i^2}{2}\right)
\label{eq206} 
\end{eqnarray}
Here, the path followed by the polymerase while copying from either
parental sequence is parametrized by the random variables $\alpha_i=\pm 1$,
with $\langle \alpha_i \rangle = 0$ and 
$\langle\alpha_i\alpha_j\rangle=\delta_{ij}$.
From Eq.\ (\ref{eq206}), we obtain the corresponding expression for the
average composition of the child sequence, $u=(N-2n)/N$
\begin{eqnarray}
u = \frac{1}{N}\sum_{i=1}^{N}\left(\frac{1+\alpha_i}{2}s_i^1
+\frac{1-\alpha_i}{2}s_i^2 \right)
\label{eq207}
\end{eqnarray}
From Eq.\ (\ref{eq207}), we obtain the average
\begin{eqnarray}
\langle u \rangle_{\alpha}=\frac{1}{N}\sum_{i=1}^{N}\frac{s_i^1+s_i^2}{2}
=\frac{u_1 + u_2}{2}
\label{eq208}
\end{eqnarray}
To obtain the variance, we calculate 
\begin{eqnarray}
\langle u^2\rangle_{\alpha} &= &
\frac{1}{N^2}\sum_{i,j=1}^{N}\langle\left(\frac{1+\alpha_i}{2}s_i^1
+\frac{1-\alpha_i}{2}s_i^2\right)\left(\frac{1+\alpha_j}{2}s_j^1
+\frac{1-\alpha_j}{2}s_j^2\right)  \rangle_{\alpha}\nonumber\\
&=&\frac{1}{4N^2 }\sum_{i,j=1}^{N}\left(s_i^1+s_i^2\right)
\left(s_j^1+s_j^2\right)
+\frac{1}{4N^2}\sum_{i=1}^{N}\langle 
\left(\alpha_i s_i^1 - \alpha_i s_i^2 \right)^2 \rangle_{\alpha}
\nonumber\\
&=&\langle u \rangle_{\alpha}^2 + \frac{1}{4N^2}\sum_{i=1}^{N}
\langle (s_i^1-s_i^2)^2\rangle
\label{eq209}
\end{eqnarray}
Therefore, we obtain the variance as
\begin{eqnarray}
\langle(\delta u)^2\rangle_{\alpha} = \frac{1}{4N^2}N\cdot 4\cdot 2
\frac{1+u}{2}\frac{1-u}{2}=\frac{1-u^2}{2N}
\label{eq210}
\end{eqnarray}
Hence, in the large N Gaussian limit, the recombination coefficients
are given by the distribution
\begin{eqnarray}
R_{u_1,u_2}^u \sim
\frac{e^{-N\left[(u_1+u_2)/2-u\right]^2/(1-u_*^2)}}
{\sqrt{\pi(1-u_*^2)/N}}
\label{eq211}
\end{eqnarray}
where $f_m = f(u_*)$.

For $p_c < 1/2$, making the ansatz
that correlations between spins at different
sites remain $\mathcal{O}(N^{-1})$,
the additional contribution to
$\langle(\delta u)^2\rangle_{\alpha}$ is 
$
[1/(4 N^2)] \sum_{i \ne j}
(1-2 p_c)^{\vert i - j \vert } 
(s_i^1 - s_i^2) 
(s_j^1 - s_j^2) 
= 
[1/(2 N^2)] \sum_{k>0} \sum_{i }
(1-2 p_c)^k
(s_i^1 - s_i^2) 
(s_{i+k}^1 - s_{i+k}^2) 
\sim 
[1/(2 N)] \sum_{k>0} 
(1-2 p_c)^k
[\langle s^1 - s^2 \rangle^2 + \mathcal{O}(1/N)]
\sim 
[1/(4 p_c N)] 
[\mathcal{O} (1/\sqrt{N})^2 + \mathcal{O}(1/N)]
\sim 
{\rm const}/N^2
$, and the large $N$ limit becomes that of the $p_c = 1/2$ case.

\section{}
\label{appendix6a}
We consider the saddle point condition for
recombination in the parallel model.
First, we look for the saddle-point condition with respect to the
fields $\bar{\xi}_{c}$, $\bar{\phi}_{c}$
\begin{eqnarray}
\frac{\delta}{\delta\bar{\xi}_{c}}\left(\frac{-S_{c}}{N t}\right)
=-\xi_{c}+\frac{2\bar{\xi}_{c}+u\bar{\phi}_{c}}
{2\left[\bar{\xi}_{c}(\bar{\xi}_{c}+u\bar{\phi}_{c})+
\left(\mu+\frac{\bar{\phi}_{c}}{2} \right)^{2}\right]^{1/2}}=0
\label{eq87}
\end{eqnarray}
\begin{eqnarray}
\frac{\delta}{\delta\bar{\phi}_{c}}\left(\frac{-S_{c}}{N t}\right)
=-\phi_{c}+\frac{1}{2}+
\frac{u\bar{\xi}_{c}+\mu+\frac{\bar{\phi}_{c}}{2}}
{2\left[\bar{\xi}_{c}(\bar{\xi}_{c}+u\bar{\phi}_{c})+
\left(\mu+\frac{\bar{\phi}_{c}}{2} \right)^{2}\right]^{1/2}}=0
\label{eq88}
\end{eqnarray}
Eqs.\ (\ref{eq87}) and (\ref{eq88}) become
\begin{eqnarray}
\xi_{c}=\frac{2\bar{\xi}_{c}+u\bar{\phi}_{c}}
{2\left[\bar{\xi}_{c}(\bar{\xi}_{c}+u\bar{\phi}_{c})+
\left(\mu+\frac{\bar{\phi}_{c}}{2} \right)^{2}\right]^{1/2}}
\label{eq89}
\end{eqnarray}
\begin{eqnarray}
\phi_{c}=\frac{1}{2}+
\frac{u\bar{\xi}_{c}+\mu+\frac{\bar{\phi}_{c}}{2}}
{2\left[\bar{\xi}_{c}(\bar{\xi}_{c}+u\bar{\phi}_{c})+
\left(\mu+\frac{\bar{\phi}_{c}}{2} \right)^{2}\right]^{1/2}}
\label{eq90}
\end{eqnarray}
By combining Eqs.\ (\ref{eq89}) and (\ref{eq90}), 
with the saddle-point action Eq.\ (\ref{eq86}), we obtain
Eq.\ (\ref{eq91}).

\section{}
\label{appendix7}
We consider horizontal gene transfer of blocks of length $M$ in the Eigen model.
The matrix elements of the Hamiltonian in the basis of coherent states
are given by
\begin{eqnarray}
-\frac{\langle \vec{z}_{k}|\hat{H}|\vec{z}_{k-1}\rangle}{\langle\vec{z}_{k}|
\vec{z}_{k-1}\rangle}&=&Ne^{-\mu+\frac{\mu}{N}\sum_{j=1}^{N}\vec{z}^{*}_{k}(j)
\sigma_{1}\vec{z}_{k-1}(j)}\nonumber \\
&&\times e^{-\frac{\nu}{\bar{M}}+\frac{\nu}{N}\sum_{b=0}^{N/\bar{M}-1}
\prod_{j_{b}=\bar{M}b+1}^{\bar{M}(b+1)}
\vec{z}_{k}^{*}(j_{b})D\vec{z}_{k-1}(j_{b})}
f\left[\frac{1}{N}\sum_{j=1}^{N}\vec{z}_{k}^{*}(j)\sigma_{3}
\vec{z}_{k-1}(j)\right]\nonumber\\
&&-N d\left[\frac{1}{N}\sum_{j=1}^{N}\vec{z}_{k}^{*}(j)\sigma_{3}
\vec{z}_{k-1}(j)\right]
\label{eq113}
\end{eqnarray}
We introduce the auxiliary fields
\begin{eqnarray}
\xi_{k}=\frac{1}{N}\sum_{j=1}^{N}\vec{z}_{k}^{*}(j)\sigma_{3}\vec{z}_{k-1}(j)
\label{eq114}
\end{eqnarray}
\begin{eqnarray}
\eta_{k}=\frac{1}{N}\sum_{j=1}^{N}\vec{z}^{*}_{k}(j)\sigma_{1}\vec{z}_{k-1}(j)
\label{eq115}
\end{eqnarray}
and the corresponding conjugate fields $\bar{\xi}_{k}$, $\bar{\eta}_{k}$ 
to enforce the constraints via 
Laplace representations of the Dirac delta functions. 
Therefore, Eq.\ (\ref{eq112}) becomes
\begin{eqnarray}
e^{-\hat{H}t}&=&\lim_{M\rightarrow\infty}\int\left[\mathcal{D}\vec{z}^{*}
\mathcal{D}\vec{z}\right]|\vec{z}_{M}\rangle\langle\vec{z}_{0}|
\int\left[\prod_{k=1}^{M}\frac{i\epsilon N d\bar{\xi}_{k}d\xi_{k}}{2\pi}
\frac{i\epsilon N d\bar{\eta}_{k}d\eta_{k}}{2\pi}\right]\nonumber\\
&&\times e^{-1/2\sum_{k=1}^{M}[\vec{z}_{k}^{*}(j)\cdot\vec{z}_{k}(j)+
\vec{z}^{*}_{k-1}(j)\cdot\vec{z}_{k-1}(j)-
2\vec{z}_{k}^{*}(j)\cdot\vec{z}_{k-1}(j)]}\nonumber\\
&&\times 
e^{\epsilon\sum_{k=1}^{M}\sum_{j=1}^{N}\vec{z}_{k}^{*}(j)(\bar{\xi}_{k}
\sigma_{3}+\bar{\eta}_{k}\sigma_{1})\vec{z}_{k-1}(j)}
e^{-\epsilon N\sum_{k=1}^{M}[\bar{\xi}_{k}\xi_{k}
+\bar{\eta}_{k}\eta_{k}]}
\nonumber\\
&&\times  e^{\epsilon N\sum_{k=1}^{M}[e^{-\mu(1-\eta_{k})-\nu/\bar{M}
+\frac{\nu}{N}\sum_{b=0}^{N/\bar{M}-1}\prod_{j_{b}=\bar{M}b+1}^{\bar{M}(b+1)}
\vec{z}_{k}^{*}(j_{b})D\vec{z}_{k-1}(j_{b})}f(\xi_{k})-d(\xi_{k})]}
\label{eq116}
\end{eqnarray}
At this point, a perturbation theory analysis similar to the case of the
horizontal gene transfer of finite  blocks in the Kimura model
leads us to conclude that to within
error $\mathcal{O}(\bar{M}/N)$ at each order in perturbation theory, it
is possible to substitute the recombination term by
\begin{eqnarray}
\frac{\nu}{\bar{M}}\left(\frac{1}{N}
\sum_{j=1}^{N}\vec{z}_{k}^{*}(j)D\vec{z}_{k-1}(j)\right)^{\bar{M}}
\label{eq117}
\end{eqnarray}
Then, it is convenient to introduce the auxiliary field
\begin{eqnarray}
\phi_{k}=\frac{1}{N}\sum_{j=1}^{N}\vec{z}_{k}^{*}(j)D\vec{z}_{k-1}(j)
\label{eq118}
\end{eqnarray}
and the corresponding $\bar{\phi}_{k}$ field to enforce the constraint
through a Laplace representation of the Dirac delta function.
The partition function is obtained from the trace of the evolution operator
in Eq.\ (\ref{eq116})
\begin{eqnarray}
Z = {\rm{Tr}}\left[e^{-\hat{H}t}\hat{P}\right]
=\int_{0}^{2\pi}\left[\prod_{j=1}^{N}
\frac{d\lambda_{j}}{2\pi}e^{-i\lambda_{j}}\right]\lim_{M\rightarrow\infty}
\int\left[\prod_{k=0}^{M}\mathcal{D}\vec{z}^{*}_{k}\mathcal{D}
\vec{z}_{k}\right]
\left. e^{-S[\vec{z}^{*},\vec{z}]}\right|_{\vec{z}_{0}
=e^{i\lambda}\vec{z}_{M}}
\label{eq119}
\end{eqnarray}
Thus, we obtain
\begin{eqnarray}
Z &=&\lim_{M\rightarrow\infty}
\int\left[\mathcal{D}\bar{\xi}\mathcal{D}\xi\mathcal{D}\bar{\eta}
\mathcal{D}\eta\mathcal{D}\bar{\phi}\mathcal{D}\phi\right]
e^{-\epsilon N \sum_{k=1}^{M}[\bar{\xi}_{k}\xi_{k}+
\bar{\eta}_{k}\eta_{k}+\bar{\phi}_{k}\phi_{k}]}
 e^{\epsilon N \sum_{k=1}^{M}[e^{-\mu(1-\eta_{k})-\nu/\bar{M}
+\frac{\nu}{\bar{M}}\phi_{k}^{\bar{M}}}f(\xi)-d(\xi)]}\nonumber\\
&&\times \int_{0}^{2\pi}\left[\frac{d\lambda_{j}}{2\pi}e^{-i\lambda_{j}}\right]
\int\left[\prod_{k=1}^{M}
\mathcal{D}\vec{z}_{k}^{*}\mathcal{D}\vec{z}_{k}\right]
\left . e^{-\sum_{j=1}^{N}\sum_{k,l=1}^{M}\vec{z}_{k}^{*}(j)S_{kl}(j)
\vec{z}_{l}(j)}\right|_{\vec{z}_{M}=e^{i\lambda_{j}}\vec{z}_{0}}
\nonumber\\
\label{eq120}
\end{eqnarray}
The matrix $S(j)$ in Eq.\ (\ref{eq120}) is defined by
\begin{eqnarray}
S(j)=\left(\begin{array}{ccccc}I & 0 & 0 & \ldots & -e^{i\lambda_{j}}
A_{1}\\
-A_{2} & I & 0 & \ldots & 0\\ 0 & -A_{3} & I & \ldots & 0\\
\vdots &\ldots &\ldots & \ldots & 0\\0 & \ldots & 0 & -A_{M} & I\end{array}
\right) \label{eq121}
\end{eqnarray}
Here $A_{k}=I+\epsilon(\bar{\xi}_{k}\sigma_{3}+\bar{\eta}_{k}\sigma_{1}+
\bar{\phi}_{k}D)$.

After calculating the Gaussian integral over the coherent states fields, we
obtain
\begin{eqnarray}
&&\lim_{M\rightarrow\infty}
\int_{0}^{2\pi}\prod_{j=1}^{N}\frac{d\lambda_{j}}{2\pi}e^{-i\lambda_{j}}
\int\left[\prod_{k=1}^{M}\mathcal{D}\vec{z}^{*}_{k}\mathcal{D}\vec{z}_{k}
\right]e^{-\sum_{j=1}^{N}\sum_{k=1}^{M}\vec{z}^{*}_{k}(j)S_{kl}(j)
\vec{z}_{l}(j)}\nonumber\\
&=&\lim_{M\rightarrow\infty}
\int_{0}^{2\pi}\prod_{j=1}^{N}\frac{d\lambda_{j}}{2\pi}
e^{-i\lambda_{j}}\left[\det S(j)\right]^{-1}\nonumber\\
&=&\lim_{M\rightarrow\infty}
\int_{0}^{2\pi}\prod_{j=1}^{N}\frac{d\lambda_{j}}{2\pi}e^{-i\lambda_{j}}
e^{-{\rm{Tr}}
\ln[I-e^{i\lambda_{j}}\hat{T}\exp(\epsilon\sum_{k=1}^{M}\bar{\xi}_{k}
\sigma_{3}+\bar{\eta}_{k}\sigma_{1}+\bar{\phi}_{k}D)]}\nonumber\\
&=&\lim_{M\rightarrow\infty}\prod_{j=1}^{N}{\rm{Tr}}
\hat{T}e^{\epsilon\sum_{k=1}^{M}(\bar{\xi}_{k}\sigma_{3}+
\bar{\eta}_{k}\sigma_{1}+\bar{\phi}_{k}D)}=Q^{N}
\nonumber\\
\label{eq122}
\end{eqnarray}
where
\begin{eqnarray}
Q={\rm{Tr}}\:\hat{T}e^{\int_{0}^{t}dt'(\bar{\xi}\sigma_{3}+\bar{\eta}\sigma_{1}
+\bar{\phi}D)}
\label{eq123}
\end{eqnarray}
With this result the partition function
in Eq.\ (\ref{eq120}) becomes Eq.\ (\ref{eq124}).

\section{}
\label{appendix8}
We consider the saddle-point equations for horizontal gene transfer of blocks
of length $M$ in the Eigen model:
\begin{eqnarray}
\frac{\delta}{\delta\bar{\xi}_{c}}\left(\frac{-S_{c}}{Nt}\right)
=-\xi_{c}+\frac{\bar{\xi}_{c}+\frac{u}{2}\bar{\phi}_{c}}
{\left[\bar{\xi}_{c}(\bar{\xi}_{c}+u\bar{\phi}_{c})+
\left(\bar{\eta}_{c}+\frac{\bar{\phi}_{c}}{2}\right)^{2}\right]^{1/2}}=0
\label{eq128}
\end{eqnarray}

\begin{eqnarray}
\frac{\delta}{\delta\bar{\phi}_{c}}\left(\frac{-S_{c}}{Nt} \right)=
-\phi_{c} +\frac{1}{2}+
\frac{u\bar{\xi}_{c}+\bar{\eta}_{c}+\frac{\bar{\phi}_{c}}{2}}
{2\left[\bar{\xi}_{c}(\bar{\xi}_{c}+u\bar{\phi}_{c})
+\left(\bar{\eta}_{c}+\frac{\bar{\phi}_{c}}{2}\right)^{2}\right]^{1/2}}
=0
\label{eq129}
\end{eqnarray}

\begin{eqnarray}
\frac{\delta}{\delta\phi_{c}}\left(\frac{-S_{c}}{Nt}\right)
=-\bar{\phi}_{c}+\nu\phi_{c}^{\bar{M}-1}
e^{-\mu(1-\eta_{c})-\frac{\nu}{\bar{M}}+\frac{\nu}{\bar{M}}
\phi_{c}^{\bar{M}-1}}f(\xi_{c})=0
\label{eq130}
\end{eqnarray}

\begin{eqnarray}
\frac{\delta}{\delta\bar{\eta}_{c}}\left(\frac{-S_{c}}{Nt}\right)
=-\eta_{c}+\frac{\bar{\eta}_{c}
+\frac{\bar{\phi}_{c}}{2}}
{2\left[\bar{\xi}_{c}(\bar{\xi}_{c}+u\bar{\phi}_{c})
+\left(\bar{\eta}_{c}+\frac{\bar{\phi}_{c}}{2}\right)^{2}\right]^{1/2}}=0
\label{eq131}
\end{eqnarray}

\begin{eqnarray}
\frac{\delta}{\delta\eta_{c}}\left(\frac{-S_{c}}{Nt} \right)=-\bar{\eta}_{c}
+\mu e^{-\mu(1-\eta_{c})-\frac{\nu}{\bar{M}}
+\frac{\nu}{\bar{M}}\phi_{c}^{\bar{M}}}f(\xi_{c})=0
\label{eq132}
\end{eqnarray}

We obtain the following identities
\begin{eqnarray}
\xi_{c}=\frac{\bar{\xi}_{c}+u\bar{\phi}_{c}/2}
{\left[\bar{\xi}_{c}(\bar{\xi}_{c}+u\bar{\phi}_{c})+(\bar{\eta}_{c}
+\bar{\phi}_{c}/2)^{2}\right]^{1/2} }
\label{eq133}
\end{eqnarray}
\begin{eqnarray}
\eta_{c}=\frac{\bar{\eta}_{c}
+\bar{\phi}_{c}/2}{\left[\bar{\xi}_{c}(\bar{\xi}_{c}+u\bar{\phi}_{c})
+(\bar{\eta}_{c}+\bar{\phi}_{c}/2)^{2}\right]^{1/2}}
\label{eq134}
\end{eqnarray}
\begin{eqnarray}
\bar{\eta}_{c}=\mu e^{-\mu(1-\eta_{c})-\frac{\nu}{\bar{M}}
(1-\phi_{c}^{\bar{M}})}f(\xi_{c})
\label{eq135}
\end{eqnarray}
\begin{eqnarray}
\phi_{c}=\frac{1}{2}+\frac{1}{2}\frac{u\bar{\xi}_{c}+\bar{\eta}_{c}
+\bar{\phi}_{c}/2}{\left[\bar{\xi}_{c}(\bar{\xi}_{c}+u\bar{\phi}_{c})
+(\bar{\eta}_{c}+\bar{\phi}_{c}/2)^{2}\right]^{1/2} }
\label{eq136}
\end{eqnarray}
\begin{eqnarray}
\bar{\phi}_{c}=\nu\phi_{c}^{\bar{M}-1}e^{-\mu(1-\eta_{c})
-\frac{\nu}{\bar{M}}(1-\phi_{c})}f(\xi_{c})
\label{eq137}
\end{eqnarray}
Combining Eq.\ (\ref{eq135}) and Eq.\ (\ref{eq137}), we obtain
\begin{eqnarray}
\nu\bar{\eta}_{c}\phi_{c}^{\bar{M}-1}=\mu\bar{\phi}_{c}
\label{eq138}
\end{eqnarray}
From the system of Eqs.\ (\ref{eq133})--(\ref{eq138}), it can be shown that
\begin{eqnarray}
-\bar{\xi}_{c}\xi_{c}-\bar{\eta}_{c}\eta_{c}-\bar{\phi}_{c}\phi_{c}+
\frac{\ln Q_{c}}{t}=0
\label{eq139}
\end{eqnarray} 

\section{}
\label{appendix9}
We consider horizontal gene transfer of blocks of variable length
in the Eigen model.
The Hamiltonian matrix elements in the coherent states basis are given,
to $\mathcal{O}(N^{-1})$, by
\begin{eqnarray}
-\frac{\langle\vec{z}_{k}|\hat{H}|\vec{z}_{k-1}\rangle}
{\langle\vec{z}_{k}|\vec{z}_{k-1}\rangle}&=&
N e^{-\mu+\frac{\mu}{N}\sum_{j=1}^{N}\vec{z}_{k}^{*}(j)D\vec{z}_{k-1}(j)}
\nonumber\\
&&\times \left(1-\frac{\nu}{\langle\bar{M}\rangle}
+\frac{\nu}{\langle\bar{M}\rangle} e^{-\langle\bar{M}\rangle 
+ \frac{\langle\bar{M}\rangle}{N}
\sum_{j=1}^{N}\vec{z}_{k}^{*}(j)D\vec{z}_{k-1}(j)}\right)\nonumber\\
&&\times 
f\left[\frac{1}{N}\sum_{j=1}^{N}\vec{z}_{k}^{*}(j)
\sigma_{3}\vec{z}_{k-1}(j)\right]
-N d\left[\frac{1}{N}\sum_{j=1}^{N}
\vec{z}_{k}^{*}(j)\sigma_{3}\vec{z}_{k-1}(j) \right]
\nonumber\\
\label{eq149}
\end{eqnarray}

We introduce the auxiliary fields
\begin{eqnarray}
\xi_{k}=\frac{1}{N}\sum_{j=1}^{N}\vec{z}_{k}^{*}(j)\sigma_{3}\vec{z}_{k-1}(j)
\label{eq150}
\end{eqnarray}
\begin{eqnarray}
\eta_{k}=\frac{1}{N}\sum_{j=1}^{N}\vec{z}_{k}^{*}(j)\sigma_{1}\vec{z}_{k-1}(j)
\label{eq151}
\end{eqnarray}
\begin{eqnarray}
\phi_{k}=\frac{1}{N}\sum_{j=1}^{N}\vec{z}_{k}^{*}(j)D\vec{z}_{k-1}(j)
\label{eq152}
\end{eqnarray}
and the corresponding $\bar{\xi}_{k}$, $\bar{\eta}_{k}$, $\bar{\phi}_{k}$
to enforce the constraints via Laplace representations of the Dirac
delta functions. From Eq.\ (\ref{eq148}), we obtain
\begin{eqnarray}
e^{-\hat{H}t}&=&\lim_{M\rightarrow\infty}\int\left[\mathcal{D}\vec{z}^{*}
\mathcal{D}\vec{z}\right]\int\left[\prod_{k=1}^{M}
\frac{i\epsilon Nd\bar{\xi}_{k}d\xi_{k}}{2\pi}
\frac{i\epsilon N d\bar{\eta}_{k}d\eta_{k}}{2\pi}
\frac{i\epsilon N d\bar{\phi}_{k}
d\phi_{k}}{2\pi}\right]|\vec{z}_{M}\rangle\langle\vec{z}_{0}|\nonumber\\
&&\times  e^{\sum_{k=1}^{M}\sum_{j=1}^{N}\left\{-1/2[\vec{z}_{k}^{*}(j)\cdot
\vec{z}_{k}(j)+\vec{z}_{k-1}^{*}(j)\cdot\vec{z}_{k-1}(j)-
2\vec{z}_{k}^{*}(j)\cdot\vec{z}_{k-1}(j)]+\epsilon
[\vec{z}_{k}^{*}(j)(\bar{\xi}_{k}\sigma_{3}+\bar{\eta}_{k}\sigma_{1}
+\bar{\phi}_{k}D)
\vec{z}_{k-1}(j)]\right\}}\nonumber\\
&&\times  e^{\epsilon N\sum_{k=1}^{M}\{-\bar{\xi}_{k}\xi_{k}-
\bar{\phi}_{k}\phi_{k}-\bar{\eta}_{k}\eta_{k}+e^{-\mu(1-\eta_{k})}
[1-\frac{\nu}{\langle\bar{M}\rangle}
+\frac{\nu}{\langle\bar{M}\rangle} 
e^{-\langle\bar{M} \rangle(1-\phi_{k})}]f(\xi_{k})-d(\xi_{k})\}}
\label{eq153}
\end{eqnarray}

We obtain the partition function from the trace of the evolution
operator Eq.\ (\ref{eq153})
\begin{eqnarray}
Z = {\rm{Tr}}\left[e^{-\hat{H}t}\hat{P}\right]=
\lim_{M\rightarrow\infty}\int_{0}^{2\pi}
\left[\prod_{j=1}^{N}\frac{d\lambda_{j}}{2\pi}e^{-i\lambda_{j}}\right]
\int\left[\prod_{k=1}^{M}\mathcal{D}\vec{z}_{k}^{*}
\mathcal{D}\vec{z}_{k}\right]\left.e^{-S[\vec{z}^{*},\vec{z}]}
\right|_{\vec{z}_{0}=e^{i\lambda}\vec{z}_{M}}
\label{eq154}
\end{eqnarray}
By inserting Eq.\ (\ref{eq153}), we obtain
\begin{eqnarray}
Z &=& \lim_{M\rightarrow\infty}
\int\left[\mathcal{D}\bar{\xi}\mathcal{D}\xi\mathcal{D}\bar{\eta}
\mathcal{D}\eta\mathcal{D}\bar{\phi}\mathcal{D}\phi\right]
e^{\epsilon N\sum_{k=1}^{M}(-\bar{\xi}_{k}\xi_{k}-\bar{\eta}_{k}\eta_{k}
-\bar{\phi}_{k}\phi_{k})}\nonumber\\
&&\times e^{\epsilon N\sum_{k=1}^{M}\{e^{-\mu(1-\eta_{k})}[1
-\frac{\nu}{\langle\bar{M}\rangle}
+\frac{\nu}{\langle\bar{M}\rangle} 
e^{-\langle\bar{M}\rangle (1-\phi_{k})}]f(\xi_{k})-d(\xi_{k})\}}
\nonumber\\
&&\times\int_{0}^{2\pi}\left[\prod_{j=1}^{N}\frac{d\lambda_{j}}{2\pi}
e^{-i\lambda_{j}}\right]\int\left[\prod_{k=1}^{M}
\mathcal{D}\vec{z}_{k}^{*}\mathcal{D}\vec{z}_{k}\right]
\left.e^{-\sum_{j=1}^{N}\sum_{k,l=1}^{M}\vec{z}_{k}^{*}(j)S_{kl}(j)
\vec{z}_{l}(j)}\right|_{\vec{z}_{M}=e^{i\lambda}\vec{z}_{0}}
\label{eq155}
\end{eqnarray}
The matrix $S(j)$ in Eq.\ (\ref{eq155}) is defined by
\begin{eqnarray}
S(j)=\left(\begin{array}{ccccc}I & 0 & 0 & \ldots & -e^{i\lambda_{j}}
A_{1}\\
-A_{2} & I & 0 & \ldots & 0\\ 0 & -A_{3} & I & \ldots & 0\\
\vdots &\ldots &\ldots & \ldots & 0\\0 & \ldots & 0 & -A_{M} & I\end{array}
\right) \label{eq156}
\end{eqnarray}
Here $A_{k}=I+\epsilon(\bar{\xi}_{k}\sigma_{3}+\bar{\eta}_{k}\sigma_{1}+
\bar{\phi}_{k}D)$.

After calculating the Gaussian integral over the coherent states fields,
we obtain
\begin{eqnarray}
&&\lim_{M\rightarrow\infty}
\int_{0}^{2\pi}\prod_{j=1}^{N}\frac{d\lambda_{j}}{2\pi}e^{-i\lambda_{j}}
\int\left[\prod_{k=1}^{M}\mathcal{D}\vec{z}_{k}^{*}\mathcal{D}\vec{z}_{k}
\right]e^{-\sum_{j=1}^{N}\sum_{k=1}^{M}\vec{z}_{k}^{*}(j)S_{kl}(j)
\vec{z}_{l}(j)}\nonumber\\
&=&\lim_{M\rightarrow\infty}
\int_{0}^{2\pi}\prod_{j=1}^{N}\frac{d\lambda_{j}}{2\pi}
e^{-i\lambda_{j}}[\det S(j)]^{-1}\nonumber\\
&=&\int_{0}^{2\pi}\prod_{j=1}^{N}\frac{d\lambda_{j}}{2\pi}
e^{-i\lambda_{j}}e^{-{\rm{Tr}}\ln[I-e^{i\lambda_{j}}\hat{T}
\exp(\epsilon\sum_{k=1}^{M}\bar{\xi}_{k}\sigma_{3}+\bar{\eta}_{k}\sigma_{1}
+\bar{\phi}_{k}D)]}\nonumber\\
&=&\lim_{M\rightarrow\infty}
\prod_{j=1}^{N}{\rm{Tr}}\: 
\hat{T}e^{\epsilon\sum_{k=1}^{M}(\bar{\xi}_{k}\sigma_{3}
+\bar{\eta}_{k}\sigma_{1}+\bar{\phi}_{k}D)}=Q^{N}
\label{eq157}
\end{eqnarray}
where, 
\begin{eqnarray}
Q = {\rm{Tr}}\:
\hat{T}e^{\int_{0}^{t}dt'(\bar{\xi}\sigma_{3}+\bar{\eta}\sigma_{1}
+\bar{\phi}D)}
\label{eq158}
\end{eqnarray}
With this result  the partition
function in Eq.\ (\ref{eq155}) becomes Eq.\ (\ref{eq159}).

\section{}
\label{appendix10}
We consider recombination in the Eigen model.
The matrix elements of the Hamiltonian operator in the coherent
states basis are given, to order $\mathcal{O}(N)$, by
\begin{eqnarray}
-\frac{\langle\vec{z}_{k}|\hat{H}|\vec{z}_{k-1}\rangle}
{\langle\vec{z}_{k}|\vec{z}_{k-1}\rangle}
&=&N e^{-\mu}e^{\frac{\mu}{N}\sum_{j=1}^{N}\vec{z}_{k}^{*}(j)\sigma_{1}
\vec{z}_{k-1}(j)}\nonumber\\
&&\times \left[1-\nu+\nu g(\{\vec{z}_{k}^{*}(j)D_{j}^{l}\vec{z}_{k-1}(j)\})
\right]f\left[\frac{1}{N}\sum_{j=1}^{N}\vec{z}_{k}^{*}(j)\sigma_{3}
\vec{z}_{k-1}(j)\right]
\nonumber\\
&&- N d\left[\frac{1}{N}\sum_{j=1}^{N}\vec{z}_{k}^{*}(j)
\sigma_{3}\vec{z}_{k-1}(j)\right]
\label{eq170}
\end{eqnarray}
Here we notice that the function 
$g(\{\vec{z}_{k}^{*}(j)D_{j}^{l}\vec{z}_{k-1}(j)\})$ is the same 
as in Eq.\ (\ref{eq70}). Therefore, the same analysis presented
through Eqs.\ (\ref{eq70}) -- (\ref{eq71}) regarding the
singular behavior of the function $g$ applies for the 
Eigen model as well. Hence, in the large $N$ limit, we have
$g\left(\frac{1}{N}\sum_{j=1}^{N}\vec{z}_{k}^{*}(j)D\vec{z}_{k-1}(j)\right)$,
with $D = \langle D_{j}^{l} \rangle$ being again the matrix
defined in Eq.\ (\ref{eq84}). 

We introduce the auxiliary fields
\begin{eqnarray}
\xi_{k}=\frac{1}{N}\sum_{j=1}^{N}\vec{z}_{k}^{*}(j)\sigma_{3}\vec{z}_{k-1}(j)
\label{eq172}
\end{eqnarray}

\begin{eqnarray}
\eta_{k}=\frac{1}{N}\sum_{j=1}^{N}\vec{z}_{k}^{*}(j)
\sigma_{1}\vec{z}_{k-1}(j)
\label{eq173}
\end{eqnarray}

\begin{eqnarray}
\phi_{k}=\frac{1}{N}\sum_{j=1}^{N}\vec{z}_{k}^{*}(j)D\vec{z}_{k-1}(j)
\label{eq174}
\end{eqnarray}
and the corresponding conjugate fields $\bar{\xi}_{k}$, $\bar{\eta}_{k}$ 
and $\bar{\phi}_{k}$ to enforce the constraints via Laplace representations 
of the Dirac delta functions.
Thus, we have
\begin{eqnarray}
e^{-\hat{H}t}&=&\lim_{M\rightarrow\infty}\int[\mathcal{D}\vec{z}^{*}
\mathcal{D}\vec{z}]\int\left[\prod_{k=1}^{M}\frac{i\epsilon N 
d\bar{\xi}_{k}d\xi_{k}}{2\pi}\frac{i\epsilon N d\bar{\eta}_{k}
d\eta_{k}}{2\pi}\frac{i\epsilon N d\bar{\phi}_{k}d\phi_{k}}{2\pi}\right]
\nonumber\\
&&\times|\vec{z}_{M}\rangle\langle\vec{z}_{0}|
e^{-\frac{1}{2}\sum_{k=1}^{M}\sum_{j=1}^{N}
[\vec{z}_{k}^{*}(j)\cdot\vec{z}_{k}(j)
+\vec{z}_{k-1}^{*}(j)\cdot\vec{z}_{k-1}(j)-2\vec{z}_{k}^{*}(j)
\cdot\vec{z}_{k-1}(j)]}\nonumber\\
&&\times  e^{\epsilon\sum_{k=1}^{M}\sum_{j=1}^{N}\vec{z}_{k}^{*}(j)
[\bar{\xi}_{k}\sigma_{3}+\bar{\eta}_{k}\sigma_{1}+\bar{\phi}_{k}D]
\vec{z}_{k-1}(j)}
e^{-\epsilon N\sum_{k=1}^{M}[\bar{\xi}_{k}\xi_{k}+\bar{\eta}_{k}\eta_{k}
+\bar{\phi}_{k}\phi_{k}]}
\nonumber\\
&&\times  e^{\epsilon N\sum_{k=1}^{M}[e^{-\mu(1-\eta_{k})}(1-\nu+\nu 
g(\phi_{k})f(\xi_{k})-d(\xi_{k})]}\nonumber\\
\label{eq175}
\end{eqnarray}

The partition function is expressed by
\begin{eqnarray}
Z = {\rm{Tr}}[e^{-\hat{H}t}\hat{P}]=\int_{0}^{2\pi}\left[\prod_{j=1}^{N}
\frac{d\lambda_{j}}{2\pi}e^{-i\lambda_{j}}\right]
 \lim_{M\rightarrow\infty}
\int\left[\prod_{k=1}^{M}\mathcal{D}\vec{z}^{*}_{k}\mathcal{D}
\vec{z}_{k}\right]\left.e^{-S[\vec{z}^{*},\vec{z}]}
\right|_{\vec{z}_{0}=e^{i\lambda}\vec{z}_{M}}
\label{eq176}
\end{eqnarray}
By inserting Eq.\ (\ref{eq175}), we obtain
\begin{eqnarray}
Z &=&\lim_{M\rightarrow\infty}
\int[\mathcal{D}\xi\mathcal{D}\bar{\xi}\mathcal{D}\bar{\eta}\mathcal{D}\eta
\mathcal{D}\bar{\phi}\mathcal{D}\phi]\nonumber\\
&&\times  e^{-\epsilon N\sum_{k=1}^{M}[\bar{\xi}_{k}\xi_{k}+\bar{\eta}_{k}
\eta_{k}+\bar{\phi}_{k}\phi_{k}]}
e^{\epsilon N\sum_{k=1}^{M}[e^{-\mu(1-\eta_{k})}(1-\nu
+\nu g(\phi_{k}))f(\xi_{k})-d(\xi_{k})]}\nonumber\\
&&\times  
\int_{0}^{2\pi}\prod_{j=1}^{N}\frac{d\lambda_{j}}{2\pi}e^{-i\lambda_{j}}
\int\left[\prod_{k=1}^{M}\mathcal{D}\vec{z}_{k}^{*}
\mathcal{D}\vec{z}_{k}\right]\left.e^{-\sum_{j=1}^{N}\sum_{k,l=1}^{M}
\vec{z}_{k}^{*}(j)S_{kl}(j)\vec{z}_{l}(j)}\right|_{\vec{z}_{0}=e^{i\lambda}
\vec{z}_{M}}
\label{eq177}
\end{eqnarray}
The Gaussian integral can be performed over the coherent state fields,
to obtain the representation in Eq.\ (\ref{eq178}).
Here, the one-dimensional Ising trace is defined by
\begin{eqnarray}
Q = {\rm{Tr}}\:
\hat{T}e^{\int_{0}^{t}dt'(\bar{\xi}\sigma_{3}+\bar{\eta}\sigma_{1}
+\bar{\phi}D)}
\label{eq180}
\end{eqnarray}

\section{}
\label{appendix12}
We analyze the effect of introducing different schemes of
horizontal gene transfer in the parallel model.

For the parallel model in the presence of horizontal gene transfer
with blocks of size $\bar{M}=1$,
we obtain
\begin{eqnarray}
\left.\frac{du}{d\nu}\right|_{\nu\rightarrow 0}
=\frac{u_{0}\xi_{0}+\sqrt{1-\xi_{0}^{2}}-1}{2 f'(u_{0})}
\label{eqA12_1}
\end{eqnarray}
Here, $(\xi_{0},u_{0})$ represents the solution for $\nu=0$, i.e., they
are obtained from the system
\begin{eqnarray}
\mathcal{F}[\xi] = f(\xi) + \mu\sqrt{1-\xi^{2}} - \mu
\label{eqA12_2}
\end{eqnarray}
\begin{eqnarray}
\left.\frac{\partial\mathcal{F}}{\partial\xi}\right|_{\xi=\xi_{0}}=0
=f'(\xi_{0})-\frac{\mu\xi_{0}}{\sqrt{1-\xi_{0}^{2}}}
\label{eqA12_3}
\end{eqnarray}
\begin{eqnarray}
f_{m} = f(u_{0}) = \mathcal{F}[\xi_{0}] = f(\xi_{0}) + \mu\sqrt{1-\xi_{0}^{2}}
-\mu
\label{eqA12_4}
\end{eqnarray}

From Eq.\ (\ref{eqA12_4}), we obtain $u_{0}$ from the inverse function
\begin{eqnarray}
u_{0} = f^{-1}[\mathcal{F}[\xi_{0}]] = 
f^{-1}[f(\xi_{0})+\mu\sqrt{1-\xi_{0}^{2}}-\mu]
\label{eqA12_5}
\end{eqnarray}

Let us Taylor-expand Eq.\ (\ref{eqA12_5}) near $x = f(\xi_{0})$,
\begin{eqnarray}
u_{0} = f^{-1}[x] + (f^{-1})^{'}[x]\delta x + (f^{-1})^{''}[x]
\frac{(\delta x)^{2}}{2}
\label{eqA12_6}
\end{eqnarray}
with $\delta x = \mu(\sqrt{1-\xi_{0}^{2}}-1)$. Here, we use the
inverse function theorem to obtain the derivatives
\begin{eqnarray}
(f^{-1})^{'}[x] &=& \frac{1}{f^{'}(f^{-1}[x])}=\frac{1}{f'(\xi_{0})}
\nonumber\\
(f^{-1})^{''}[x] &=& \frac{-f^{''}(f^{-1}[x])}{(f^{'}(f^{-1}[x]))^{3}}
=-\frac{f^{''}(\xi_{0})}{(f^{'}(\xi_{0}))^{3}}
\label{eqA12_7}
\end{eqnarray}

Hence, Eq.\ (\ref{eqA12_6}) becomes
\begin{eqnarray}
u_{0} = \xi_{0} + \frac{\delta x}{f'(\xi_{0})} 
- \frac{f^{''}(\xi_{0})}{(f^{'}(\xi_{0}))^{3}}\frac{(\delta x)^{2}}{2}
\label{eqA12_8}
\end{eqnarray} 

From Eq.\ (\ref{eqA12_3}), we have
\begin{eqnarray}
\frac{\delta x}{f'(\xi_{0})} = 
\frac{\mu(\sqrt{1-\xi_{0}^{2}}-1)}{\frac{\mu\xi_{0}}{\sqrt{1-\xi_{0}^{2}}}}
= \frac{1-\xi_{0}^{2}-\sqrt{1-\xi_{0}^{2}}}{\xi_{0}}
\label{eqA12_9}
\end{eqnarray}

From Eq.\ (\ref{eqA12_8}) into Eq.\ (\ref{eqA12_7}), after multiplying
by $\xi_{0}$, we have
\begin{eqnarray}
u_{0}\xi_{0} &=& \xi_{0}^{2} + \xi_{0}\frac{\delta x}{f^{'}(\xi_{0})}
-\xi_{0}\frac{f^{''}(\xi_{0})}{(f^{'}(\xi_{0}))^{3}}\frac{(\delta x)^{2}}{2}
\nonumber\\
&=&\xi_{0}^{2} + \xi_{0}\frac{(1-\xi_{0}^{2}-\sqrt{1-\xi_{0}^{2}})}{\xi_{0}}
-\frac{\xi_{0}}{f'(\xi_{0})}\frac{f^{''}(\xi_{0})}{(f^{'}(\xi_{0}))^{2}}
\frac{(\delta x)^{2}}{2}\nonumber\\
&=&1 - \sqrt{1-\xi_{0}^{2}}-f^{''}(\xi_{0})
\frac{(\delta x)^{2}}{2(f^{'}(\xi_{0}))^{2}}
\frac{\sqrt{1-\xi_{0}^{2}}}{\mu}
\label{eqA12_10}
\end{eqnarray}
Therefore, we finally obtain
\begin{eqnarray}
u_{0}\xi_{0} + \sqrt{1-\xi_{0}^{2}}-1 =
-\frac{f^{''}(\xi_{0})}{2}\frac{(\delta x)^{2}}{(f^{'}(\xi_{0}))^{2}}
\frac{\sqrt{1-\xi_{0}^{2}}}{\mu}
\label{eqA12_11}
\end{eqnarray}
The sign of this expression is clearly determined by $-f^{''}(\xi_{0})$,
and hence after Eq.\ (\ref{eqA12_1}) we obtain the condition
\begin{eqnarray}
\left.\frac{du}{d\nu}\right|_{\nu\rightarrow 0}=
\left\{\begin{array}{ccc}>0 & \rm{if} & f^{''}(\xi_{0})<0\\
< 0 & \rm{if} & f^{''}(\xi_{0})>0 \end{array}\right.
\label{eqA12_12}
\end{eqnarray}
From Eq.\ (\ref{eqA12_12}), we conclude that horizontal gene
transfer
will enhance selection towards the fittest individuals
when negative epistasis is present [$f^{''}(u)<0$],
while it will introduce an additional load against
selection, with the corresponding deleterious effect
on the mean fitness, when positive epistasis is
present [$f^{''}(u)>0$]. This result proves that the
mutational deterministic hypothesis holds for horizontal
gene transfer of blocks of size $\bar{M}=1$ in the parallel model.

For the case of horizontal gene transfer of blocks $\bar{M}>1$, we obtain
the equation
\begin{eqnarray}
\left.\frac{du}{d\nu}\right|_{\nu\rightarrow 0} 
= \frac{\left[1 + \frac{u_{0}\xi_{0}-1
+\sqrt{1-\xi_{0}^{2}}}{2}\right]^{\bar{M}}-1}{\bar{M}f^{'}(u_{0})}
\label{eqA12_M1}
\end{eqnarray}
We notice by expanding the binomial up to first order, that the leading term
in Eq.\ (\ref{eqA12_M1}) is
\begin{eqnarray}
\left.\frac{du}{d\nu}\right|_{\nu\rightarrow 0} \sim 
\frac{u_{0}\xi_{0}-1+\sqrt{1-\xi_{0}^{2}}}{2 f^{'}(u_{0})}
\label{eqA12_M2}
\end{eqnarray}
which is identical to Eq.\ (\ref{eqA12_1}), and hence the analysis
presented for the case $\bar{M}=1$ also applies for $\bar{M}>1$,
in particular Eq.\ (\ref{eqA12_12}).

For the process of horizontal gene transfer with multiple-size
blocks, with average $\langle\bar{M}\rangle$, we obtain
the equation
\begin{eqnarray}
\left.\frac{du}{d\nu}\right|_{\nu\rightarrow 0} 
= \frac{e^{\frac{\langle\bar{M}\rangle}{2}(u_{0}\xi_{0}-1
+\sqrt{1-\xi_{0}^{2}})}-1}{\langle\bar{M}\rangle f^{'}(u_{0})}
\label{eqA12_M3}
\end{eqnarray}
By expanding the exponential at first order, we obtain
that the leading term in this case is also Eq.\ (\ref{eqA12_M2}),
which is identical to Eq.\ (\ref{eqA12_1}). Therefore, the analysis
presented for $\bar{M}=1$, and in particular Eq.\ (\ref{eqA12_12})
applies in this case as well.

In conclusion, we proved that the mutational deterministic hypothesis,
expressed in quantitative form by Eq.\ (\ref{eqA12_12}),
holds for the different forms of horizontal gene transfer
discussed in our work for the parallel model.

\section{}
\label{appendix12_b}
We analyze the effect of introducing different schemes of
horizontal gene transfer in the Eigen model.

For the Eigen model in the presence of
horizontal gene transfer, and for zero degradation
rate $d(u)=0$, we obtain the equation
\begin{eqnarray}
\left.\frac{du}{d\nu}\right|_{\nu\rightarrow 0} 
= \frac{u_{0}\xi_{0} + \sqrt{1-\xi_{0}^{2}}-1}{2 f^{'}(u)}
e^{-\mu[1-\sqrt{1-\xi_{0}^{2}}]}f(\xi_{0})
\label{eqA12_14}
\end{eqnarray}
The sign of this derivative is determined by the combination
$u_{0}\xi_{0}+\sqrt{1-\xi_{0}^{2}}-1$, where $(\xi_{0},u_{0})$ represents
the solution for $\nu=0$, i.e. they are obtained from the
system
\begin{eqnarray}
\mathcal{F}[\xi] = f(\xi)e^{-\mu(1-\sqrt{1-\xi^{2}})}
\label{eqA12_15}
\end{eqnarray}
\begin{eqnarray}
\left.\frac{\partial\mathcal{F}}{\partial\xi}\right|_{\xi=\xi_{0}}=0=
\left(f'(\xi_{0})-\frac{\mu\xi_{0}}{\sqrt{1-\xi_{0}^{2}}}\right)
e^{-\mu[1-\sqrt{1-\xi_{0}^{2}}]}
\label{eqA12_16}
\end{eqnarray}
\begin{eqnarray}
f_{m}=f(u_{0})=\mathcal{F}[\xi_{0}]=e^{-\mu(1-\sqrt{1-\xi_{0}^{2}})}f(\xi_{0})
\label{eqA12_17}
\end{eqnarray}
By inverting Eq.\ (\ref{eqA12_17}), we obtain $u_{0}$
\begin{eqnarray}
u_{0} = f^{-1}[\mathcal{F}[\xi_{0}]]=f^{-1}[f(\xi_{0})
e^{-\mu(1-\sqrt{1-\xi_{0}^{2}})}]
\label{eqA12_18}
\end{eqnarray}
We expand Eq.\ (\ref{eqA12_18}) near $x = f(\xi_{0})$,
by applying identities Eqs.\ (\ref{eqA12_6}--\ref{eqA12_9})
\begin{eqnarray}
u_{0} = \xi_{0}+\frac{\delta x}{f^{'}(\xi_{0})}
-\frac{f^{''}(\xi_{0})}{[f^{'}(\xi_{0})]^{3}}\frac{(\delta x)^{2}}{2}
\label{eqA12_19}
\end{eqnarray}
with
$\delta x = \left[e^{-\mu(1-\sqrt{1-\xi_{0}^{2}})}-1 \right]f(\xi_{0})
\sim -\mu[1-\sqrt{1-\xi_{0}^{2}}]f(\xi_{0})$. From Eq.\ (\ref{eqA12_16}),
we have
\begin{eqnarray}
\frac{\delta x}{f^{'}(\xi_{0})}=
\frac{\mu[\sqrt{1-\xi_{0}^{2}}-1]
f(\xi_{0})}{\frac{\mu\xi_{0}}{\sqrt{1-\xi_{0}^{2}}}f(\xi_{0})}=
\frac{1-\xi_{0}^{2}-\sqrt{1-\xi_{0}^{2}}}{\xi_{0}}
\label{eqA12_20}
\end{eqnarray}
From Eq.\ (\ref{eqA12_20}) into Eq.\ (\ref{eqA12_19}), after multiplying
by $\xi_{0}$ we find
\begin{eqnarray}
u_{0}\xi_{0} &=& \xi_{0}^{2}+\xi_{0}\frac{1-\xi_{0}^{2}-\sqrt{1-\xi_{0}^{2}}}
{\xi_{0}}-\xi_{0}\frac{(\delta x)^{2}}{2}
\frac{f^{''}(\xi_{0})}{[f^{'}(\xi_{0})]^{3}}\nonumber\\
&=& 1 - \sqrt{1-\xi_{0}^{2}} - 
f^{''}(\xi_{0})\frac{\xi_{0}(\delta x)^{2}}{[f^{'}(\xi_{0})]^{3}}
\label{eqA12_21}
\end{eqnarray}
Hence, we obtain
\begin{eqnarray}
u_{0}\xi_{0} + \sqrt{1-\xi_{0}^{2}} - 1 = 
-f^{''}(\xi_{0})\frac{\xi_{0}(\delta x)^{2}}{[f^{'}(\xi_{0})]^{3}}
\label{eqA12_22}
\end{eqnarray}
Clearly, the sign of this expression is determined by the sign
of $-f^{''}(\xi_{0})$, and hence after Eq.\ (\ref{eqA12_14}) we obtain
the condition
\begin{eqnarray}
\left.\frac{du}{d\nu}\right|_{\nu\rightarrow 0} = \left\{
\begin{array}{ccc}>0 & \rm{if} & f^{''}(\xi_{0})<0\\
< 0 & \rm{if} & f^{''}(\xi_{0})>0 \end{array}\right.
\label{eqA12_23}
\end{eqnarray}
which proves that the mutational deterministic
hypothesis holds for horizontal gene transfer of blocks of size
$\bar{M}=1$ in the Eigen model.

For the case of horizontal gene transfer of blocks of size $\bar{M}>1$,
we obtain the equation
\begin{eqnarray}
\left.\frac{du}{d\nu}\right|_{\nu\rightarrow 0} 
= \frac{\left[1+\frac{u_{0}\xi_{0}-1
+\sqrt{1-\xi_{0}^{2}}}{2}\right]^{\bar{M}}-1}{\bar{M} f^{'}(u_{0})}
e^{-\mu(1-\sqrt{1-\xi_{0}^{2}})}f(\xi_{0})
\label{eqA12_24}
\end{eqnarray}

By expanding the binomial in the numerator of Eq.\ (\ref{eqA12_24})
up to first order, we notice that the leading term is given by
\begin{eqnarray}
\left.\frac{du}{d\nu}\right|_{\nu\rightarrow 0}\sim
\frac{u_{0}\xi_{0}-1+\sqrt{1-\xi_{0}^{2}}}{2 f^{'}(u_{0})}
e^{-\mu(1-\sqrt{1-\xi_{0}^{2}})}f(\xi_{0})
\label{eqA12_25}
\end{eqnarray}
which is identical to Eq.\ (\ref{eqA12_14}). Therefore, the analysis
presented for the case $\bar{M}=1$, and in particular 
Eq.\ (\ref{eqA12_23}) applies for $\bar{M}>1$ as well.

When considering the process of horizontal gene transfer of
blocks of multiple size with average $\langle\bar{M}\rangle$,
we obtain the equation
\begin{eqnarray}
\left.\frac{du}{d\nu}\right|_{\nu\rightarrow 0}
=\frac{e^{\frac{\langle\bar{M}\rangle}{2}(u_{0}\xi_{0}-1
+\sqrt{1-\xi_{0}^{2}})}-1}{\langle\bar{M}\rangle f^{'}(u_{0})}f(\xi_{0})
e^{-\mu(1-\sqrt{1-\xi_{0}^{2}})}
\label{eqA12_26}
\end{eqnarray}

By expanding the exponential in Eq.\ (\ref{eqA12_26}) up to first
order, we notice that the leading term is given by Eq.\ (\ref{eqA12_25})
in this case as well, which is identical to Eq.\ (\ref{eqA12_14}). Therefore,
the analysis presented for the process with $\bar{M}=1$, and
in particular Eq.\ (\ref{eqA12_23}), applies for the process
of horizontal gene transfer of multiple size blocks as well.

Summarizing, we proved that the mutational deterministic hypothesis,
expressed quantitatively in Eq.\ (\ref{eqA12_23}), holds for the
different forms of horizontal gene transfer studied in this
work for the Eigen model.

\section{}
\label{appendix14}

For the case of two-parent recombination in the parallel
model, we find that the
phase structure is defined by two fitness functions.
A low $\nu$-dependent phase S1, defined as the maximum in $\xi$ of
\begin{eqnarray}
\mathcal{F}_{\nu}^{(1)}[\xi] = f(\xi) + \mu(\sqrt{1-\xi^{2}}-1) - \nu
\label{eqA14_1}
\end{eqnarray}
The maximum of this expression, attained at $\xi_{0}$,
is obtained from the equation
\begin{eqnarray}
\frac{\partial}{\partial\xi}\mathcal{F}_{\nu}^{(1)}[\xi_{0}]=f^{'}[\xi_{0}]
-\frac{\mu\xi_{0}}{\sqrt{1-\xi_{0}^{2}}}
\label{eqA14_2}
\end{eqnarray}
We notice that the value $\xi_{0}$ is the same as in the absence
of recombination, when $\nu=0$. Therefore, from the self-consistency
condition, we obtain for this phase
\begin{eqnarray}
f_{m}^{(1)} = \mathcal{F}_{\nu}^{(1)}[\xi_{0}] = \mathcal{F}_{0}[\xi_{0}] 
- \nu = f(u_{\nu})
\label{eqA14_3}
\end{eqnarray}
Here, we have denoted $u_{\nu}$ as the value of the average composition
in phase S1, when the recombination rate is $\nu$. Correspondingly,
we also have from Eq.\ (\ref{eqA14_3}) the exact relation
\begin{eqnarray}
f(u_{\nu}) = f(u_{0}) - \nu
\label{eqA14_4}
\end{eqnarray} 
with $f(u_{0}) = \mathcal{F}_{0}[\xi_{0}]$ and $u_{0}$ the average composition
in the absence of recombination, when $\nu=0$.

Let us define as $u_{*}$ the value of the average composition
at the S2 phase, which is independent of the recombination rate. 
The value $u_{*}$ is obtained as the solution of the
non-linear equation
\begin{eqnarray}
f^{'}(u_{*}) = \frac{2\mu u_{*}}{1-u_{*}^{2}}
\label{eqA14_5}
\end{eqnarray}

We consider in Eq.\ (\ref{eqA14_4}) the value $\nu=\nu^{*}$ at which
the average fitness of the S1 and S2 phases are identical, 
as the condition $u_{\nu^{*}}=u_{*}$,
\begin{eqnarray}
\nu^{*} = f(u_{0}) - f(u_{*})
\label{eqA14_6}
\end{eqnarray}
In Eq.\ (\ref{eqA14_6}), let us consider the Taylor expansion 
of $f(u_{*})$ near $u_{0}$, up to
first order in $\epsilon = u_{*} - u_{0}$,
\begin{eqnarray}
\nu^{*} = -\epsilon f^{'}(u_{*}) + \mathcal{O}(\epsilon^{2})
\label{eqA14_7} 
\end{eqnarray}

We expand Eq.\ (\ref{eqA14_5}) near $u_{0}$ at first order in
$\epsilon = u_{*} - u_{0}$,
\begin{eqnarray}
f^{'}(u_{0}) + \epsilon f^{''}(u_{0}) &=& 
\frac{2\mu(u_{0}+\epsilon)}{1-(u_{0}+\epsilon)^{2}}\sim 
\frac{2\mu(u_{0}+\epsilon)}{1-u_{0}^2}
\left[1-\frac{2u_{0}}{1-u_{0}^2}\epsilon \right]^{-1}\nonumber\\
&=& \frac{2\mu u_{0}}{1-u_{0}^2} + 
2\mu\frac{1+u_{0}^2}{(1 - u_{0}^2)^2}\epsilon + \mathcal{O}(\epsilon^{2})
\label{eqA14_8}
\end{eqnarray}

We solve explicitly for $\epsilon$ in Eq.\ (\ref{eqA14_8}), and combine
with Eq.\ (\ref{eqA14_7}), to obtain an expression for $\nu^{*}$
\begin{eqnarray}
\nu^{*} = \frac{f^{'}(u_{0})\left[f^{'}(u_{0})
-\frac{2\mu u_{0}}{1-u_{0}^{2}}\right]}{f^{''}(u_{0})-2\mu
\frac{1+u_{0}^{2}}{(1-u_{0}^{2})^2}}
\label{eqA14_9}
\end{eqnarray}

Let us now analyze the sign of $\nu^{*}$ as a function of the
sign of the curvature of the fitness function, as defined by $f^{''}$.
We consider the Laurent series of $f(u)$ for small $u$. That is,
\begin{eqnarray}
f(u) &=& k u^{\alpha}\nonumber\\
f^{'}(u) &=& k\alpha u^{\alpha-1}\nonumber\\
f^{''}(u) &=& k\alpha(\alpha-1) u^{\alpha-2}
\label{eqA14_10}
\end{eqnarray}
where $\alpha > 0$ to satisfy the monotonically increasing
condition. This family of polynomials provides a representation
of arbitrary, monotonically increasing functions for small $u_{0}$.
 
The case $\alpha = 0$, corresponding to a constant
identical fitness for all sequence types in the population, 
possesses the trivial solution after Eq.\ (\ref{eqA14_2}) $\xi_{0}=0$,
which implies $u_{0}=0$,
and after Eq.\ (\ref{eqA14_5}) $u_{*}=0$. Thus a single non-selective
phase is observed for this case, both in the presence and in the absence of
recombination.

From Eq.\ (\ref{eqA14_10}), we have $f^{''}<0$ for $\alpha<1$,
$f^{''}>0$ for $\alpha>1$ and $f^{''}=0$ at $\alpha=1$. We analyze
these possible cases separately.    
From Eq.\ (\ref{eqA14_10}) into Eq.\ (\ref{eqA14_9}), we have
\begin{eqnarray}
\nu^{*} = \frac{k\alpha u_{0}^{\alpha}(k\alpha u_{0}^{\alpha}-
\frac{2\mu u_{0}^{2}}{1-u_{0}^{2}})}{k\alpha(\alpha-1)u_{0}^{\alpha}
-2\mu u_{0}^{2}\frac{1+u_{0}^{2}}{(1-u_{0}^{2})^{2}}}
\label{eqA14_11}
\end{eqnarray}

Case 1: $\alpha < 1$, $f^{''}<0$.

The denominator in Eq.\ (\ref{eqA14_11}) is clearly negative, since
$\alpha - 1 < 0$ in this case.

The numerator, for $u_{0} \ll 1$
\begin{eqnarray}
k\alpha u_{0}^{\alpha} - \frac{2\mu u_{0}^{2}}{1-u_{0}^{2}} 
\sim k\alpha u_{0}^{\alpha} - 2\mu u_{0}^{2} > 0
\label{eqA14_12}
\end{eqnarray}

Therefore, in this case $\nu^{*}=\frac{(>0)}{(<0)}<0$, and
hence $u_{*} - u_{0} > 0$.

Case 2: $1 < \alpha < 2$, $f^{''}>0$.

The denominator in Eq.\ (\ref{eqA14_11}), for $u_{0} \ll 1$
and $\alpha - 1 > 0$,
\begin{eqnarray}
k\alpha(\alpha - 1)u_{0}^{\alpha} - 2\mu u_{0}^{2}
\frac{1+u_{0}^{2}}{(1-u_{0}^{2})^2} \sim k\alpha (\alpha-1)u_{0}^{\alpha}
-2\mu u_{0}^{2} > 0
\label{eqA14_13}
\end{eqnarray}

The numerator is also positive, by the same argument as 
in Eq.\ (\ref{eqA14_12}). Therefore, in this case 
$\nu^{*}=\frac{(>0)}{(>0)}>0$, and hence $u_{*} < u_{0}$.

Case 3: $\alpha > 2$, $f^{''}>0$. 

The denominator in Eq. (\ref{eqA14_11}), for $u_{0} \ll 1$ and
$\alpha - 1 > 0$,
\begin{eqnarray}
k\alpha(\alpha - 1)u_{0}^{\alpha} - 2\mu u_{0}^{2}
\frac{1+u_{0}^{2}}{(1-u_{0}^{2})^2} \sim k\alpha (\alpha-1)u_{0}^{\alpha}
-2\mu u_{0}^{2} < 0
\label{eqA14_14}
\end{eqnarray}

The numerator is
\begin{eqnarray}
k\alpha u_{0}^{\alpha} - \frac{2\mu u_{0}^{2}}{1-u_{0}^{2}} 
\sim k\alpha u_{0}^{\alpha} - 2\mu u_{0}^{2} < 0
\label{eqA14_15}
\end{eqnarray}

Therefore, in this case $\nu^{*}=\frac{(<0)}{(<0)}>0$, and hence
$u_{*} - u_{0} < 0$.

For $\alpha = 1$, we obtain an exact solution from
Eq.\ (\ref{eqA14_2}), $u_{0} = \sqrt{1+\mu^{2}/k^{2}}-\mu/k$. This
result in Eq.\ (\ref{eqA14_11}) yields $\nu^{*}=0$, and thus
$u_{*} = u_{0}$ for this particular case.

For $\alpha=2$, we have the analytical solution presented
in Eqs.\ (\ref{eq99}),
\begin{eqnarray}
u_{*} - u_{0} = \sqrt{1-2\frac{\mu}{k}} - \left(1-\frac{\mu}{k}\right)
= \sqrt{\left(1-\frac{\mu}{k} \right)^{2}-\frac{\mu^{2}}{k^{2}}}-
\left(1-\frac{\mu}{k} \right) < 0
\label{eqA14_16}
\end{eqnarray}
with $\nu^{*} = \frac{\mu^2}{2k}>0$.

Summarizing, we proved that
\begin{eqnarray}
u_{*} - u_{0} = \left\{\begin{array}{cc}>0, & f^{''} < 0\\
<0, & f^{''} > 0 \end{array} \right.
\label{eqA14_17}
\end{eqnarray}

This result proves the mutational deterministic hypothesis for two-parent
recombination in the parallel model.

\section{}
\label{appendix15}

For the case of two-parent recombination in the Eigen model, we find that the
phase structure is defined by two fitness functions.
A low $\nu$-dependent phase S1, defined as the maximum in $\xi$ of
\begin{eqnarray}
\mathcal{F}_{\nu}^{(1)}[\xi] = (1- \nu)e^{-\mu[1-\sqrt{1-\xi^{2}}]}
\label{eqA15_1}
\end{eqnarray}
The maximum of this expression, attained at $\xi_{0}$,
is obtained from the equation
\begin{eqnarray}
\frac{\partial}{\partial\xi}\mathcal{F}_{\nu}^{(1)}[\xi_{0}]&=&0\nonumber\\
f^{'}(\xi_{0})&=&\frac{\mu\xi_{0}}{\sqrt{1-\xi_{0}^{2}}}f(\xi_{0})\nonumber\\
\left[\ln f(\xi_{0})\right]^{'} &=& \frac{\mu\xi_{0}}{\sqrt{1-\xi_{0}^{2}}}
\label{eqA15_2}
\end{eqnarray}
We notice that the value $\xi_{0}$ is the same as in the absence
of recombination, when $\nu=0$. Therefore, from the self-consistency
condition, we obtain for this phase
\begin{eqnarray}
f_{m}^{(1)} = \mathcal{F}_{\nu}^{(1)}[\xi_{0}] = 
(1-\nu)\mathcal{F}_{0}[\xi_{0}] = f(u_{\nu})
\label{eqA15_3}
\end{eqnarray}
Here, we have denoted $u_{\nu}$ as the value of the average composition
in phase S1, when the recombination rate is $\nu$. Correspondingly,
we also have from Eq. (\ref{eqA15_3}) the exact relation
\begin{eqnarray}
f(u_{\nu}) = (1-\nu)f(u_{0}) 
\label{eqA15_4}
\end{eqnarray} 
with $f(u_{0}) = \mathcal{F}_{0}[\xi_{0}]$ and $u_{0}$ the average composition
in the absence of recombination, when $\nu=0$.

Let us define as $u_{*}$ the value of the average composition
at the S2 phase, which is independent of the recombination rate. 
The value $u_{*}$ is obtained as the solution of the
non-linear equation 
\begin{eqnarray}
f^{'}(u_{*}) &=& \frac{2\mu u_{*}}{1-u_{*}^{2}}f(u_{*})\nonumber\\
\left[\ln f(u_{*})\right]^{'} &=& \frac{2\mu u_{*}}{1-u_{*}^{2}}
\label{eqA15_5}
\end{eqnarray}

We consider in Eq.\ (\ref{eqA15_3}) the value $\nu = \nu^{*}$ at which
the average fitness of the two phases are equal, as the
condition $u_{\nu^{*}}=u_{*}$,
\begin{eqnarray}
1-\nu^{*} = \frac{f(u_{*})}{f(u_{0})}
\label{eqA15_6}
\end{eqnarray}

We take the logarithm of this expression, and Taylor expand up to
first order in $\epsilon = u_{*}-u_{0}$,
\begin{eqnarray}
\ln(1 - \nu^{*}) = \ln[f(u_{0}+\epsilon)] - \ln[f(u_{0})]\nonumber\\
-\nu^{*} = \epsilon[\ln f(u_{0})]^{'}
\label{eqA15_7}
\end{eqnarray}

We expand Eq.\ (\ref{eqA15_5}) near $u_{0}$ at first order in 
$\epsilon = u_{*} - u_{0}$, 
\begin{eqnarray}
[\ln f(u_{0})]^{'} + \epsilon [\ln f(u_{0})]^{''} &=&
\frac{2\mu (u_{0}+\epsilon)}{1-(u_{0}+\epsilon)^{2}}\nonumber\\
&=& \frac{2\mu(u_{0}+\epsilon)}{1-u_{0}^{2}}
\left[1-\frac{2 u_{0}}{1-u_{0}^{2}}\epsilon\right]^{-1}
+ \mathcal{O}(\epsilon^2)\nonumber\\
&=&\frac{2\mu u_{0}}{1-u_{0}^{2}}
+2\mu\frac{1+u_{0}^{2}}{(1-u_{0}^{2})^{2}}\epsilon + \mathcal{O}(\epsilon^2)
\label{eqA15_8}
\end{eqnarray}

We solve explicitly for $\epsilon$ in Eq.\ (\ref{eqA15_8}), and combine
with Eq.\ (\ref{eqA15_7}), to obtain an expression for $\nu^{*}$
\begin{eqnarray}
\nu^{*} = [\ln f(u_{0})]^{'}\frac{[\ln f(u_{0})]^{'}
-\frac{2\mu u_{0}}{1-u_{0}^{2}}}{[\ln f(u_{0})]^{''}
-\frac{2\mu(1+u_{0}^{2})}{(1-u_{0}^{2})^{2}}}
\label{eqA15_9}
\end{eqnarray}

The analysis follows the same lines as in the parallel model case.
That is, we analyze the sign of $\nu^{*}$ after Eq.\ (\ref{eqA15_9}).
We consider a family of polynomials
$f(u) = k u^{\alpha} + k_{0}$, which for $u_{0}\ll 1$
\begin{eqnarray}
\ln f(u) &=& \ln\left(1 + \frac{k}{k_{0}}u^{\alpha}\right) + \ln (k_{0})
\sim \frac{k}{k_{0}}u^{\alpha} + \ln(k_{0})\nonumber\\
\left[\ln f(u)\right]^{'} &=& \alpha\frac{k}{k_{0}}u^{\alpha - 1}\nonumber\\
\left[\ln f(u)\right]^{''} &=& \alpha(\alpha - 1)\frac{k}{k_{0}}u^{\alpha-2}
\label{eqA15_10}
\end{eqnarray} 
with $\alpha > 0$ to satisfy the monotonically increasing condition. This
family of polynomials provides a representation of 
smooth and monotonically increasing functions for small $u_{0}$.

The case $\alpha = 0$ corresponds to a constant identical fitness
for all sequence types in the population, and possesses the
trivial solution after Eq. (\ref{eqA15_2}) $\xi_{0}=0$, which implies
$u_{0}=0$, and after Eq. (\ref{eqA15_5}) $u_{*}=0$. Therefore, a single
non-selective phase is observed for this case, both in the presence
and in the absence of recombination.

From Eq.\ (\ref{eqA15_10}), we have $f^{''}<0$ for $\alpha < 1$, $f^{''}>0$
for $\alpha>1$ and $f^{''}=0$ at $\alpha=1$. We analyze these possible
cases separately. From Eq.\ (\ref{eqA15_10}) into Eq.\ (\ref{eqA15_9}),
we have
\begin{eqnarray}
\nu^{*} = \frac{\frac{k}{k_{0}}
\alpha u_{0}^{\alpha}\left(\frac{k}{k_{0}}\alpha u_{0}^{\alpha}-
\frac{2\mu u_{0}^{2}}{1-u_{0}^{2}} \right)}{\frac{k}{k_{0}}
\alpha(\alpha-1)u_{0}^{\alpha}
-2\mu u_{0}^{2}\frac{1+u_{0}^{2}}{(1-u_{0}^{2})^{2}}}
\label{eqA15_11}
\end{eqnarray}

Case 1: $\alpha < 1$, $f^{''}<0$.

The denominator in Eq.\ (\ref{eqA15_11}) is clearly negative, since
$\alpha - 1<0$ in this case.

The numerator, for $u_{0}\ll 1$
\begin{eqnarray}
\frac{k}{k_{0}}\alpha u_{0}^{\alpha} - \frac{2\mu u_{0}^{2}}{1-u_{0}^{2}}\sim
\frac{k}{k_{0}}\alpha u_{0}^{\alpha} - 2\mu u_{0}^{2}>0
\label{eqA15_12}
\end{eqnarray}

Therefore, in this case $\nu^{*}=\frac{(>0)}{(<0)}<0$, and hence 
$u_{*}-u_{0}>0$.

Case 2: $1 < \alpha < 2$, $f^{''}>0$.

The denominator in Eq.\ (\ref{eqA15_11}), for $u_{0}\ll 1$ and
$\alpha-1>0$,
\begin{eqnarray}
\frac{k}{k_{0}}\alpha(\alpha-1)u_{0}^{\alpha}
-2\mu u_{0}^{2}\frac{1+u_{0}^{2}}{(1-u_{0}^{2})^{2}}\sim
\frac{k}{k_{0}}\alpha(\alpha-1)u_{0}^{\alpha}-2\mu u_{0}^{2}>0
\label{eqA15_13}
\end{eqnarray}

The numerator is also positive, by the same argument as in 
Eq.\ (\ref{eqA15_12}). Therefore, in this case $\nu^{*}=\frac{(>0)}{(>0)}$,
and hence $u_{*}<u_{0}$.

Case 3: $\alpha>2$, $f^{''}>0$.

The denominator in Eq.\ (\ref{eqA15_11}), for $u_{0}\ll 1$ and
$\alpha-1>0$,
\begin{eqnarray}
\frac{k}{k_{0}}\alpha(\alpha-1)u_{0}^{\alpha}-2\mu u_{0}^{2}
\frac{1+u_{0}^{2}}{(1-u_{0}^{2})^{2}}\sim 
\frac{k}{k_{0}}\alpha(\alpha-1)u_{0}^{\alpha}
-2\mu u_{0}^{2}<0
\label{eqA15_14}
\end{eqnarray}

The numerator is
\begin{eqnarray}
\frac{k}{k_{0}}\alpha u_{0}^{\alpha} - \frac{2\mu u_{0}^{2}}{1-u_{0}^{2}}
\sim \frac{k}{k_{0}}\alpha u_{0}^{\alpha} - 2\mu u_{0}^{2}<0
\label{eqA15_15}
\end{eqnarray}

Therefore, in this case $\nu^{*}=\frac{(<0)}{(<0)}>0$, and hence
$u_{*}-u_{0}<0$.

For $\alpha=1$, we find that for $u_{*}\ll 1$ and $u_{0}\ll 1$, 
$u_{*} = \frac{k}{2\mu k_{0}} + \mathcal{O}\left(\frac{k}{2\mu k_{0}}\right)^2$,
$\xi_{0} = \frac{k}{\mu k_{0}} + \mathcal{O}\left(\frac{k}{2\mu k_{0}}\right)^2$and $u_{0} = \frac{k}{2\mu k_{0}} + \mathcal{O}\left(\frac{k}{2\mu k_{0}}\right)^2$. Therefore, $u_{*} - u_{0} = 0$ and $\nu^{*}=0$ in this case.

For $\alpha = 2$, we have the exact solution expressed in 
Eqs.\ (\ref{eq187}), (\ref{eq187b}). The region of parameters space
where phases S1 and S2 intersect is $2\frac{\mu k_{0}}{k}<1$. We 
analyze these formulas considering that 
$u_{*} < 1$ and $u_{0} < 1$. It is convenient to
define in this case the small parameter $\epsilon = 2\frac{\mu k_{0}}{k}< 1$.
From Eq.\ (\ref{eqA15_5}), we have 
\begin{eqnarray}
u_{*}=\frac{1}{1+\mu} - \mathcal{O}(\epsilon)
\label{eqA15_16}
\end{eqnarray}
Expanding Eq.\ (\ref{eq187})
up to first order in $\epsilon$, we obtain the result
\begin{eqnarray}
u_{0} = \sqrt{2\frac{\sqrt{1+\mu^2}-1}{\mu^2}} - \mathcal{O}(\epsilon)
\label{eqA15_17}
\end{eqnarray}
Therefore, for $\epsilon \ll 1$, from Eq.\ (\ref{eqA15_17}) and 
Eq.\ (\ref{eqA15_16}), when $\alpha=2$, $u_{*}<u_{0}$,
and hence $\nu^{*}>0$.

Summarizing, we have shown that
\begin{eqnarray}
u_{*}-u_{0}=\left\{\begin{array}{cc}>0, & f^{''}<0\\<0, & f^{''}>0\end{array} 
\right.
\label{eqA15_18}
\end{eqnarray}

This result proves the mutational deterministic hypothesis for two-parent
recombination in the Eigen model.

\bibliography{recombination}

\begin{thebibliography}{60}
\expandafter\ifx\csname natexlab\endcsname\relax\def\natexlab#1{#1}\fi
\expandafter\ifx\csname bibnamefont\endcsname\relax
  \def\bibnamefont#1{#1}\fi
\expandafter\ifx\csname bibfnamefont\endcsname\relax
  \def\bibfnamefont#1{#1}\fi
\expandafter\ifx\csname citenamefont\endcsname\relax
  \def\citenamefont#1{#1}\fi
\expandafter\ifx\csname url\endcsname\relax
  \def\url#1{\texttt{#1}}\fi
\expandafter\ifx\csname urlprefix\endcsname\relax\def\urlprefix{URL }\fi
\providecommand{\bibinfo}[2]{#2}
\providecommand{\eprint}[2][]{\url{#2}}

\bibitem[{\citenamefont{Cohen et~al.}(2005)\citenamefont{Cohen, Kesslerm, and
  Levine}}]{cohen05}
\bibinfo{author}{\bibfnamefont{E.}~\bibnamefont{Cohen}},
  \bibinfo{author}{\bibfnamefont{D.~A.} \bibnamefont{Kesslerm}},
  \bibnamefont{and} \bibinfo{author}{\bibfnamefont{H.}~\bibnamefont{Levine}},
  \bibinfo{journal}{Phys. Rev. Lett.} \textbf{\bibinfo{volume}{94}},
  \bibinfo{pages}{098102} (\bibinfo{year}{2005}).

\bibitem[{\citenamefont{Muller}(1964)}]{Muller64}
\bibinfo{author}{\bibfnamefont{H.~J.} \bibnamefont{Muller}},
  \bibinfo{journal}{Mutation Research} \textbf{\bibinfo{volume}{1}},
  \bibinfo{pages}{2} (\bibinfo{year}{1964}).

\bibitem[{\citenamefont{Lawrence}(1997)}]{Lawrence97}
\bibinfo{author}{\bibfnamefont{J.~G.} \bibnamefont{Lawrence}},
  \bibinfo{journal}{Trends Microbiol.} \textbf{\bibinfo{volume}{5}},
  \bibinfo{pages}{355} (\bibinfo{year}{1997}).

\bibitem[{\citenamefont{Patten et~al.}(1997)\citenamefont{Patten, Howard, and
  Stemmer}}]{Patten97}
\bibinfo{author}{\bibfnamefont{P.~A.} \bibnamefont{Patten}},
  \bibinfo{author}{\bibfnamefont{R.~J.} \bibnamefont{Howard}},
  \bibnamefont{and} \bibinfo{author}{\bibfnamefont{W.~P.~C.}
  \bibnamefont{Stemmer}}, \bibinfo{journal}{Curr. Opin. Biotechnol.}
  \textbf{\bibinfo{volume}{8}}, \bibinfo{pages}{724} (\bibinfo{year}{1997}).

\bibitem[{\citenamefont{Lutz and Benkovic}(2000)}]{Lutz00}
\bibinfo{author}{\bibfnamefont{S.}~\bibnamefont{Lutz}} \bibnamefont{and}
  \bibinfo{author}{\bibfnamefont{S.~J.} \bibnamefont{Benkovic}},
  \bibinfo{journal}{Curr. Opin. Biotechnol.} \textbf{\bibinfo{volume}{11}},
  \bibinfo{pages}{319} (\bibinfo{year}{2000}).

\bibitem[{\citenamefont{Otto and Lenormand}(2002)}]{Otto02}
\bibinfo{author}{\bibfnamefont{S.~P.} \bibnamefont{Otto}} \bibnamefont{and}
  \bibinfo{author}{\bibfnamefont{T.}~\bibnamefont{Lenormand}},
  \bibinfo{journal}{Nature Rev. Genet.} \textbf{\bibinfo{volume}{3}},
  \bibinfo{pages}{252} (\bibinfo{year}{2002}).

\bibitem[{\citenamefont{Arjan et~al.}(2007)\citenamefont{Arjan, de~Visser, and
  Elena}}]{Arjan07}
\bibinfo{author}{\bibfnamefont{J.}~\bibnamefont{Arjan}},
  \bibinfo{author}{\bibfnamefont{G.~M.} \bibnamefont{de~Visser}},
  \bibnamefont{and} \bibinfo{author}{\bibfnamefont{S.~F.} \bibnamefont{Elena}},
  \bibinfo{journal}{Nature {R}ev. {G}enet.} \textbf{\bibinfo{volume}{8}},
  \bibinfo{pages}{139} (\bibinfo{year}{2007}).

\bibitem[{\citenamefont{Misevic et~al.}(2006)\citenamefont{Misevic, Ofria, and
  Lenski}}]{Misevic06}
\bibinfo{author}{\bibfnamefont{D.}~\bibnamefont{Misevic}},
  \bibinfo{author}{\bibfnamefont{C.}~\bibnamefont{Ofria}}, \bibnamefont{and}
  \bibinfo{author}{\bibfnamefont{R.~E.} \bibnamefont{Lenski}},
  \bibinfo{journal}{{P}roc. {R}. {S}oc. {B}} \textbf{\bibinfo{volume}{273}},
  \bibinfo{pages}{457} (\bibinfo{year}{2006}).

\bibitem[{\citenamefont{Kondrashov}(1982)}]{Kondrashov82}
\bibinfo{author}{\bibfnamefont{A.~S.} \bibnamefont{Kondrashov}},
  \bibinfo{journal}{{G}enet. {R}es.} \textbf{\bibinfo{volume}{42}},
  \bibinfo{pages}{325} (\bibinfo{year}{1982}).

\bibitem[{\citenamefont{Kondrashov}(1993)}]{Kondrashov93}
\bibinfo{author}{\bibfnamefont{A.~S.} \bibnamefont{Kondrashov}},
  \bibinfo{journal}{J. {H}ered.} \textbf{\bibinfo{volume}{84}},
  \bibinfo{pages}{372} (\bibinfo{year}{1993}).

\bibitem[{\citenamefont{Azevedo et~al.}(2006)\citenamefont{Azevedo, Lohaus,
  Srinivasan, Dang, and Burch}}]{Azevedo06}
\bibinfo{author}{\bibfnamefont{R.~B.~R.} \bibnamefont{Azevedo}},
  \bibinfo{author}{\bibfnamefont{R.}~\bibnamefont{Lohaus}},
  \bibinfo{author}{\bibfnamefont{S.}~\bibnamefont{Srinivasan}},
  \bibinfo{author}{\bibfnamefont{K.~K.} \bibnamefont{Dang}}, \bibnamefont{and}
  \bibinfo{author}{\bibfnamefont{C.~L.} \bibnamefont{Burch}},
  \bibinfo{journal}{Nature} \textbf{\bibinfo{volume}{440}}, \bibinfo{pages}{87}
  (\bibinfo{year}{2006}).

\bibitem[{\citenamefont{Phillips et~al.}(2000)\citenamefont{Phillips, Otto, and
  Whitlock}}]{Phillips}
\bibinfo{author}{\bibfnamefont{P.~C.} \bibnamefont{Phillips}},
  \bibinfo{author}{\bibfnamefont{S.~P.} \bibnamefont{Otto}}, \bibnamefont{and}
  \bibinfo{author}{\bibfnamefont{M.~C.} \bibnamefont{Whitlock}}, in
  \emph{\bibinfo{booktitle}{Beyond the average}}, edited by
  \bibinfo{editor}{\bibfnamefont{J.~B.} \bibnamefont{Wolf}},
  \bibinfo{editor}{\bibfnamefont{E.~D.} \bibnamefont{{Brodie III}}},
  \bibnamefont{and} \bibinfo{editor}{\bibfnamefont{M.~J.} \bibnamefont{Wade}}
  (\bibinfo{publisher}{{O}xford {U}niversity {P}ress}, \bibinfo{year}{2000}),
  ISBN-0-19-512806-0, chap. \bibinfo{chapter}{The evolutionary importance of
  gene interactions and variability of epistatic effects}.

\bibitem[{\citenamefont{Kimura and Maruyama}(1966)}]{Kimura66}
\bibinfo{author}{\bibfnamefont{M.}~\bibnamefont{Kimura}} \bibnamefont{and}
  \bibinfo{author}{\bibfnamefont{T.}~\bibnamefont{Maruyama}},
  \bibinfo{journal}{Genetics} \textbf{\bibinfo{volume}{54}},
  \bibinfo{pages}{1337} (\bibinfo{year}{1966}).

\bibitem[{\citenamefont{Kondrashov}(1988)}]{Kondrashov88}
\bibinfo{author}{\bibfnamefont{A.~S.} \bibnamefont{Kondrashov}},
  \bibinfo{journal}{Nature} \textbf{\bibinfo{volume}{336}},
  \bibinfo{pages}{435} (\bibinfo{year}{1988}).

\bibitem[{\citenamefont{Kouyos et~al.}(2007)\citenamefont{Kouyos, Silander, and
  Bonhoeffer}}]{Kouyos07}
\bibinfo{author}{\bibfnamefont{R.~D.} \bibnamefont{Kouyos}},
  \bibinfo{author}{\bibfnamefont{O.~K.} \bibnamefont{Silander}},
  \bibnamefont{and}
  \bibinfo{author}{\bibfnamefont{S.}~\bibnamefont{Bonhoeffer}},
  \bibinfo{journal}{Trends {E}col. {E}vol.} \textbf{\bibinfo{volume}{22}},
  \bibinfo{pages}{310} (\bibinfo{year}{2007}).

\bibitem[{\citenamefont{Rice and Chippindale}(2001)}]{Rice01}
\bibinfo{author}{\bibfnamefont{W.~R.} \bibnamefont{Rice}} \bibnamefont{and}
  \bibinfo{author}{\bibfnamefont{A.~K.} \bibnamefont{Chippindale}},
  \bibinfo{journal}{Science} \textbf{\bibinfo{volume}{294}},
  \bibinfo{pages}{555} (\bibinfo{year}{2001}).

\bibitem[{\citenamefont{Kouyos et~al.}(2006)\citenamefont{Kouyos, Otto, and
  Bonhoeffer}}]{Kouyos06}
\bibinfo{author}{\bibfnamefont{R.~D.} \bibnamefont{Kouyos}},
  \bibinfo{author}{\bibfnamefont{S.~P.} \bibnamefont{Otto}}, \bibnamefont{and}
  \bibinfo{author}{\bibfnamefont{S.}~\bibnamefont{Bonhoeffer}},
  \bibinfo{journal}{Genetics} \textbf{\bibinfo{volume}{173}},
  \bibinfo{pages}{589} (\bibinfo{year}{2006}).

\bibitem[{\citenamefont{Liberman and Feldman}(2005)}]{Liberman05}
\bibinfo{author}{\bibfnamefont{U.}~\bibnamefont{Liberman}} \bibnamefont{and}
  \bibinfo{author}{\bibfnamefont{M.~W.} \bibnamefont{Feldman}},
  \bibinfo{journal}{Theor. {P}opul. {B}iol.} \textbf{\bibinfo{volume}{67}},
  \bibinfo{pages}{141} (\bibinfo{year}{2005}).

\bibitem[{\citenamefont{Liberman et~al.}(2007)\citenamefont{Liberman, Puniyani,
  and Feldman}}]{Liberman07}
\bibinfo{author}{\bibfnamefont{U.}~\bibnamefont{Liberman}},
  \bibinfo{author}{\bibfnamefont{A.}~\bibnamefont{Puniyani}}, \bibnamefont{and}
  \bibinfo{author}{\bibfnamefont{M.~W.} \bibnamefont{Feldman}},
  \bibinfo{journal}{Theor. {P}opul. {B}iol.} \textbf{\bibinfo{volume}{71}},
  \bibinfo{pages}{230} (\bibinfo{year}{2007}).

\bibitem[{\citenamefont{Liberman and Feldman}(2008)}]{Liberman08}
\bibinfo{author}{\bibfnamefont{U.}~\bibnamefont{Liberman}} \bibnamefont{and}
  \bibinfo{author}{\bibfnamefont{M.}~\bibnamefont{Feldman}},
  \bibinfo{journal}{Theor. {P}opul. {B}iol.} pp. \bibinfo{pages}{307--316}
  (\bibinfo{year}{2008}).

\bibitem[{\citenamefont{Malmberg}(1977)}]{Malmberg77}
\bibinfo{author}{\bibfnamefont{R.~L.} \bibnamefont{Malmberg}},
  \bibinfo{journal}{Genetics} \textbf{\bibinfo{volume}{86}},
  \bibinfo{pages}{607} (\bibinfo{year}{1977}).

\bibitem[{\citenamefont{Bonhoeffer et~al.}(2004)\citenamefont{Bonhoeffer,
  Chappey, Parkin, Whitcomb, and Petropoulos}}]{Bonhoeffer04}
\bibinfo{author}{\bibfnamefont{S.}~\bibnamefont{Bonhoeffer}},
  \bibinfo{author}{\bibfnamefont{C.}~\bibnamefont{Chappey}},
  \bibinfo{author}{\bibfnamefont{N.~T.} \bibnamefont{Parkin}},
  \bibinfo{author}{\bibfnamefont{J.~M.} \bibnamefont{Whitcomb}},
  \bibnamefont{and} \bibinfo{author}{\bibfnamefont{C.~J.}
  \bibnamefont{Petropoulos}}, \bibinfo{journal}{Science}
  \textbf{\bibinfo{volume}{306}}, \bibinfo{pages}{1547} (\bibinfo{year}{2004}).

\bibitem[{\citenamefont{Wloch et~al.}(2001)\citenamefont{Wloch, Borts, and
  Korona}}]{Wloch01}
\bibinfo{author}{\bibfnamefont{D.~M.} \bibnamefont{Wloch}},
  \bibinfo{author}{\bibfnamefont{R.~H.} \bibnamefont{Borts}}, \bibnamefont{and}
  \bibinfo{author}{\bibfnamefont{R.}~\bibnamefont{Korona}},
  \bibinfo{journal}{J. {E}vol. {B}iol.} \textbf{\bibinfo{volume}{14}},
  \bibinfo{pages}{310} (\bibinfo{year}{2001}).

\bibitem[{\citenamefont{Eigen and Schuster}(1971)}]{Eigen71}
\bibinfo{author}{\bibfnamefont{M.}~\bibnamefont{Eigen}} \bibnamefont{and}
  \bibinfo{author}{\bibfnamefont{P.}~\bibnamefont{Schuster}},
  \bibinfo{journal}{Naturwissenschaften} \textbf{\bibinfo{volume}{58}},
  \bibinfo{pages}{465} (\bibinfo{year}{1971}).

\bibitem[{\citenamefont{Eigen et~al.}(1988)\citenamefont{Eigen, McCaskill, and
  Schuster}}]{Eigen88}
\bibinfo{author}{\bibfnamefont{M.}~\bibnamefont{Eigen}},
  \bibinfo{author}{\bibfnamefont{J.}~\bibnamefont{McCaskill}},
  \bibnamefont{and} \bibinfo{author}{\bibfnamefont{P.}~\bibnamefont{Schuster}},
  \bibinfo{journal}{J. Phys. Chem.} \textbf{\bibinfo{volume}{92}},
  \bibinfo{pages}{6881} (\bibinfo{year}{1988}).

\bibitem[{\citenamefont{Eigen et~al.}(1989)\citenamefont{Eigen, McCaskill, and
  Schuster}}]{Eigen89}
\bibinfo{author}{\bibfnamefont{M.}~\bibnamefont{Eigen}},
  \bibinfo{author}{\bibfnamefont{J.}~\bibnamefont{McCaskill}},
  \bibnamefont{and} \bibinfo{author}{\bibfnamefont{P.}~\bibnamefont{Schuster}},
  \bibinfo{journal}{Adv.Chem.Phys.} \textbf{\bibinfo{volume}{75}},
  \bibinfo{pages}{149} (\bibinfo{year}{1989}).

\bibitem[{\citenamefont{Biebricher and Eigen}(2005)}]{Biebricher05}
\bibinfo{author}{\bibfnamefont{C.~K.} \bibnamefont{Biebricher}}
  \bibnamefont{and} \bibinfo{author}{\bibfnamefont{M.}~\bibnamefont{Eigen}},
  \bibinfo{journal}{Virus Res.} \textbf{\bibinfo{volume}{107}},
  \bibinfo{pages}{117} (\bibinfo{year}{2005}).

\bibitem[{\citenamefont{Crow and Kimura}(1970)}]{Kimura70}
\bibinfo{author}{\bibfnamefont{J.~F.} \bibnamefont{Crow}} \bibnamefont{and}
  \bibinfo{author}{\bibfnamefont{M.}~\bibnamefont{Kimura}},
  \emph{\bibinfo{title}{An introduction to population genetics theory}}
  (\bibinfo{publisher}{Harper and Row}, \bibinfo{address}{New York},
  \bibinfo{year}{1970}).

\bibitem[{\citenamefont{Baake and Wagner}(2001)}]{Baake01}
\bibinfo{author}{\bibfnamefont{E.}~\bibnamefont{Baake}} \bibnamefont{and}
  \bibinfo{author}{\bibfnamefont{H.}~\bibnamefont{Wagner}},
  \bibinfo{journal}{Genet. Res. Camb.} \textbf{\bibinfo{volume}{78}},
  \bibinfo{pages}{93} (\bibinfo{year}{2001}).

\bibitem[{\citenamefont{Tarazona}(1992)}]{Tarazona92}
\bibinfo{author}{\bibfnamefont{P.}~\bibnamefont{Tarazona}},
  \bibinfo{journal}{Phys. Rev. A} \textbf{\bibinfo{volume}{45}},
  \bibinfo{pages}{6038} (\bibinfo{year}{1992}).

\bibitem[{\citenamefont{Leuthausser}(1987)}]{Leuthausser87}
\bibinfo{author}{\bibfnamefont{I.}~\bibnamefont{Leuthausser}},
  \bibinfo{journal}{J. Stat.Phys.} \textbf{\bibinfo{volume}{48}},
  \bibinfo{pages}{343} (\bibinfo{year}{1987}).

\bibitem[{\citenamefont{Franz and Peliti}(1997)}]{Franz97}
\bibinfo{author}{\bibfnamefont{S.}~\bibnamefont{Franz}} \bibnamefont{and}
  \bibinfo{author}{\bibfnamefont{L.}~\bibnamefont{Peliti}},
  \bibinfo{journal}{J. Phys. A: Math. Gen.} \textbf{\bibinfo{volume}{30}},
  \bibinfo{pages}{4481} (\bibinfo{year}{1997}).

\bibitem[{\citenamefont{Park and Deem}(2006)}]{Park06}
\bibinfo{author}{\bibfnamefont{J.-M.} \bibnamefont{Park}} \bibnamefont{and}
  \bibinfo{author}{\bibfnamefont{M.~W.} \bibnamefont{Deem}},
  \bibinfo{journal}{J. Stat.Phys.} \textbf{\bibinfo{volume}{123}},
  \bibinfo{pages}{975} (\bibinfo{year}{2006}).

\bibitem[{\citenamefont{Saakian et~al.}(2006)\citenamefont{Saakian,
  {Mu\~{n}oz}, Hu, and Deem}}]{Saakian06}
\bibinfo{author}{\bibfnamefont{D.~B.} \bibnamefont{Saakian}},
  \bibinfo{author}{\bibfnamefont{E.}~\bibnamefont{{Mu\~{n}oz}}},
  \bibinfo{author}{\bibfnamefont{C.-K.} \bibnamefont{Hu}}, \bibnamefont{and}
  \bibinfo{author}{\bibfnamefont{M.~W.} \bibnamefont{Deem}},
  \bibinfo{journal}{Phys. Rev. E} \textbf{\bibinfo{volume}{73}},
  \bibinfo{pages}{041913} (\bibinfo{year}{2006}).

\bibitem[{\citenamefont{Domingo et~al.}(1978)\citenamefont{Domingo, Sabo,
  Taniguchi, and Weissman}}]{Domingo78}
\bibinfo{author}{\bibfnamefont{E.}~\bibnamefont{Domingo}},
  \bibinfo{author}{\bibfnamefont{D.}~\bibnamefont{Sabo}},
  \bibinfo{author}{\bibfnamefont{T.}~\bibnamefont{Taniguchi}},
  \bibnamefont{and} \bibinfo{author}{\bibfnamefont{C.}~\bibnamefont{Weissman}},
  \bibinfo{journal}{Cell} \textbf{\bibinfo{volume}{13}}, \bibinfo{pages}{735}
  (\bibinfo{year}{1978}).

\bibitem[{\citenamefont{Domingo et~al.}(2005)\citenamefont{Domingo, Escarmis,
  Lazaro, and Manrubia}}]{Domingo05}
\bibinfo{author}{\bibfnamefont{E.}~\bibnamefont{Domingo}},
  \bibinfo{author}{\bibfnamefont{C.}~\bibnamefont{Escarmis}},
  \bibinfo{author}{\bibfnamefont{E.}~\bibnamefont{Lazaro}}, \bibnamefont{and}
  \bibinfo{author}{\bibfnamefont{S.~C.} \bibnamefont{Manrubia}},
  \bibinfo{journal}{Virus Res.} \textbf{\bibinfo{volume}{107}},
  \bibinfo{pages}{129} (\bibinfo{year}{2005}).

\bibitem[{\citenamefont{Ortin et~al.}(1980)\citenamefont{Ortin, Najera, Lopez,
  Davila, and Domingo}}]{Ortin80}
\bibinfo{author}{\bibfnamefont{J.}~\bibnamefont{Ortin}},
  \bibinfo{author}{\bibfnamefont{R.}~\bibnamefont{Najera}},
  \bibinfo{author}{\bibfnamefont{C.}~\bibnamefont{Lopez}},
  \bibinfo{author}{\bibfnamefont{M.}~\bibnamefont{Davila}}, \bibnamefont{and}
  \bibinfo{author}{\bibfnamefont{E.}~\bibnamefont{Domingo}},
  \bibinfo{journal}{Gene} \textbf{\bibinfo{volume}{11}}, \bibinfo{pages}{319}
  (\bibinfo{year}{1980}).

\bibitem[{\citenamefont{Domingo et~al.}(1985)\citenamefont{Domingo,
  Martinez-Salas, Sobrino, de~la Torre, Portela, Ortin, Lopez-Galindez, na,
  Villanueva, Najera et~al.}}]{Domingo85}
\bibinfo{author}{\bibfnamefont{E.}~\bibnamefont{Domingo}},
  \bibinfo{author}{\bibfnamefont{E.}~\bibnamefont{Martinez-Salas}},
  \bibinfo{author}{\bibfnamefont{F.}~\bibnamefont{Sobrino}},
  \bibinfo{author}{\bibfnamefont{J.~C.} \bibnamefont{de~la Torre}},
  \bibinfo{author}{\bibfnamefont{A.}~\bibnamefont{Portela}},
  \bibinfo{author}{\bibfnamefont{J.}~\bibnamefont{Ortin}},
  \bibinfo{author}{\bibfnamefont{C.}~\bibnamefont{Lopez-Galindez}},
  \bibinfo{author}{\bibfnamefont{P.~P.-B.} \bibnamefont{na}},
  \bibinfo{author}{\bibfnamefont{N.}~\bibnamefont{Villanueva}},
  \bibinfo{author}{\bibfnamefont{R.}~\bibnamefont{Najera}},
  \bibnamefont{et~al.}, \bibinfo{journal}{Gene} \textbf{\bibinfo{volume}{40}},
  \bibinfo{pages}{1} (\bibinfo{year}{1985}).

\bibitem[{\citenamefont{Eigen}(2002)}]{Eigen02}
\bibinfo{author}{\bibfnamefont{M.}~\bibnamefont{Eigen}},
  \bibinfo{journal}{Proc. Natl. Acad. Sci. USA} \textbf{\bibinfo{volume}{99}},
  \bibinfo{pages}{13374} (\bibinfo{year}{2002}).

\bibitem[{\citenamefont{Graci et~al.}(2007)\citenamefont{Graci, Harki,
  Korneeva, Edathil, Too, Franco, Smidansky, Paul, Peterson, Brown
  et~al.}}]{Graci07}
\bibinfo{author}{\bibfnamefont{J.~D.} \bibnamefont{Graci}},
  \bibinfo{author}{\bibfnamefont{D.~A.} \bibnamefont{Harki}},
  \bibinfo{author}{\bibfnamefont{V.~S.} \bibnamefont{Korneeva}},
  \bibinfo{author}{\bibfnamefont{J.~P.} \bibnamefont{Edathil}},
  \bibinfo{author}{\bibfnamefont{K.}~\bibnamefont{Too}},
  \bibinfo{author}{\bibfnamefont{D.}~\bibnamefont{Franco}},
  \bibinfo{author}{\bibfnamefont{E.~D.} \bibnamefont{Smidansky}},
  \bibinfo{author}{\bibfnamefont{A.~V.} \bibnamefont{Paul}},
  \bibinfo{author}{\bibfnamefont{B.~R.} \bibnamefont{Peterson}},
  \bibinfo{author}{\bibfnamefont{D.~M.} \bibnamefont{Brown}},
  \bibnamefont{et~al.}, \bibinfo{journal}{J. Virol.}
  \textbf{\bibinfo{volume}{81}}, \bibinfo{pages}{11256} (\bibinfo{year}{2007}).

\bibitem[{\citenamefont{Loeb et~al.}(1999)\citenamefont{Loeb, Essigmann,
  Kazazi, Zhang, and Rose}}]{Loeb99}
\bibinfo{author}{\bibfnamefont{L.~A.} \bibnamefont{Loeb}},
  \bibinfo{author}{\bibfnamefont{J.~M.} \bibnamefont{Essigmann}},
  \bibinfo{author}{\bibfnamefont{F.}~\bibnamefont{Kazazi}},
  \bibinfo{author}{\bibfnamefont{J.}~\bibnamefont{Zhang}}, \bibnamefont{and}
  \bibinfo{author}{\bibfnamefont{K.~D.} \bibnamefont{Rose}},
  \bibinfo{journal}{Proc. Natl. Acad. Sci. USA} \textbf{\bibinfo{volume}{96}},
  \bibinfo{pages}{1492} (\bibinfo{year}{1999}).

\bibitem[{\citenamefont{Loeb and Mullins}(2000)}]{Loeb00}
\bibinfo{author}{\bibfnamefont{L.~A.} \bibnamefont{Loeb}} \bibnamefont{and}
  \bibinfo{author}{\bibfnamefont{J.~I.} \bibnamefont{Mullins}},
  \bibinfo{journal}{{AIDS} {R}es. {H}um. {R}etroviruses}
  \textbf{\bibinfo{volume}{16}}, \bibinfo{pages}{1} (\bibinfo{year}{2000}).

\bibitem[{\citenamefont{Bull et~al.}(2007)\citenamefont{Bull, Sanjuan, and
  Wilke}}]{Bull07}
\bibinfo{author}{\bibfnamefont{J.~J.} \bibnamefont{Bull}},
  \bibinfo{author}{\bibfnamefont{R.}~\bibnamefont{Sanjuan}}, \bibnamefont{and}
  \bibinfo{author}{\bibfnamefont{C.~O.} \bibnamefont{Wilke}},
  \bibinfo{journal}{J. Virol.} \textbf{\bibinfo{volume}{81}},
  \bibinfo{pages}{2930} (\bibinfo{year}{2007}).

\bibitem[{\citenamefont{Baake et~al.}(1997)\citenamefont{Baake, Baake, and
  Wagner}}]{Baake97}
\bibinfo{author}{\bibfnamefont{E.}~\bibnamefont{Baake}},
  \bibinfo{author}{\bibfnamefont{M.}~\bibnamefont{Baake}}, \bibnamefont{and}
  \bibinfo{author}{\bibfnamefont{H.}~\bibnamefont{Wagner}},
  \bibinfo{journal}{Phys. Rev. Lett.} \textbf{\bibinfo{volume}{78}},
  \bibinfo{pages}{559} (\bibinfo{year}{1997}).

\bibitem[{\citenamefont{Baake et~al.}(1998)\citenamefont{Baake, Baake, and
  Wagner}}]{Baake98}
\bibinfo{author}{\bibfnamefont{E.}~\bibnamefont{Baake}},
  \bibinfo{author}{\bibfnamefont{M.}~\bibnamefont{Baake}}, \bibnamefont{and}
  \bibinfo{author}{\bibfnamefont{H.}~\bibnamefont{Wagner}},
  \bibinfo{journal}{Phys. Rev. E} \textbf{\bibinfo{volume}{57}},
  \bibinfo{pages}{1191} (\bibinfo{year}{1998}).

\bibitem[{\citenamefont{Saakian and Hu}(2006)}]{Saakian06b}
\bibinfo{author}{\bibfnamefont{D.~B.} \bibnamefont{Saakian}} \bibnamefont{and}
  \bibinfo{author}{\bibfnamefont{C.-K.} \bibnamefont{Hu}},
  \bibinfo{journal}{Proc. Natl. Acad. Sci. USA} \textbf{\bibinfo{volume}{103}},
  \bibinfo{pages}{4935} (\bibinfo{year}{2006}).

\bibitem[{\citenamefont{Saakian et~al.}(2004)\citenamefont{Saakian, Hu, and
  Kachatryan}}]{Saakian04a}
\bibinfo{author}{\bibfnamefont{D.~B.} \bibnamefont{Saakian}},
  \bibinfo{author}{\bibfnamefont{C.-K.} \bibnamefont{Hu}}, \bibnamefont{and}
  \bibinfo{author}{\bibfnamefont{H.}~\bibnamefont{Kachatryan}},
  \bibinfo{journal}{Phys. Rev. E} \textbf{\bibinfo{volume}{70}},
  \bibinfo{pages}{041908} (\bibinfo{year}{2004}).

\bibitem[{\citenamefont{Saakian and Hu}(2004)}]{Saakian04b}
\bibinfo{author}{\bibfnamefont{D.~B.} \bibnamefont{Saakian}} \bibnamefont{and}
  \bibinfo{author}{\bibfnamefont{C.-K.} \bibnamefont{Hu}},
  \bibinfo{journal}{Phys. Rev. E} \textbf{\bibinfo{volume}{70}},
  \bibinfo{pages}{021913} (\bibinfo{year}{2004}).

\bibitem[{\citenamefont{Park and Deem}(2007)}]{Park07}
\bibinfo{author}{\bibfnamefont{J.-M.} \bibnamefont{Park}} \bibnamefont{and}
  \bibinfo{author}{\bibfnamefont{M.~W.} \bibnamefont{Deem}},
  \bibinfo{journal}{Phys. Rev. Lett.} \textbf{\bibinfo{volume}{98}},
  \bibinfo{pages}{05101} (\bibinfo{year}{2007}).

\bibitem[{\citenamefont{Boerlijst et~al.}(1996)\citenamefont{Boerlijst,
  Bonhoeffer, and Nowak}}]{Boerlijst96}
\bibinfo{author}{\bibfnamefont{M.~C.} \bibnamefont{Boerlijst}},
  \bibinfo{author}{\bibfnamefont{S.}~\bibnamefont{Bonhoeffer}},
  \bibnamefont{and} \bibinfo{author}{\bibfnamefont{M.~A.} \bibnamefont{Nowak}},
  \bibinfo{journal}{P. Roy. Soc. London B} \textbf{\bibinfo{volume}{263}},
  \bibinfo{pages}{1577} (\bibinfo{year}{1996}).

\bibitem[{\citenamefont{Jacobi and Nordahl}(2006)}]{Jacobi06}
\bibinfo{author}{\bibfnamefont{M.~N.} \bibnamefont{Jacobi}} \bibnamefont{and}
  \bibinfo{author}{\bibfnamefont{M.}~\bibnamefont{Nordahl}},
  \bibinfo{journal}{Theor. {P}opul. {B}iol.} \textbf{\bibinfo{volume}{70}},
  \bibinfo{pages}{479} (\bibinfo{year}{2006}).

\bibitem[{\citenamefont{Lee and Cardy}(1995)}]{Lee95}
\bibinfo{author}{\bibfnamefont{B.~P.} \bibnamefont{Lee}} \bibnamefont{and}
  \bibinfo{author}{\bibfnamefont{J.}~\bibnamefont{Cardy}}, \bibinfo{journal}{J.
  Stat.Phys.} \textbf{\bibinfo{volume}{80}}, \bibinfo{pages}{971}
  (\bibinfo{year}{1995}).

\bibitem[{\citenamefont{Mattis and Glasser}(1998)}]{Mattis98}
\bibinfo{author}{\bibfnamefont{D.~C.} \bibnamefont{Mattis}} \bibnamefont{and}
  \bibinfo{author}{\bibfnamefont{M.~L.} \bibnamefont{Glasser}},
  \bibinfo{journal}{Rev. Mod. Phys.} \textbf{\bibinfo{volume}{70}},
  \bibinfo{pages}{979} (\bibinfo{year}{1998}).

\bibitem[{\citenamefont{Peliti}(1985)}]{Peliti85}
\bibinfo{author}{\bibfnamefont{L.}~\bibnamefont{Peliti}}, \bibinfo{journal}{J.
  Physique} \textbf{\bibinfo{volume}{46}}, \bibinfo{pages}{1469}
  (\bibinfo{year}{1985}).

\bibitem[{not({\natexlab{a}})}]{note}
\bibinfo{note}{There was a typo in [49] for the sharp peak fitness case: the
  formula reads $P_0 = 1 - \mu/A$ [instead of $u$, which is unity to $O(1/N)$].
  Similarly for the Eigen model $P_0 = (A e^{- \mu} - A_0)/(A-A_0)$ (rather
  than $u$, which is unity).}

\bibitem[{not({\natexlab{b}})}]{note2}
\bibinfo{note}{The alternative choice $u\xi_{c} < u^{2}$, combined with the
  self-consistency condition, leads to the equation $\max_{-1\leq\xi_{c}\leq
  1}f(\xi_{c}) = f(u)$, whose unique solution for the quadratic fitness is
  $|\xi_{c}|=|u|=1$, in contradiction with the assumption $u\xi_{c}<u^{2}$.
  Hence, we necessarily have $u\xi_{c} \ge u^{2}$.}

\bibitem[{\citenamefont{Bortz et~al.}(1995)\citenamefont{Bortz, Kalos, and
  Lebowitz}}]{Lebowitz75}
\bibinfo{author}{\bibfnamefont{A.~B.} \bibnamefont{Bortz}},
  \bibinfo{author}{\bibfnamefont{M.~H.} \bibnamefont{Kalos}}, \bibnamefont{and}
  \bibinfo{author}{\bibfnamefont{J.~L.} \bibnamefont{Lebowitz}},
  \bibinfo{journal}{J. Comput. Phys.} \textbf{\bibinfo{volume}{17}},
  \bibinfo{pages}{10} (\bibinfo{year}{1995}).

\bibitem[{\citenamefont{Gillespie}(1976)}]{Gillespie76}
\bibinfo{author}{\bibfnamefont{D.~T.} \bibnamefont{Gillespie}},
  \bibinfo{journal}{J. Comput. Phys.} \textbf{\bibinfo{volume}{22}},
  \bibinfo{pages}{403} (\bibinfo{year}{1976}).

\bibitem[{not({\natexlab{c}})}]{note3}
\bibinfo{note}{For the case $\phi_{c}=1$ there is an alternative solution
  $\eta_{c}=1$. This implies the equation
  $u\bar{\xi}_{c}=u[f'(\xi_{c})-d'(\xi_{c})]=0$, whose solution is $u=0$,
  unless there is an absolute maxima for $f(\xi_{c})-d(\xi_{c})$ in the
  interior of the region $|\xi_{c}|<1$, both of which possibilities are
  contained in Eq. (107).}

\bibitem[{\citenamefont{Sun and Deem}(2007)}]{Sun07}
\bibinfo{author}{\bibfnamefont{J.}~\bibnamefont{Sun}} \bibnamefont{and}
  \bibinfo{author}{\bibfnamefont{M.~W.} \bibnamefont{Deem}},
  \bibinfo{journal}{Phys. Rev. Lett.} \textbf{\bibinfo{volume}{99}},
  \bibinfo{pages}{228107} (\bibinfo{year}{2007}).

\end{thebibliography}
\end{document}